\documentclass[review]{siamonline1116}

\usepackage{lipsum}
\usepackage{amsfonts}
\usepackage{graphicx}
\usepackage{epstopdf}
\usepackage{algorithmic}

\ifpdf
\DeclareGraphicsExtensions{.eps,.pdf,.png,.jpg}
\else
\DeclareGraphicsExtensions{.eps}
\fi

\numberwithin{theorem}{section}

\newcommand{\TheTitle}{Scaled Gaussian Stochastic Process for Computer Model Calibration and Prediction}
\newcommand{\TheAuthors}{Mengyang Gu and Long Wang}

\headers{\TheTitle}{\TheAuthors}

\title{{\TheTitle}\thanks{
}}

\author{
	Mengyang Gu\thanks{Department of Applied Mathematics and Statistics, Johns Hopkins University, Baltimore, MD (\email{mengyang.gu@jhu.edu},
		\email{long.wang@jhu.edu}).}
	\and
	Long Wang\footnotemark[2]
}

\usepackage{amsopn}

\usepackage{amssymb, amsfonts, bm}
\usepackage{mathtools, mathrsfs}
\usepackage{relsize} 
\usepackage{hyperref}

\usepackage{graphicx,epstopdf} 
\usepackage[caption=false]{subfig} 

\newtheorem{example}{Example}

\DeclareMathOperator{\E}{E}

\DeclareMathOperator{\argmin}{argmin}

\newcommand{\GaSP}{\text{GaSP}}

 \newcommand{\noop}[1]{}
\newcommand{\deltabm}{\bm{\delta}}
\newcommand{\gammabm}{\bm{\gamma}}
\newcommand{\deltabmz}{\bm{\delta}_z}

\newcommand{\Lambdabm}{\bm{\Lambda}}
\newcommand{\mubm}{\bm{\mu}}
\newcommand{\thetabm}{\bm{\theta}}
\newcommand{\Thetabm}{\bm{\Theta}}
\newcommand{\xibm}{\bm{\xi}}
\newcommand{\zerobm}{\bm{0}}

\newcommand{\Bbf}{\mathbf{B}}
\newcommand{\Ibf}{\mathbf{I}}
\newcommand{\rbf}{\mathbf{r}}
\newcommand{\Rbf}{\mathbf{R}}
\newcommand{\Ubf}{\mathbf{U}}
\newcommand{\xbf}{\mathbf{x}}

\newcommand{\Xcal}{\mathcal{X}}

\newcommand{\inv}[1]{\parenth{#1}^{-1}}
\newcommand{\parenth}[1]{\left(#1\right)}
\newcommand{\sqbracket}[1]{\left[#1\right]}
\newcommand{\trans}[1]{\parenth{#1}^T}

\newcommand{\Vol}{\text{Vol}}

\newcommand{\xiinX}{{\xibm\in\Xcal}}

\newcommand{\smallintdelta}{\mathsmaller{\int}_\xiinX\delta(\xibm)^2d\xibm=z}

\newcommand{\sumdelta}{\sum_{i=1}^{N_C}\delta(\xbf_i^C)^2\Delta x=z}
\newcommand{\sumdeltat}{\sum_{i=1}^{N_C}\delta(\xbf_i^C)^2\Delta x=t}

\newcommand{\bzero}{b_0(\Thetabm)}
\newcommand{\bone}{b_1(\deltabmz,\Thetabm)}
\newcommand{\bzeroa}{b_0^a(\Thetabm)}
\newcommand{\bonea}{b_1^a(\bm \delta^a_z,\Thetabm)}

\usepackage{color,soul}
\soulregister\cite7
\soulregister\citep7
\soulregister\ref7
\soulregister\pageref7

\ifpdf
\hypersetup{
 pdftitle={\TheTitle},
 pdfauthor={\TheAuthors}
}
\fi

\begin{document}

\maketitle


\begin{abstract}
	We consider the problem of calibrating an imperfect computer model using experimental data. To compensate the misspecification of the computer model and make more accurate predictions, a discrepancy function is often included and modeled via a Gaussian stochastic process (GaSP). The calibrated computer model alone, however, sometimes fits the experimental data poorly, as the calibration parameters become unidentifiable. In this work, we propose the scaled Gaussian stochastic process (S-GaSP), a novel stochastic process that bridges the gap between two predominant methods, namely the $L_2$ calibration and the GaSP calibration. It is shown that our approach performs well in both calibration and prediction. A computationally feasible approach is introduced for this new model under the Bayesian paradigm. Compared with the GaSP calibration, the S-GaSP calibration enables the calibrated computer model itself to predict the reality well, based on the posterior distribution of the calibration parameters. Numerical comparisons of the simulated and real data are provided to illustrate the connections and differences between the proposed S-GaSP and other alternative approaches.
\end{abstract}



\section{Introduction} 
\label{sec:Intro}

Computer models or simulators are increasingly used to reproduce the behavior of complex systems in physics, engineering and human processes. These models are essentially computer implementations of mathematical models to generate outputs based on a collection of inputs, such as initial conditions or model parameters. Some model parameters, however, are unknown or unobservable in experiments. One of the fundamental tasks in uncertainty quantification is to adjust the unknown parameters until the outputs of the model fit the observed data, often referred as the \textit{model calibration} or \textit{inverse problem} \cite{kennedy2001bayesian}.

Assume a set of field data from experiments is collected at $\mathbf x_i$, denoted as $y^F(\mathbf x_i)$ for $i=1,...,n$. The computer model outputs, defined as $f^M(\mathbf x, \bm \theta)$, are evaluated at the variable input $\mathbf x \in \mathcal X$ and calibration parameter $\bm \theta$. For simplicity, $\mathcal X$ is assumed to be a bounded rectangle in $\mathbb R^{p_x} $ and $\bm \theta \in \mathbb R^{p_{\theta}} $. If the computer model has no bias to the reality, meaning that the field data is a noisy realization of the computer model for some set of parameters, the calibration is to choose $\bm \theta$ that minimizes the distance between the field data and outputs of the computer model.

In practice, a perfect computer model to the reality is rarely the case. It is common to address the model misspecification by a discrepancy function, such that the reality can be represented as $y^R(\mathbf x)=f^M(\mathbf x, \bm \theta)+\delta(\mathbf x)$, where $y^R(\cdot)$ and $\delta(\cdot)$ denote the reality and discrepancy function, respectively. It leads to the following statistical model for calibration,
\begin{equation}
y^F(\mathbf x)=f^{M}(\mathbf x, \bm \theta)+\delta(\mathbf x)+\epsilon,
\label{equ:model_calibration}
\end{equation}
where $\epsilon$ is a zero-mean noise. This model is investigated in a vast amount of literature \cite{bayarri2007framework, higdon2008computer, liu2009modularization,qian2008bayesian}. Unfortunately, many following-up studies found that the use of \cref{equ:model_calibration} sometimes results in an identifiability problem of $\bm \theta$ \cite{arendt2012quantification, arendt2012improving, plumlee2016bayesian,tuo2016theoretical}.

To some extent, this identifiability problem is inherently rooted in \cref{equ:model_calibration}. Suppose that the field data is noise-free and the discrepancy function is defined as $\delta_{\bm \theta}(\mathbf x)= y^F(\mathbf x)- f^{M}(\mathbf x, \bm \theta)$, then the reality can always be modeled perfectly well regardless of the choice of $ \bm{\theta} $. Much effort has been made to eliminate the identifiability issue recently. We briefly review two popular approaches in \Cref{subsec:L2} and \Cref{subsec:GaSP}.

\subsection{$L_2$ calibration}
\label{subsec:L2}
In \cite{tuo2016theoretical} and \cite{tuo2015efficient}, the ``optimal" calibration parameter $ \bm \theta $ is defined as the one that minimizes the $L_2$ norm of the discrepancy function (henceforth the $L_2$ calibration), i.e.,
\begin{equation}
{\bm \theta}_{L_2} =\underset{\bm \theta }{\argmin} || \delta_{\bm \theta}(\cdot) ||_{L_2(\mathcal X)}= \underset{\bm \theta}{\argmin} \left\{\int_{\mathbf x\in \mathcal X} \left[ y^R(\mathbf x)- f^{M}(\mathbf x, \bm \theta) \right]^2d \mathbf x\right\}^{1/2}. 
\label{equ:L2_loss_minimizer}
\end{equation}

Since $ y^R(\cdot) $ is not observable due to the noise in the experimental data, \cite{tuo2015efficient} proposes to first obtain an estimate $ \hat{y}^R(\cdot) $ of the reality $ y^R(\cdot) $ via a Gaussian stochastic stochastic process and then plug it into \cref{equ:L2_loss_minimizer} to get the $L_2$ calibration estimator $ \hat{\thetabm}_{L_2} $. This approach is $ \sqrt{n} $-consistent and semi-parametric efficient, which provides an optimal estimator for $ \thetabm_{L_2} $. Besides the nice theoretical properties, the $ L_2 $ calibration forces the computer model to explain the variability of the reality as much as possible, and the calibrated computer model often fits the reality well, if the $ \hat{y}^R(\cdot) $ is an accurate estimator of $y^R(\cdot) $ given a finite number of observations.

A related method to $ L_2 $ calibration is the least squares (LS) estimator, which minimizes the squared error between the experimental data and computer model, i.e.,
\begin{equation}
	\hat {\bm \theta}_{LS} = \underset{\bm \theta }{\argmin} \sum^n_{i=1} \left[y^F(\mathbf x_i) - f^M(\mathbf x_i, \bm \theta)\right]^2. 
	\label{equ:least_squares}
\end{equation}
It is shown in \cite{tuo2015efficient} and \cite{wong2017frequentist} that $\hat {\bm \theta}_{LS}$ also converges to ${\bm \theta}_{L_2}$ in probability under some mild conditions, but it is generally less efficient than the $ L_2 $ calibration unless the computer model is perfect. The LS calibration is also used in \cite{wong2017frequentist} as a plug-in estimator for estimating the discrepancy function via a nonparametric regression.



Both approaches have limitations in predicting the reality when the number of observations is not large. Specifically, the approach in \cite{tuo2015efficient} estimates the reality $ y^R(\cdot) $ without the computer model, which contradicts the basic assumption on the usefulness of the computer model in reproducing the reality. On the other hand, the approach in \cite{wong2017frequentist} estimates $ \thetabm $ without penalizing the complexity in residuals, which sometimes makes it hard to capture the residuals $y^R(\cdot)-f^M(\cdot, \hat {\bm \theta}_{LS} )$ by the nonparametric regression model. A simulated example is given in Section~\ref{subsec:simulation} to illustrate the differences between our approach and these approaches.

\subsection{GaSP calibration}
\label{subsec:GaSP}

In \cite{kennedy2001bayesian}, the discrepancy function is specified as a Gaussian stochastic process (GaSP) (henceforth the GaSP calibration). As modeling the discrepancy through a GaSP provides a flexible sampling model for the experimental data, the prediction of the unobserved field data, based on {both} the calibrated computer model and discrepancy function, is empirically better than using the calibrated computer model alone. Despite this benefit, modeling the discrepancy function through a GaSP sometimes overwhelms the effects of the computer model when the residuals between the experimental data and computer model outputs are correlated, resulting in an identifiability issue of the calibration parameters. 






Here we provide an example to illustrate the identifiability problem in the GaSP calibration when the data is correlated. Consider a simple case where the computer model contains only a mean parameter, i.e., $ f^M(x, \theta) = \theta $, with the true parameter being $ \theta^*=0 $. Assume one obtains $n=200$ observations, denoted as $ \mathbf y^F=(y^F(x_1),..., y^F(x_{200}))^T $, sampled from a zero-mean multivariate normal distribution at $ x_i = (i-1)/199$ for $ i = 1, ..., 200 $
\begin{equation}
	\mathbf y^F\sim MN(\mathbf 0, \sigma^2_{\delta} \mathbf R), 
	\label{equ:multi_norm_y}
\end{equation}
with {$ \sigma_\delta^2 = 1 $} and the $(i,j)$ term of $\mathbf R$ being $R_{i,j}=\exp[- (|x_i-x_j|/\gamma^{\delta})^{1.9}]$ for some $ \gamma^\delta $. Here the 
symbol $\delta$ is used to indicate the covariance is used for modeling the discrepancy function. For simplicity, assume no prior information is available and only two values of the calibration parameter are considered, i.e., $ \theta = 0 $ and $ \theta = 1 $, corresponding to the true model and misspecified model, respectively.


\begin{figure}[tbhp]
	\centering
	\subfloat{\includegraphics[width=0.5\textwidth]{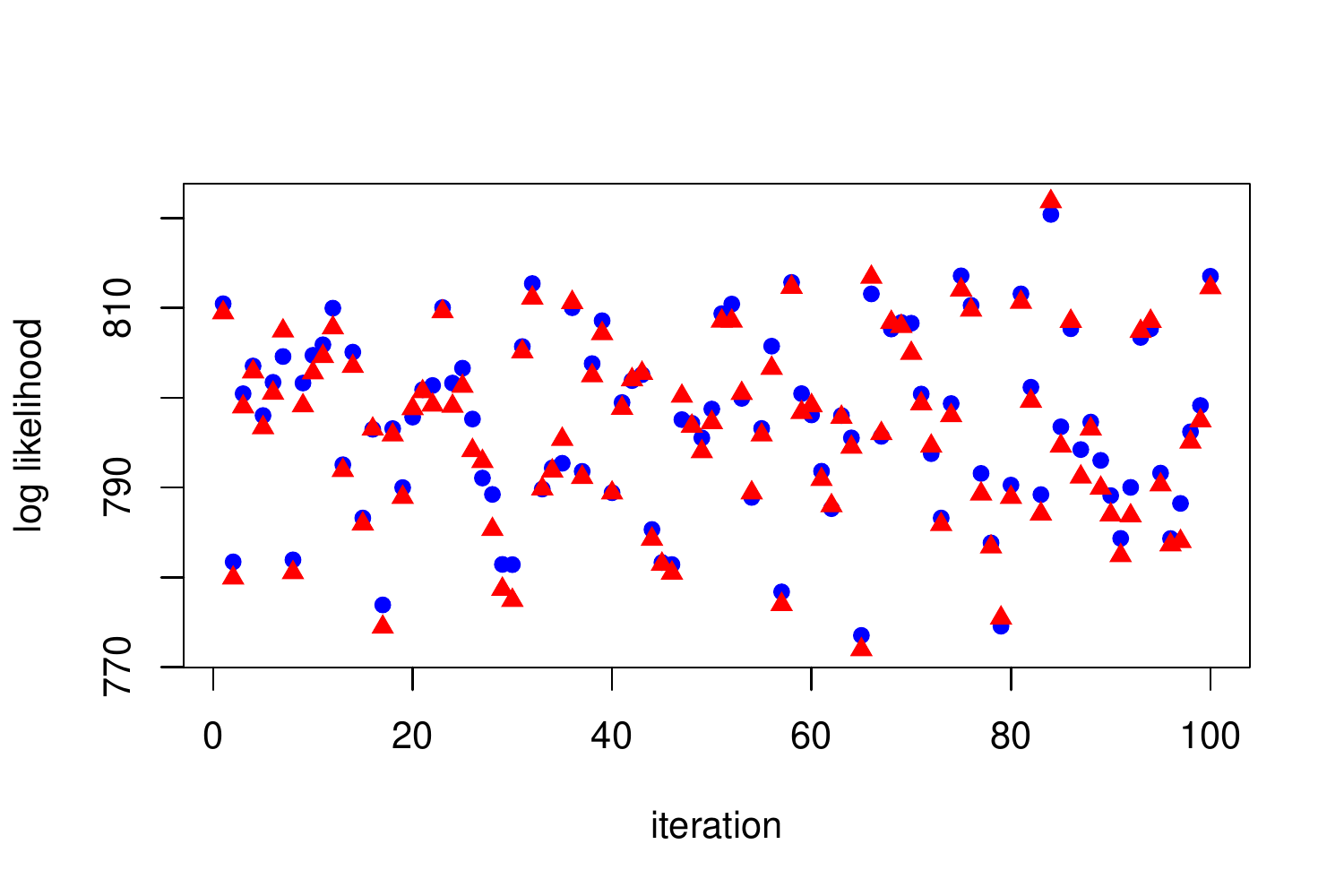}}
	\subfloat{\includegraphics[width=0.5\textwidth]{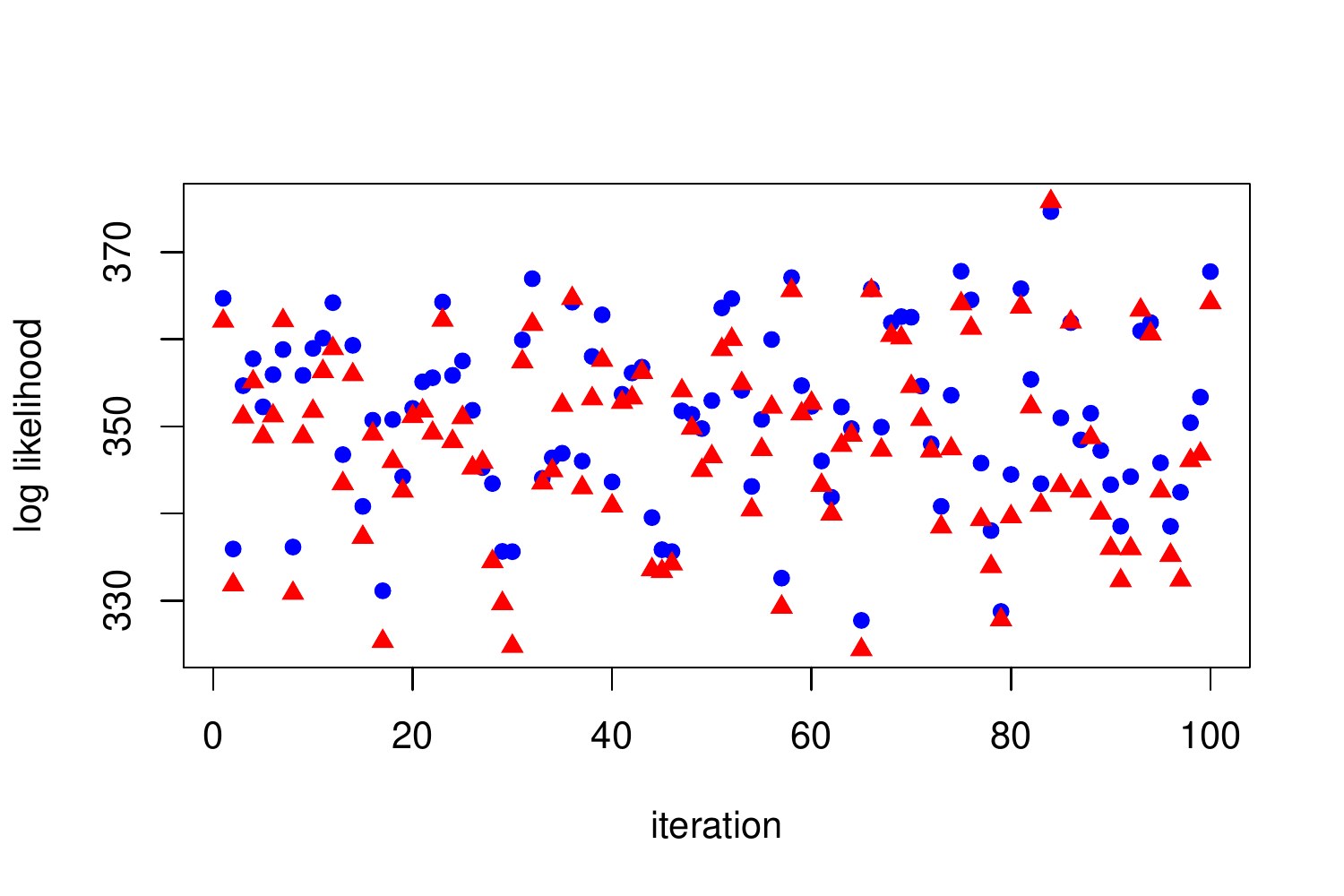}} \vspace{-.25in}\\
	\subfloat{\includegraphics[width=0.5\textwidth]{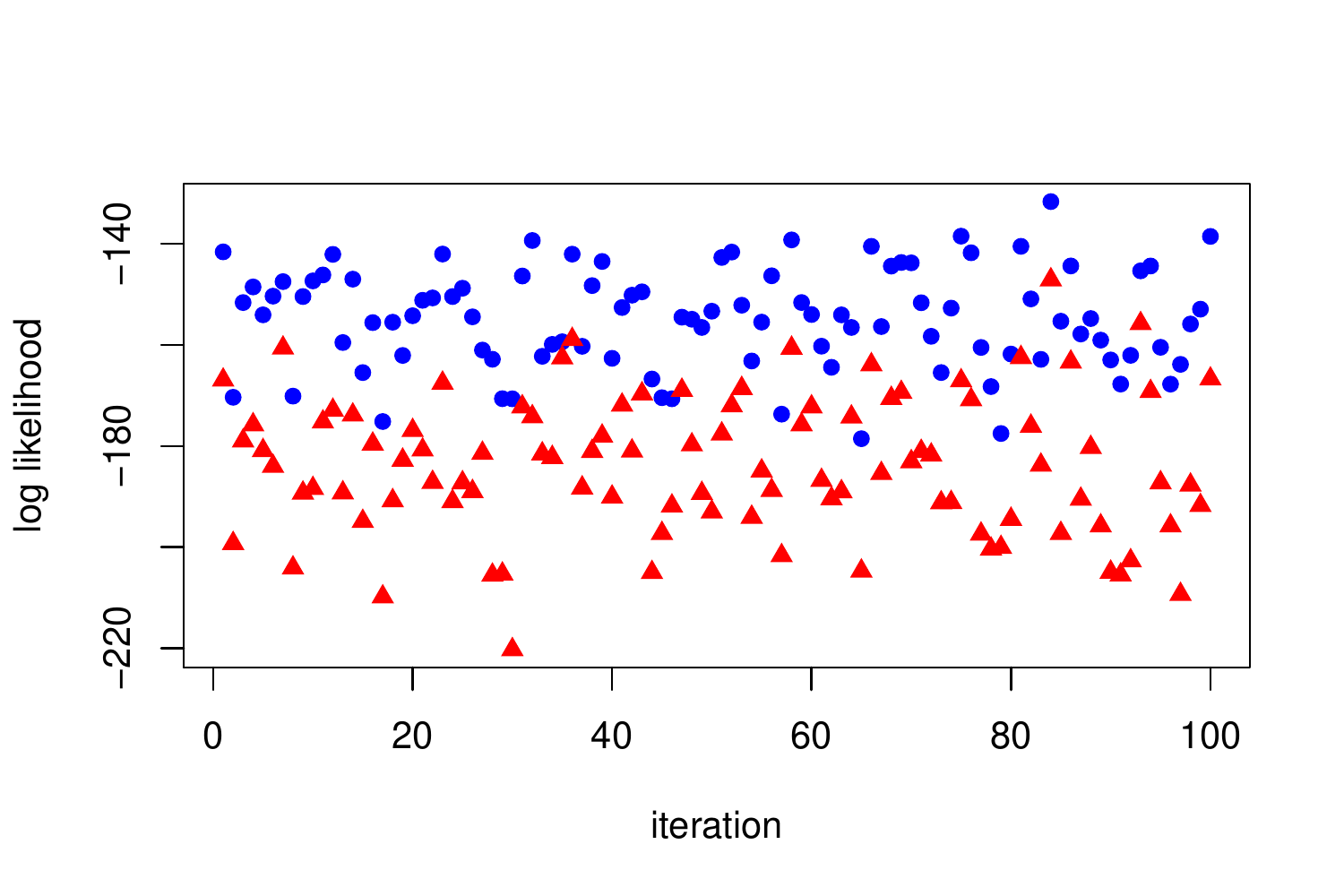}}
	\subfloat{\includegraphics[width=0.5\textwidth]{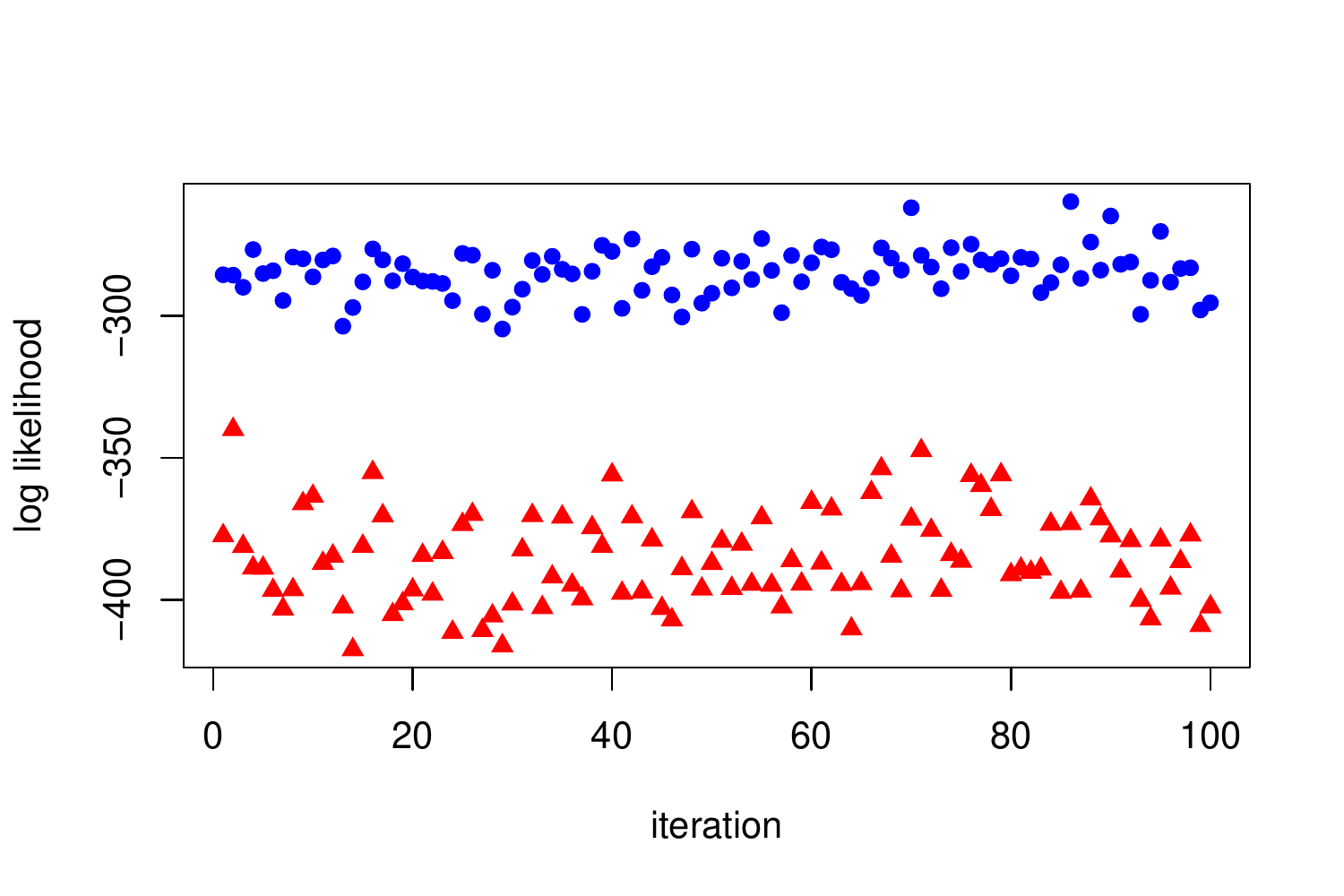}}
	\caption{The log-likelihood functions $ \ell(\theta):=\log(p(\mathbf y^F\,|\, \theta ))$ at different values of correlation. The observations are sampled from a zero-mean multivariate normal distribution with the $(i,j)$ entry of the covariance being $\exp[- (|x_i-x_j|/\gamma^{\delta})^{1.9}]$. A hundred simulations are implemented in each case with $\gamma^{\delta}=\{1,1/10,1/100\}$ for the upper left, upper right and lower left panels, respectively. The independent case is graphed in the lower right panel. The blue dots are $\ell(\theta=0)$ (true model) and the red triangles are  $\ell(\theta=1)$ (misspecified model) in each simulation. The average differences of the log-likelihood between the true model and misspecified model are $0.93$, $3.50$, $28.91$ and $100.41$ for these four cases.}
	\label{fig:samples_GP}
\end{figure}






Under different $ \gamma^\delta $, the natural logarithm of the likelihood (log-likelihood) functions of these two models from 100 simulations are graphed in \cref{fig:samples_GP}. When the data is correlated, the log-likelihoods between these two models are extremely close, making them barely identifiable. Specifically, for $ \gamma^{\delta}=1 $ and $ \gamma^{\delta}=1/10 $, the average log-likelihoods of the true model are only $0.93$ and $3.50$ larger than the misspecified model, respectively. In contrast, it is easy to show the expected difference of the log-likelihood between $\theta=0$ and $\theta=1$ is $1$ when there are $n=2$ independent samples. The average difference of the log-likelihoods between the true and misspecified models with $n=200$ correlated observations under $\gamma^{\delta}=1$ is even smaller than the one with $n=2$ independent samples.


Although the GaSP model provides a flexible sampling model of the reality, the likelihood function of the GaSP model is very flat on the space of $\theta$ when the observations are moderately correlated. Consequently, two different $ \theta $ with almost the same probability masses could have very different $L_2$ losses (the squared $L_2$ norm of the discrepancy function). {The motivation of this study is to develop a stochastic process that not only is flexible enough for modeling the reality but also maintains an adequate amount of probability mass on the small $ L_2 $ loss, especially when data are correlated.}

\subsection{Our contribution}
\label{subsec:contribution}
In this work, we introduce a new stochastic process, called the scaled Gaussian stochastic process (S-GaSP), that reconciles the differences between the $L_2$ calibration and GaSP calibration. We derive a closed-form likelihood of the new process, which makes the computation feasible. Under the Bayesian framework, we provide the full uncertainty quantification of the calibration parameters through their posterior distributions.   

We evaluate the results of the calibration by the following two criteria.
\begin{itemize}
		\item Using both the calibrated computer model and discrepancy function, the proposed approach should predict the reality precisely.
		\item Using the calibrated computer model alone, the proposed approach should fit the data and predict the reality well.
\end{itemize} 

The first criterion requires an appropriate model for the discrepancy function, since we rely on the discrepancy function to improve the prediction when the computer model is misspecified. The second criterion is from the $L_2$ calibration, as the computer model tends to be better calibrated and more interpretable, when it can predict the reality reasonably well.

The interpretability of the calibration parameter in the computer model depends on the specific background of the scientific problem, but an interpretable calibration parameter should allow the computer model to predict the reality relatively well even without the discrepancy function. The computer model, along with some linear order discrepancy terms, is interpretable by the scientists, whereas the nonlinear effects from the discrepancy function can be hard to interpret. Therefore, we define the interpretability of the calibration parameters by measuring the predictive power {of} the calibrated computer model without a discrepancy function. We demonstrate below that the proposed S-GaSP calibration performs better than the previous approaches based on these two criteria. The method introduced in this work is implemented in a new R package on CRAN \cite{Gu2018RobustCalibrationpackage}.

The remainder of the paper is organized as follows. We introduce the new calibration approach in \Cref{sec:model} with a focus on the connection to the previous approaches. The closed-form marginal and predictive distributions of our new process are also derived. The method is extended to slow computer models in \Cref{sec:calibrate_slow}. Detailed discussions on parameter estimation and computation are covered in \Cref{sec:computation}. {Simulated and real data are studied in \Cref{sec:numerical}}. We conclude this work in \Cref{sec:conclusion}.

\section{Calibration and prediction by the scaled Gaussian stochastic process}
\label{sec:model}

We introduce a new approach for computer model calibration in this section. The scaled Gaussian stochastic process is introduced in \Cref{subsec:S-GaSP}. In \Cref{subsec:choice_Z}, we discuss the default choice of a key random variable in the new stochastic process, along with the {connections} to the previous approaches. \Cref{subsec:finite_S-GaSP} provides an efficient way to compute the marginal and predictive distributions. The comparison to the orthogonal Gaussian stochastic process is illustrated in \Cref{subsec:O-GaSP}. The proofs of the lemmas are provided in \cref{sec:appendix}. 

\subsection{Scaled Gaussian stochastic process}
\label{subsec:S-GaSP}

We propose using the following model for calibration, 
\begin{equation}
	\label{equ:scaled_GP}
	\begin{split}
		&y^F(\mathbf x)=f^{M}(\mathbf x, \bm \theta)+\mu^{\delta}(\mathbf x)+ \delta_z(\mathbf x) +\epsilon, \\
		& \delta_z(\mathbf x) = \left\{\delta(\mathbf x) \mid { \mathsmaller{\int}_{\bm{\xi}\in\mathcal X} }  \delta( \bm \xi )^2 d\bm \xi=Z \right\}, \\
		&\delta(\cdot) \sim \text{GaSP}(0,  \sigma^2_{\delta} c^{\delta}(\cdot, \cdot)),\\
		&Z\sim p_{\delta_z}(\cdot), \, \epsilon \sim N(0, \sigma^2_0 ).
	\end{split}
\end{equation}
We call $\delta_z(\cdot)$ follows the {scaled Gaussian stochastic process} (S-GaSP). The {innovation} of the above model comes from the random variable $Z$, whose distribution is discussed in \Cref{subsec:choice_Z}. Given $Z=z$, the new process $\delta_z(\cdot)$ is the GaSP $\delta(\cdot)$ constrained at the space of $\int_{\bm{\xi}\in\mathcal X} \delta(\bm \xi)^2 d\bm \xi=z$. 

In \cref{equ:scaled_GP}, $\mu^{\delta}(\cdot)$ is a mean discrepancy, typically modeled by regression, 
\begin{equation}
	\mu^{\delta}(\mathbf x)= {\mathbf h^{\delta}(\mathbf x)}\bm \beta^{\delta}=\sum^{q_{\delta}}_{i=1} h^{\delta}_i(\mathbf x) \beta^{\delta}_i,
	\label{equ:mean_dis}
\end{equation}
where $\mathbf h^{\delta}(\mathbf x)=( h^{\delta}_1(\mathbf x),h^{\delta}_2(\mathbf x), ..., h^{\delta}_{q_{\delta}}(\mathbf x) )$ is a {known} $q_{\delta}$-dimensional vector of basis functions and $\bm \beta^{\delta}=(\beta^{\delta}_1,\beta^{\delta}_2,...,\beta^{\delta}_{q_{\delta}})^T$ {is an unknown $q_{\delta}$-dimensional vector} with each $\beta^{\delta}_i$ being the regression parameter of $h^{\delta}_i(\mathbf x)$ for $i=1,...,q_{\delta}$.

As the mean discrepancy only contains the intercept and linear order terms {which are easy to interpret}, we treat it as a part of the computer {model}. The mean discrepancy benefits the prediction when the computer model does not contain the intercept or {does not} properly explain the trend with regard to the variable inputs. However, $\mu^{\delta}(\cdot)$ should be zero when the intercept and trend are properly modeled in the computer model.

 We assume $\delta(\cdot)$ follows a zero-mean GaSP, meaning that the density of any $\{\mathbf x_1,...,\mathbf x_n\}$ takes the form of a multivariate normal
\begin{equation}
	(\delta(\mathbf x_1),..., \delta(\mathbf x_n) )^T \mid \mathbf R^{\delta} \sim \text{MN}( \bm 0, \sigma^2_{\delta} \mathbf R^{\delta}),
	\label{equ:multivar}
\end{equation}
where $\mathbf R^{\delta}$ is a correlation matrix such that its $(i,j)$ entry is defined as $R^{\delta}_{i,j}=c^{\delta}(\mathbf x_i, \mathbf x_j)$.


For any inputs $\mathbf x_a, \mathbf x_b \in \mathcal X$, the correlation is typically assumed to {have} a product form
\begin{equation}
	c^{\delta}(\mathbf x_a, \mathbf x_b)= \prod_{l=1}^{p_x }c^{\delta}_l( x_{al}, x_{bl}),
	\label{equ:product_c}
\end{equation} 
where each $c^{\delta}_l(\cdot,\cdot)$ is a one-dimensional correlation function for the $l^{th}$ coordinate of the input space. Power exponential correlation and Mat{\'e}rn correlation are widely used in GaSP models. {The power exponential correlation has the following form
	\begin{equation}
		c^{\delta}_l(d_l)=\exp\left\{ - \left(\frac{d_l}{\gamma^{\delta}_l}\right)^{\nu^{\delta}_l} \right\},
		\label{equ:Pow_exp}
	\end{equation}
	where $d_l=|x_{al}- x_{bl}|$ is the distance of the $l^{th}$ coordinate of the input vectors. $\nu^{\delta}_l$ is a roughness parameter typically held fixed and $\gamma^{\delta}_l$ is an unknown range parameter to be estimated. }

The Mat{\'e}rn correlation has recently gained more interest in constructing GaSP emulators \cite{Gu2018robustness}. The Mat{\'e}rn correlation with the roughness parameter $\nu^{\delta}_l=(2k+1)/2$ for $k \in \mathbb N$ has a closed-form expression. For instance, when $\nu^{\delta}_l = 5/2$, the Mat{\'e}rn correlation is as follows
\begin{equation}
	c^{\delta}_l(d_l)=\left(1+\frac{\sqrt{5}d_l}{\gamma^{\delta}_l}+\frac{5d_l^2}{3(\gamma^{\delta}_l)^2}\right)\exp\left(-\frac{\sqrt{5}d_l}{\gamma^{\delta}_l}\right),
	\label{equ:Matern}
\end{equation}
where $\gamma^{\delta}_l$ is an unknown range parameter. This covariance function is widely used in modeling spatially correlated data \cite{stein2012interpolation} and is {becoming} popular in the field of uncertainty quantification. It is also the default choice of several R packages for computer models \cite{gu2018robustgasp_package,roustant2012dicekriging}. We use this correlation function herein for the demonstration purpose, and the method is applicable to any covariance function of interest.

Let $ \bm \gamma^{\delta} = (\gamma_1^\delta, \dots, \gamma_{p_x}^\delta)^T $. The parameters in the S-GaSP model are denoted as $\bm \Theta$ {that contain $\{\bm \theta; \bm \beta^{\delta}; \bm \gamma^{\delta}; \sigma^2_{\delta}; \sigma^2_0\}$ and some possible parameters in the scaling {density} $p_{\delta_z}(\cdot)$}. Recall that $Z= {{\int}_{\mathbf x \in\mathcal X} }  \delta( \mathbf x )^2 d\mathbf x$. Any marginal distribution of the S-GaSP at $\bm \delta_z= \left(\delta_z(\mathbf x_1),...,\delta_z(\mathbf x_n) \right)^T$ for $\{\mathbf x_1,...,\mathbf x_n\}$ can be computed by marginalizing out $Z$ as follows
\begin{align}
	p_{\delta_z}\left(\bm \delta_z \mid \bm \Theta \right)\nonumber
	&=  \int_0^\infty p_{ \delta} \left(\bm \delta_z \mid Z =z,\bm\Theta \right) p_{\delta_z}(Z=z\mid\bm \Theta) dz \nonumber \\
	&= {p_{ \delta} \left( \bm \delta_z \mid \bm\Theta\right)} \int_0^\infty \frac{ p_{ \delta} \left( Z=z \mid \bm \delta_z , \bm\Theta \right) } {p_{ \delta}\left(Z =z \mid\bm\Theta \right)} p_{\delta_z}(Z=z\bm \mid\bm{\Theta}) dz,
	\label{equ:integrate_z}
\end{align}
where $p_{ \delta}(\bm \delta_z \mid\bm \Theta)$ is the multivariate normal density in \cref{equ:multivar} evaluated at $\bm \delta_z$.

By the properties of the multivariate normal distribution, it follows that   
\[
	\delta(\cdot) \mid \bm \delta_z , \bm\Theta \sim \text{GaSP} (\mu^{*\delta}(\cdot), \sigma^2_{\delta} c^{*\delta}(\cdot, \cdot)),
\]
where for any $\mathbf x^* \in \mathcal X$,
\begin{equation}
	{ { \mu}^{*\delta}(\mathbf x^*)= \mathbf r^{\delta}(\mathbf x^*)^T \left(\mathbf { R}^{\delta}\right)^{-1} \bm \delta_z },
	\label{equ:pred_mean_gasp}
\end{equation}
with $\mathbf r^{\delta}(\mathbf x^*) = ({c}^{\delta}(\mathbf x^*, \mathbf x_1), ..., {c}^{\delta}(\mathbf x^*, \mathbf x_n))^T$ and for any $\mathbf x^*_a, \mathbf x^*_b \in \mathcal X$,
\begin{equation}
	{{ c}^{*\delta}(\mathbf x^*_a, \mathbf x^*_b)= c^{\delta}(\mathbf x^*_a, \mathbf x^*_b )- \mathbf r^{\delta}(\mathbf x^*_a)^T \left(\mathbf {R}^{\delta}\right)^{-1} \mathbf r^{\delta}(\mathbf x^*_b).}
	\label{equ:c_star_gasp}
\end{equation}

It is clear that $ \delta(\cdot) \,|\, \Thetabm$ and $ \delta(\cdot)\,|\, \bm \delta_z, \Thetabm $ are both GaSPs with different means and covariance functions. We have the following lemma for computing the density in \cref{equ:integrate_z}. 
\begin{lemma}
	For  $\delta (\cdot) \sim \text{GaSP}(\mu(\cdot), \sigma^2 c(\cdot,\cdot) )$ defined on $\mathcal X$, if $||\mu(\cdot) ||_{L_2(\mathcal X) }<\infty$, one has 
	\[
	\int_{\mathbf x \in \mathcal X} \delta(\mathbf x)^2 d\mathbf x \sim \sigma^2 \sum^{\infty}_{i=1} \lambda_i \chi^2_{a_i}(1),
	\]
	where $\chi^2_{a_i}(1)$ is a non-central chi-squared distribution with 1 degree of freedom and the non-central parameter $a_i = \mu_i^2 / (\lambda_i\sigma^2)$ with $ \mu_i = \int_{\mathbf x \in \mathcal X} \mu(\mathbf x) \phi_i(\mathbf x)d\mathbf x$, $\lambda_i$ being the $i^{th}$ eigenvalue and  $ \phi_i $ being the $i^{th}$ normalized eigenfunction of  $c(\cdot,\cdot)$ with regard to the Lebesgue measure.
	\label{lemma:chi-square}
\end{lemma}

Based on \cref{lemma:chi-square}, {both} $p_{ \delta}\left(Z=z \mid \bm \delta_z, \bm\Theta \right)$ and $p_{ \delta}\left(Z=z \mid \bm\Theta \right)$ in \cref{equ:integrate_z} are {the densities} of an infinite weighted sum of non-central chi-squared distributions, which can be approximated by discretization. Direct {calculation} of the {densities} of the infinite weighted sum of chi-squared distributions, however, is computationally challenging and could lead to a large approximation {error}. Later we introduce a more robust way to evaluate the likelihood of the S-GaSP model along with the default choice of $p_{\delta_z}(\cdot)$ in the following \Cref{subsec:choice_Z}.

\subsection{Choice of $p_{\delta_z}(\cdot)$}
\label{subsec:choice_Z}
The pivotal part of the S-GaSP model is the measure of the random variable $Z=\int_{\mathbf x \in \Xcal} \delta(\mathbf x)^2 d\mathbf x$, {which is also} the $L_2$ loss between the reality and computer model. Unlike the $L_2$ calibration, we do not assume the model with the smallest $L_2$ loss is necessarily the best model. Instead, we put a prior on all the values, but favor the one with the smaller $ L_2 $ loss, since the $ L_2 $ loss is  a good indicator of a well calibrated computer model. On the other hand, the widely used GaSP calibration approach implicitly places a prior density for $Z=z$, denoted as $p_{\delta}(Z=z \mid \bm \Theta)$, given the parameters $\bm \Theta$. 
{Therefore,} to combine these two ideas, we define the default choice of $p_{\delta_z}(\cdot)$ as follows: 
\begin{equation}
	p_{\delta_z}(Z=z {\mid} \bm \Theta) =
	\frac{f_Z\left(Z=z \mid \bm \Theta\right) p_{ \delta}\left(Z=z \mid \bm \Theta\right)}
	{\int_0^\infty f_Z\left( Z=t \mid \bm \Theta\right) p_{ \delta}\left(Z = t \mid \bm \Theta \right)d t },
	\label{equ:p_z}
\end{equation}
where $f_Z(z \mid \bm \Theta)$ is a non-increasing function. In this specification, the density for $Z$ is chosen to be proportional to {the} GaSP prior for $Z$ (as it is used widely in the previous literature), but scaled by a scaling function $f_Z(\cdot \mid \bm \Theta)$ to avoid the sample path deviating too much from zero. We present the following lemma that connects the S-GaSP calibration and GaSP calibration.

\begin{lemma}[Connection between the GaSP and S-GaSP model]
	When $p_{\delta_z}(Z=z \mid \bm \Theta)$ is specified in \cref{equ:p_z} and $f_Z(Z=z \mid \bm \Theta)$ is a non-zero constant for all $z\in [0, \, +\infty)$, the S-GaSP in \cref{equ:scaled_GP} becomes the calibration model in \cref{equ:model_calibration} where $\delta(\cdot) \sim GaSP(\mu^{\delta}(\cdot), \sigma^2_{\delta}c^{\delta}(\cdot, \cdot))$ and $\epsilon\sim N(0,\sigma^2_0)$.
	\label{lemma:connection}
\end{lemma}

With the assumption in \cref{equ:p_z}, the marginal density for any {$\deltabmz$} in the S-GaSP model in \cref{equ:integrate_z} becomes
\begin{align}
	p_{{\delta_z}}(\bm \delta_z \mid \bm \Theta )=\frac{ b_1(\bm \delta_z,\bm \Theta)}{b_0(\bm \Theta)} {p_{ \delta}\left(\bm \delta_z \mid \bm\Theta\right)},
	\label{equ:integrate_z_assume_f}
\end{align}
where
\begin{align}
	\label{equ:b0}
	b_0(\bm \Theta)
	&= \int_0^\infty p_{ \delta}\left(Z = t \mid \bm \Theta \right) f_Z( t\mid \bm \Theta) d t,\\
	\label{equ:b1}
	b_1(\bm \delta_z,\bm \Theta)
	&= \int_0^\infty p_{ \delta}\left(Z=z \mid \bm \delta_z, \bm\Theta \right) f_Z(z\mid \bm \Theta) dz.
\end{align}

The {$p_{ \delta}(\bm \delta_z \mid \bm\Theta) $} is the likelihood by the GaSP model and $\bone/ \bzero $ is the weight depending on the choice of $f_Z(\cdot)$. The motivation is to give a higher weight to the $\bm \theta$ that leads to {the} smaller $L_2$ loss {conditional on} the observations and other parameters.





{For any proper density $f_Z(\cdot)$, $\bzero$ and $\bone$ can be computed by the standard Monte Carlo integration method. To illustrate, one can first draw samples $ z_1, ..., z_M $ from $ p_{ \delta}\left(Z=z \mid \bm\Theta \right) $ using \Cref{lemma:chi-square}, and then use $\sum^M_{i=1} f_Z(z_i{\mid\Thetabm})/M$ to approximate $ \bzero$. Similar strategy can be applied to approximate $\bone$. This method provides a comparatively more stable way in evaluating the likelihood in the S-GaSP model than directly computing the density of an infinite weighted sum of chi-squared distributions.}



For the demonstration purpose and computational reason, we choose the following exponential distribution for $f_Z(\cdot)$,
\begin{equation}
	{f_Z(Z=z\mid \bm \Theta) = \frac{\lambda}{2\sigma^2_{\delta} \Vol(\mathcal X)} \exp\left(-\frac {\lambda z}{2\sigma^2_{\delta} \Vol(\mathcal X)}\right),}
	\label{equ:f_Z}
\end{equation}
where $\lambda$ is a positive scaling parameter and $\Vol(\mathcal X)$ is the volume of $\mathcal X$. A larger $\lambda$ favors the sample function with a smaller $L_2$ norm, while {a} smaller $\lambda$ means the S-GaSP model behaves more similarly to a GaSP model. The magnitude of $ \lambda $ determines how similar the S-GaSP model is to these two approaches. 

With the specification of $f_Z(\cdot)$ in \cref{equ:f_Z}, $\sigma^2_{\delta}$ is still a scale parameter and the S-GaSP is equivalent to the GaSP with a transformed kernel function, stated in the following lemma.


\begin{lemma}[{Marginal} distribution of the S-GaSP model]
	\label{lemma:density_delta_z}
	Assume $p_{\delta_z}(Z=z \mid \bm \Theta)$ and $f_Z(Z=z\mid \bm \Theta)$  are specified in \cref{equ:p_z} and in \cref{equ:f_Z}, respectively. The marginal distribution of $\bm \delta_z=\left(\delta_z(\mathbf x_1),...,\delta_z(\mathbf x_n) \right)^T$ in the S-GaSP in \cref{equ:scaled_GP} is a multivariate normal distribution
	\begin{equation}
		\bm \delta_z \mid \bm \Theta \sim \text{MN}( \bm 0,  \sigma^2_{\delta} \mathbf R_z ),
		\label{equ:delta_z_marginal}
	\end{equation}
	where the covariance follows
	\begin{equation}
		{\mathbf R_z= \left(  \mathbf B + (\mathbf R^{\delta})^{-1}\right)^{-1}, }
		\label{equ:R_z}
	\end{equation}
	{and} $\mathbf B$ is an $n \times n$ real-valued matrix with the following form 
	\begin{equation}
		{\mathbf B= (\mathbf R^{\delta})^{-1} \left\{\sum^{\infty}_{i=1} \frac{\lambda}{\Vol(\mathcal X)+\lambda^*_i \lambda } \left(\int_{{\xbf\in\Xcal}} \mathbf r^{\delta}(\mathbf x) \phi^*_i(\mathbf x) d\mathbf x \right)\left( \int_{{\xbf\in\Xcal}} \mathbf r^{\delta}(\mathbf x) \phi^*_i(\mathbf x) d\mathbf x\right)^T \right\} (\mathbf R^{\delta})^{-1},}
		\label{equ:B}
	\end{equation}
	with $\lambda^*_i$ and $\phi^*_i(\mathbf x)$ being the $i^{th}$ eigenvalue and normalized eigenfunction of $c^{*\delta}(\cdot, \cdot)$ in \cref{equ:c_star_gasp} with regard to the Lebesgue measure, respectively. 
\end{lemma} 

Based on \cref{lemma:density_delta_z}, the marginal distribution of $\mathbf y^F=(y^F(\mathbf x_1), ..., y^F(\mathbf x_n))^T$ can be computed by marginalizing $ \deltabm_z $ out and it is still a multivariate normal distribution. It is worth noting that the such simplification relies on the default choices of $p_{\delta_z}(\cdot)$ in \cref{equ:p_z} and $f_Z(\cdot)$ in \cref{equ:f_Z}. The calculation of the covariance $ \Rbf_z $ in \cref{equ:R_z}, however, requires an approximation of the integral in \cref{equ:B}, which can still be quite complicated especially for predictions at many unobserved points. To further simplify the computation, we propose a feasible way for the computation by the discretized S-GaSP in the following subsection.





\subsection{Discretized S-GaSP model with constraints on finite points}
\label{subsec:finite_S-GaSP}

 We use $N_C$ distinct points $\mathbf x^C_{i} \in \mathcal X$ for $i=1,...,N_C$, to discretize the integral in the S-GaSP model in \cref{equ:scaled_GP}, such that ${ {\int}_{\bm{\xi}\in\mathcal X} }  \delta( \bm \xi )^2 d\bm \xi \approx \sum^{N_C}_{i=1} \delta( \mathbf x^C_{i} )^2 \Delta x$, with $\Delta x={ {\rm Vol}(\mathcal X) / N_C} $ and $ {\rm Vol}(\mathcal X)$ being the volume of $\mathcal X$. The S-GaSP constrained on finitely many points is defined as follows:
\begin{equation}
	\label{equ:scaled_GP_approx}
	\begin{split}
		&y^F(\mathbf x)=f^{M}(\mathbf x, \bm \theta)+\mu^{\delta}(\mathbf x)+ \delta^a_z(\mathbf x) +\epsilon, \\
		& \delta^a_z(\mathbf x) = \left\{ \delta(\mathbf x) \mid\sum^{N_C}_{i=1} \delta( \mathbf x^C_{i} )^2 \Delta x=Z \right\}, \\
		&\delta(\cdot) \sim \text{GaSP}(0,  \sigma^2_{\delta} c^{\delta}(\cdot, \cdot)),\\
		&Z\sim p_{\delta^a_z}(\cdot), \, \epsilon \sim N(0, \sigma^2_0 ).
	\end{split}
\end{equation}
We call $\delta^a_z(\cdot)$ follows the discretized S-GaSP model, where the only approximation is the discretization of the integral in \cref{equ:scaled_GP}. The constraint points could be hard to choose especially when $ p_x $ is large. A convenient choice is to have $\mathbf x^C_i=\mathbf x_i$ for $ i = 1, ..., n $, which gives a good approximation when $ n $ is large and the design of the experiments follows a ``space-filling" scheme. We show below that under the default choice of the scaling density $ p_{\delta_z}(\cdot) $, the marginal distribution of the discretized S-GaSP model is still a multivariate normal distribution.

\begin{lemma}[{Marginal} distribution of the discretized S-GaSP model]
	\label{lemma:approx_delta_z}
	Assume $p_{\delta_z}(Z=z \mid \bm \Theta)$ and $f_Z(Z=z\mid \bm \Theta)$  are specified in \cref{equ:p_z} and in \cref{equ:f_Z}, respectively. The marginal distribution of $\bm \delta^a_z=\left(\delta^a_z(\mathbf x_1),...,\delta^a_z(\mathbf x_n) \right)^T$ in the discretized S-GaSP in \cref{equ:scaled_GP_approx} is a multivariate normal distribution
	\begin{equation} 
		\bm \delta^a_z \mid \bm \Theta \sim \text{MN}( \bm 0,  \sigma^2_{\delta} \mathbf R^{a}_z ),
		\label{equ:delta_z_marginal_approx}
	\end{equation}
	where
	\begin{equation}
		\label{equ:R_a_z}
		{\mathbf R^{a}_z=\mathbf R^{\delta}- (\mathbf r^C)^T\parenth{\mathbf R^C+\frac{N_C}{\lambda }\mathbf I_{N_C}}^{-1} \mathbf r^C.}
	\end{equation}
	with $\mathbf R^C$ being an ${N_C \times N_C}$ correlation matrix with the $(i,j)$ entry $\mathbf R^C_{i,j}=c^{\delta}(\mathbf x^C_i, \mathbf x^C_j)$ and $\mathbf r^C$ being an $N_C \times n$ correlation matrix with the $(i,j)$ entry being $ r^C_{i,j}=c^{\delta}(\mathbf x^C_i, \mathbf x_j)$.
\end{lemma}

\begin{figure}[t]
	\centering
	\subfloat{\includegraphics[width=0.5\textwidth]{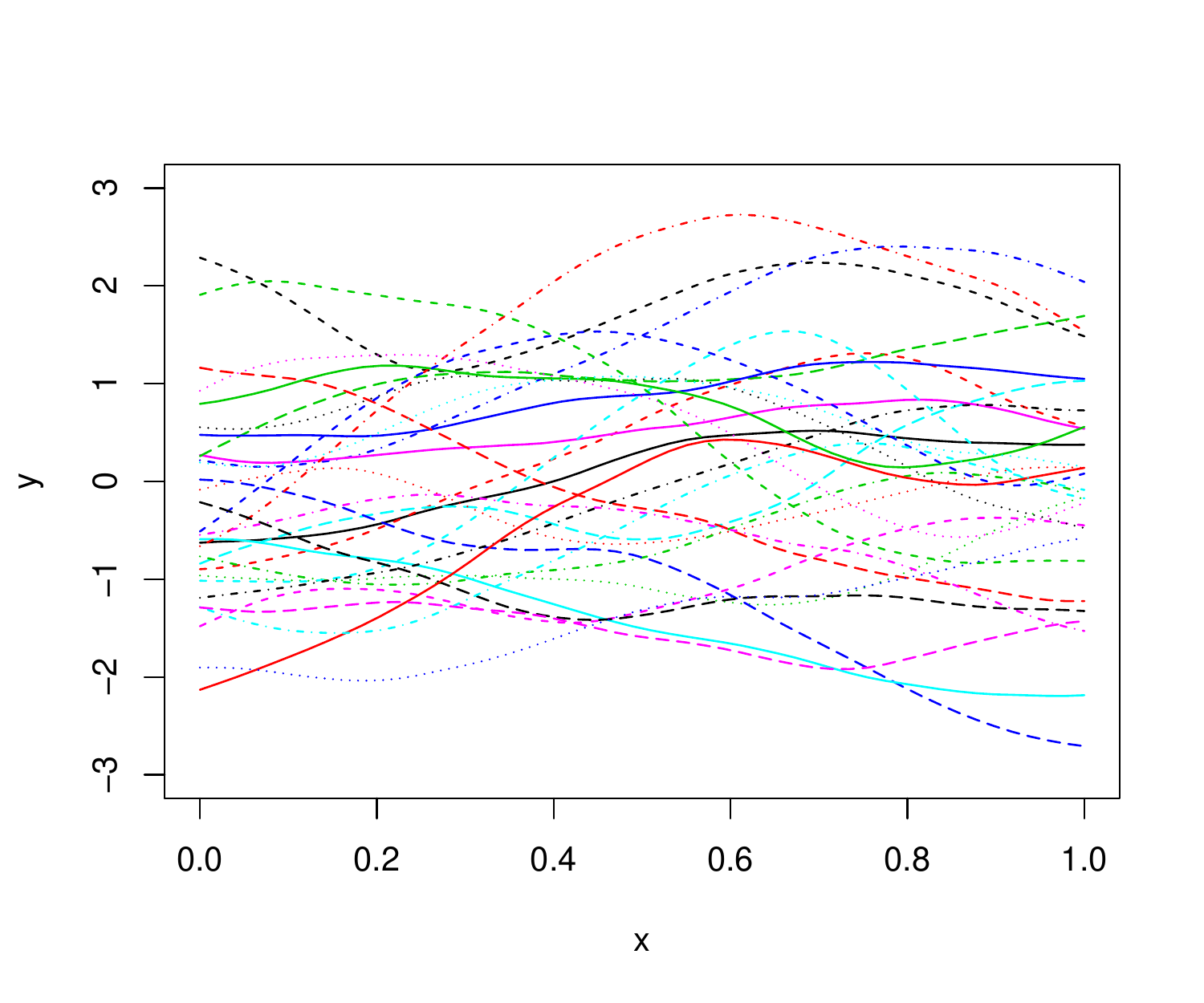}}
	\subfloat{\includegraphics[width=0.5\textwidth]{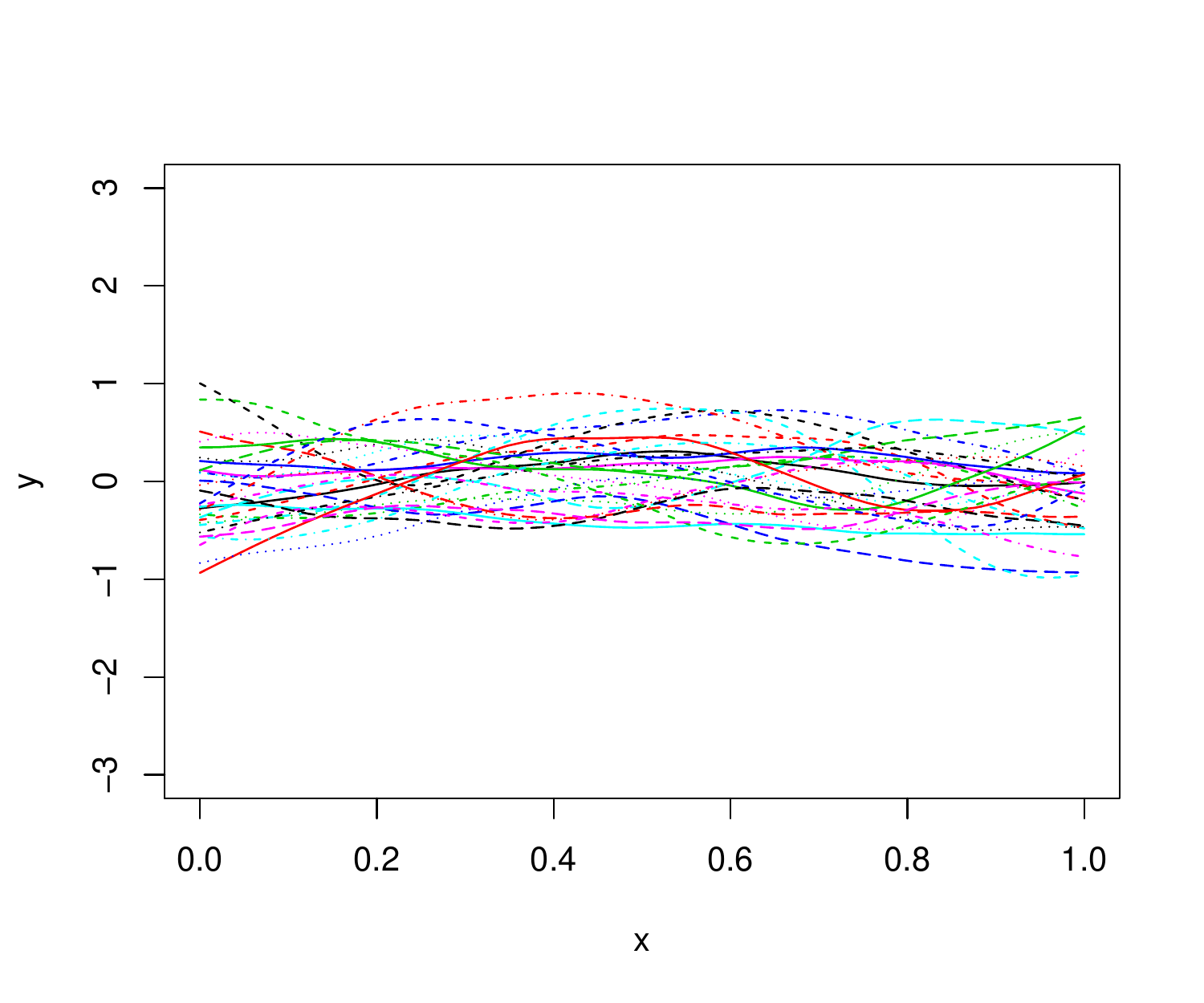}}
	\caption{Fifty samples from the GaSP and discretized S-GaSP are graphed in the left and right panels, respectively, where $x_i$ is equally spaced in $[0,\,1]$. For both processes, we let $\mu^{\delta}=0$, $\sigma^2_{\delta}=1$ and $\gamma^{\delta}=1/2$. In the discretized S-GaSP, $\mathbf x^C_i=\mathbf x_i$ for $i=1,...,N_C$, $N_C=n$ and $\lambda=n/2$ are assumed..  }
	\label{fig:samples_GP_SGP}
\end{figure}

Based on \Cref{lemma:approx_delta_z}, for any $\mathbf x_a, \mathbf x_b \in \mathcal X$, one has 
\begin{equation}
\label{equ:cov_discretized_S_GaSP}
c^a_z(\mathbf x_a, \mathbf x_b )= c^{\delta}(\mathbf x_a,\mathbf x_b ) - \mathbf r^C(\mathbf x_a)^T (\tilde {\mathbf R}^C )^{-1} \mathbf r^C(\mathbf x_b),
\end{equation}
 with
 \begin{equation}
 \label{equ:tilde_R_C}
 {\tilde {\mathbf R}^C=\mathbf R^C+N_C\mathbf I_{N_C} / \lambda}.
 \end{equation}
From \cref{equ:cov_discretized_S_GaSP}, it is worth noting that covariance of the discretized S-GaSP is equivalent to the covariance of the predictive distribution of a zero-mean GaSP given $ N_C $ noisy observations, if the variance of the  i.i.d. zero-mean Gaussian noise is $ N_C/\lambda $. Therefore, as shown in \cref{fig:samples_GP_SGP}, the samples from a zero-mean S-GaSP are more concentrated around zero than the samples from a zero-mean GaSP. The S-GaSP, however, behaves quite differently than a GaSP with a decreased variance. Simply controlling the variance of the GaSP may not solve the identifiability issue, as the problem is caused by the correlation as discussed in \Cref{sec:Intro}. A numerical example toward this direction is provided at the end of this subsection to further illustrate the differences between the GaSP and S-GaSP in calibration and prediction. 



The marginal distribution of $\bm \delta^a_z$ in \Cref{lemma:approx_delta_z} is attractable since no Monte Carlo method or eigen-decomposition of the covariance matrix is required. The marginal distribution of the field data is readily available by marginalizing out $\bm \delta^a_z$. 



\begin{theorem}[Marginal distribution of the field data]
	Assume the conditions of \cref{lemma:approx_delta_z} hold. After marginalizing out $ \delta^a_z(\cdot)$, the marginal likelihood of $\mathbf y^F=(y^F(\mathbf x_1), ..., y^F(\mathbf x_n))^T$ of the discretized S-GaSP model in \cref{equ:scaled_GP_approx} follows
	\begin{equation}
		\mathbf y^F \mid \bm \Theta \sim MN(\mathbf f^M(\mathbf x_{1:n}, \bm \theta)+\bm \mu^{\delta}, \sigma^2_{\delta}\mathbf R^a_z+\sigma^2_0 \mathbf I_n  ),
		\label{equ:marginal_lik_scaled}
	\end{equation}
	where $\mathbf f^M(\mathbf x_{1:n}, \bm \theta)=( f^M(\mathbf x_1, \bm \theta),..., f^M(\mathbf x_n, \bm \theta))^T$ are the computer model outputs evaluated at $\mathbf x_i$ and $ \bm \theta$ for $i=1,...,n$. The mean discrepancy is $\bm \mu^{\delta}= (\mu^{\delta}(\mathbf x_1), ..., \mu^{\delta}(\mathbf x_n))^T$ with each term defined in \cref{equ:mean_dis} and the correlation matrix $\mathbf R^a_z$ of the discrepancy function is defined in \cref{equ:R_a_z}.
	\label{thm:marlik_y}
\end{theorem}

 
 The following \Cref{thm:pred_dist} gives the predictive distribution of $ y^F(\mathbf x^*) $ when both the calibrated computer model and discrepancy function are used. It can be shown easily using \cref{thm:marlik_y} and the properties of the multivariate normal distribution.


\begin{theorem}[Predictive distribution]
	\label{thm:pred_dist}
	Assume the conditions of \cref{lemma:approx_delta_z} hold. The predictive distribution of the field data at a new point $\mathbf x^*$ by the discretized S-GaSP model in \cref{equ:scaled_GP_approx} follows	
	\[ y^F(\mathbf x^*) \mid \mathbf y^F, \bm \Theta \sim N( \hat \mu(\mathbf x^*), \sigma^2_{\delta} c^*+{\sigma^2_0} ), \]
	where
	\begin{align*}
		\hat \mu(\mathbf x^*)&=f^M(\mathbf x^*, \bm \theta)+ \mu^{\delta}(\mathbf x^*)+\mathbf r^a_z (\mathbf x^*)^T (\tilde {\mathbf R}^a_z)^{-1}(\mathbf y^F- \mathbf f^M(\mathbf x_{1:n}, \bm \theta)-\bm \mu^{\delta} ), \\
		c^*&= c^a_z(\mathbf x^*, \mathbf x^* )-\mathbf r^a_z (\mathbf x^*)^T (\tilde  {\mathbf R}^a_z)^{{-1}} \mathbf r^a_z(\mathbf x^*),
	\end{align*}
with	$\tilde {\mathbf R}^a_z= {\mathbf R}^a_z+\sigma^2_0 \mathbf I_n / \sigma^2_{\delta}$, $\mathbf r^a_z(\mathbf x^*)=\mathbf r^{\delta} (\mathbf x^*)-(\mathbf r^C)^T (\tilde {\mathbf R}^C )^{-1} \mathbf r^C(\mathbf x^*)$, $c^a_z(\cdot,\cdot)$ and $\tilde {\mathbf R}^C$ being defined in (\ref{equ:cov_discretized_S_GaSP}) and (\ref{equ:tilde_R_C}) respectively. 
\end{theorem}

\begin{figure}[t]
	\centering
	\subfloat{\includegraphics[width=0.33\textwidth]{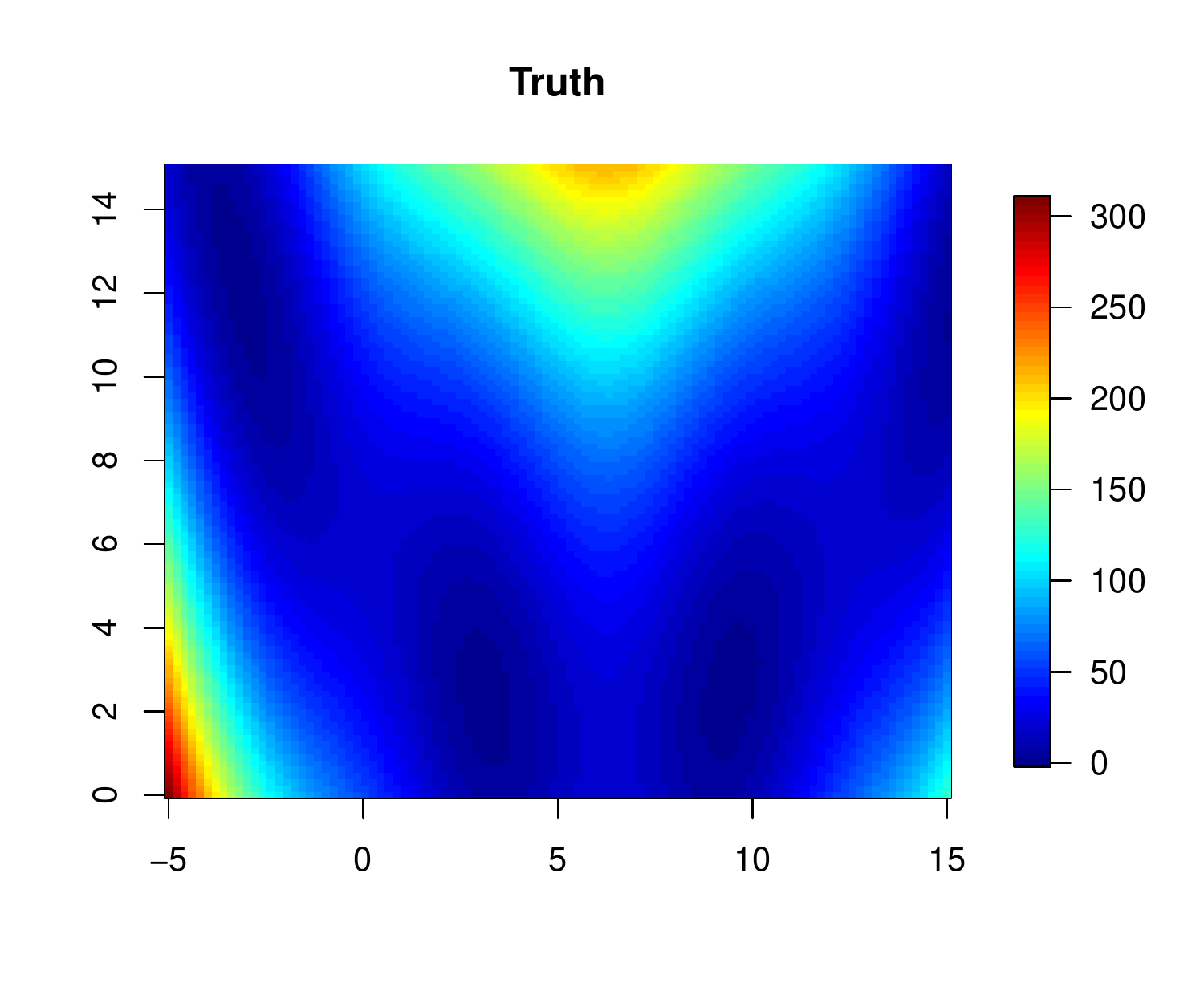}}
	\subfloat{\includegraphics[width=0.33\textwidth]{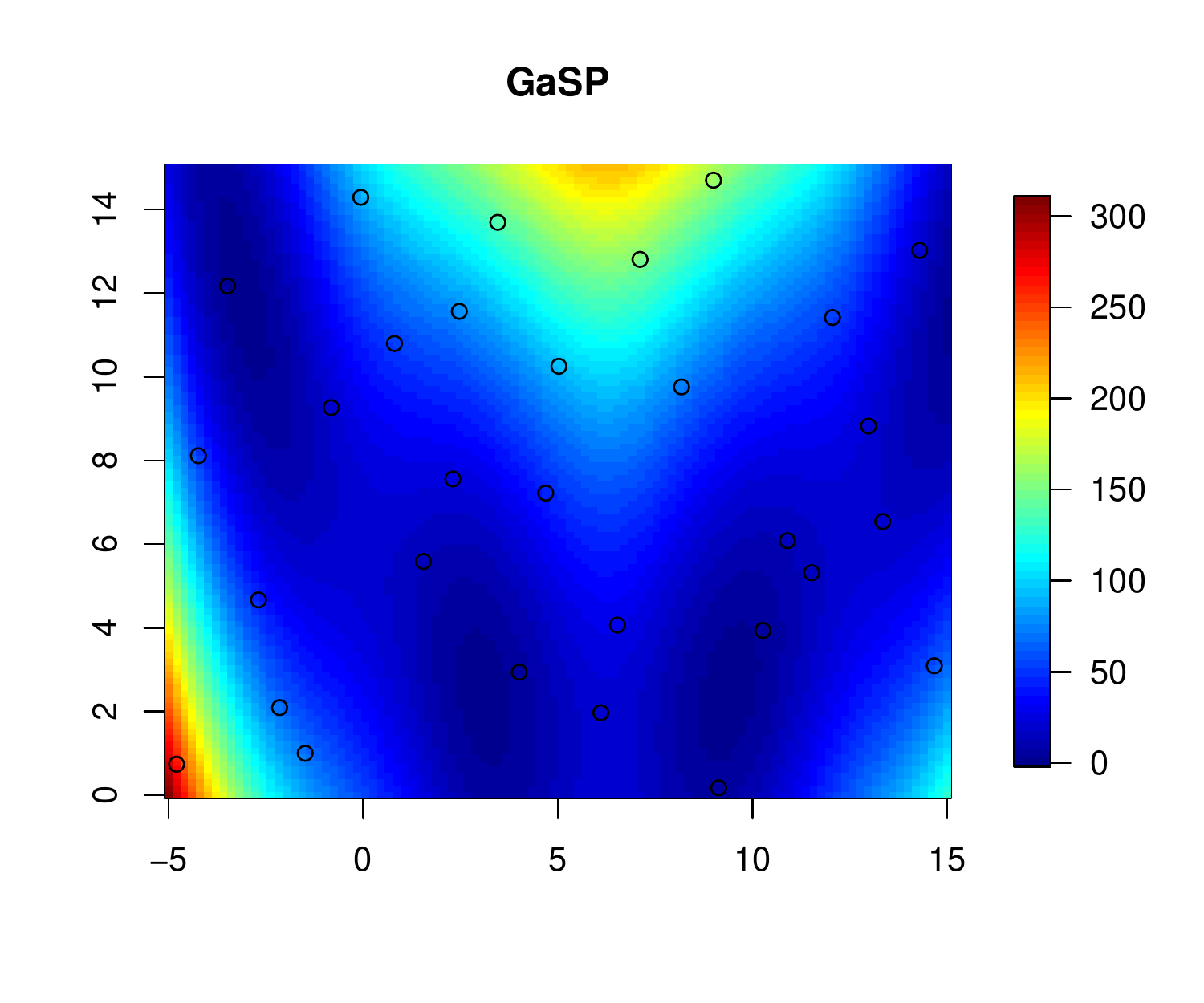}}
	\subfloat{\includegraphics[width=0.33\textwidth]{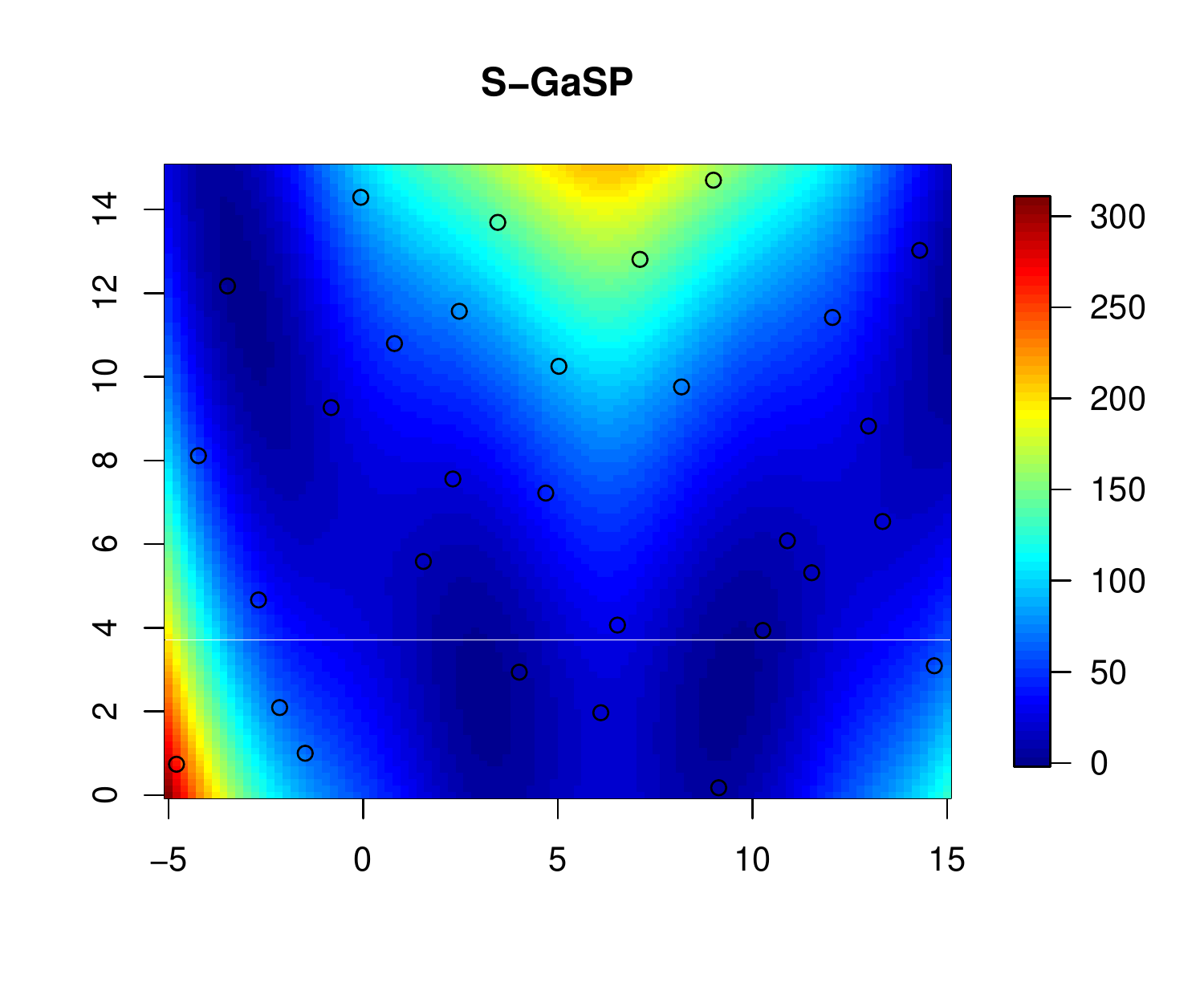}}
	\caption{Predictions of the GaSP and S-GaSP for the Brainin function \cite{forrester2008engineering}. Thirty training values of the Brainin function are plotted as black circles.  Both models take a constant mean and Mat{\'e}rn correlation in \cref{equ:Matern} with the range parameter estimated by the RobustGaSP R package \cite{gu2018robustgasp_package}. The default choices of $p_{\delta_z}(\cdot)$ in \cref{equ:p_z} and $f_Z(\cdot)$ in \cref{equ:f_Z} of the S-GaSP model are used with $\lambda=n/2$. The mean squared predictive errors for the GaSP and S-GaSP are $0.659$ and $0.587$, respectively.}
	\label{fig:pred_GaSP_SGaSP_2dim}
\end{figure}

\cref{fig:pred_GaSP_SGaSP_2dim} compares the predictions by the GaSP and S-GaSP for the Brainin function with $ p_x = 2 $ \cite{forrester2008engineering}. With the inputs sampled from the maximin Latin hypercube design \cite{santner2013design}, both models perform reasonably well, suggesting good predictive powers for the complicated nonlinear function.


%
%
%
%
%
%


The following \Cref{eg:eg_variance} further illustrates the differences between the GaSP and S-GaSP, where the reality is assumed to be the function studied in \cite{park1991tuning}.


\begin{example}
Assume $y^F(\mathbf x)=y^R( \mathbf x)+\epsilon$ where $\epsilon \sim N(0,0.01^2)$ and $y^R(\mathbf x)=\frac{2}{3}\exp(x_1+ x_2)-x_4\sin(x_3)+x_3$
with $x_i \in [0,1]$. For simplicity, assume $ f^M(\mathbf x, \theta) = \theta $ and $ \xbf_i $ for $i=1,...,50$, are drawn from the maximin Latin hypercube design \cite{santner2013design}.
\label{eg:eg_variance}
\end{example}


	\begin{table}[t]
		\caption{Predictive mean squared errors and maximum likelihood estimation (MLE) for the parameters in GaSP and S-GaSP calibration models in Example~\ref{eg:eg_variance}. MSE$_{f^M}$ is the mean squared error using the calibrated computer model to predict. MSE$_{f^M+\delta}$ is the mean squared error using the calibrated computer model and discrepancy function to predict, where the discrepancy function is specified as the GaSP and S-GaSP. The results in the first two rows are the cases when all parameters are estimated via MLE, while $\sigma^2_{\delta}$ is fixed for the results in the other rows. In the S-GaSP calibration model, $\mathbf x^C_i=\mathbf x_i$, $i=1,...,N_C$, $N_C=n$ and $\lambda=n/2$ are assumed.}	
		\label{tab:variance}
		\centering
		\begin{tabular}{lcccccc}
			\hline
			& MSE$_{f^M}$ & MSE$_{f^M+\delta}$  & $\hat \theta$ & $\hat \sigma^2_{\delta}$ & $\hat{ \bm \gamma}^{\delta}$ & $\hat \sigma^2_0$ \\
			\hline  
		 GaSP 	& $20$  & $2.8\times 10^{-4}$ & $6.6$ & $48$ & $(3.2,  3.3,  5.7, 6.6)$ & $6.5\times 10^{-5}$ \\
		 S-GaSP  & $0.84$  & $2.7\times 10^{-4}$ & $2.6$ & $90$ & $(3.7,  3.7,  6.6, 7.6)$ &$6.9\times 10^{-5}$\\
			\hline
		 GaSP, $\sigma_\delta^2=1$ 	& $1.5$  & $6.0\times 10^{-4}$ & $3.1$ & / & $(1.2,  1.3,  2.3, 2.7)$ &$6.3\times 10^{-5}$ \\
		 S-GaSP, $\sigma_\delta^2=1$  & $0.70$  & $6.4\times 10^{-4}$ & $2.3$ & / & $(1.2,  1.3,  2.2, 2.7)$ &$6.3\times 10^{-5}$ \\
			\hline
		 GaSP, $\sigma_\delta^2=10$ 	& $6.2$  & $3.3\times 10^{-4}$ & $4.6$ & / & $(2.2,  2.4,  3.9, 4.6)$& $5.8\times 10^{-5}$ \\
		 S-GaSP, $\sigma_\delta^2=10$  & $0.73$  & $3.5\times 10^{-4}$ & $2.4$ & / & $(2.2,  2.4, 3.9, 4.6)$&$5.7\times 10^{-5}$ \\
			\hline
		 GaSP, $\sigma_\delta^2=10^2$ 	& $34$  & $2.7\times 10^{-4}$ & $8.0$ & / & $(3.8,  3.7,  6.7, 7.7)$& $7.2\times 10^{-5}$ \\
		 S-GaSP, $\sigma_\delta^2=10^2$  & $0.85$  & $2.7\times 10^{-4}$ & $2.6$ & / & $(3.8,  3.7,  6.7, 7.8)$ &$7.2\times 10^{-5}$ \\
			\hline
		 GaSP, $\sigma_\delta^2=10^3$ 	& $186$  & $2.5\times 10^{-4}$ & $16$ & /& $(6.2,  5.5,  12, 14)$& $9.1\times 10^{-5}$ \\
		 S-GaSP, $\sigma_\delta^2=10^3$  & $1.4$  & $2.5\times 10^{-4}$ & $3.1$ & / & $(6.2,  5.5,  12, 14)$&$9.1\times 10^{-5}$  \\
			\hline
		\end{tabular}
	\end{table}

For \Cref{eg:eg_variance}, we are interested in estimating $\theta$ and predicting $y^F(\mathbf x^*_i)$ at the held-out $\mathbf x^*_i$, uniformly sampled from $[0, 1]^4$ for $i=1,...,1000$. The out-of-sample predictions using the GaSP and S-GaSP calibrations are provided in \Cref{tab:variance}, where the parameters are estimated by the maximum likelihood estimation (MLE) via the low-storage quasi-Newton optimization method \cite{nocedal1980updating} with 10 different initializations. 

Let MSE$_{f^M}$ denote the mean squared error using only the calibrated computer model and MSE$_{f^M + \delta}$ denote the mean squared error using both the calibrated computer model and the discrepancy function for prediction. When $ \sigma_\delta^2 $ is not given, the estimated range parameters $ \hat{\gammabm} $ are large in both models, while the MSE$_{f^M}$ is much larger in GaSP than in S-GaSP shown in the first two rows in \Cref{tab:variance}. When $ \sigma_\delta^2 $ is fixed at small values, the MSE$_{f^M}$ is small under both models since $ \hat{\gammabm} $ are small. However, the MSE$_{f^M + \delta}$ becomes larger as $ \hat{\gammabm} $ decrease. In particular, the MSE$_{f^M + \delta}$ with $ \sigma_\delta^2 = 1 $ is two times larger than the one when $ \sigma_\delta^2$ is estimated. When $ \sigma_\delta^2 $ is fixed at a large value, the MSE$_{f^M + \delta}$ decreases, while the MSE$_{f^M}$ increases in the GaSP model due to the large $ \hat{\gammabm} $. In the S-GaSP model, the MSE$_{f^M}$ is always small for all tested scenarios, because the S-GaSP calibration puts more probability mass to the discrepancy function that leads to a smaller $L_2$ loss between the reality and computer model.

\Cref{tab:variance} indicates that S-GaSP satisfies the two criteria in \Cref{subsec:contribution} by obtaining both small values in MSE$_{f^M}$ and MSE$_{f^M+\delta} $, but these two goals are not simultaneously attainable in the GaSP model. Here $\lambda$ is chosen to be $n/2$ in the S-GaSP model. The sensitivity analysis of $\lambda$ is given in the supplementary materials. In general, selecting or estimating $ \lambda $ in a principle way is still an open question.

%
%
%
%
%

\subsection{Comparison to the orthogonal Gaussian stochastic process}
\label{subsec:O-GaSP}

In this subsection, we compare our S-GaSP model to a recent approach, called the orthogonal Gaussian stochastic process (O-GaSP) introduced in \cite{plumlee2016bayesian}. Under some regularity conditions of $f^M(\cdot,\cdot)$ and assuming the existence and uniqueness of the $L_2$ minimizer $\bm \theta_{L_2}$, it is shown in \cite{plumlee2016bayesian} that the following constraint on the discrepancy function holds
\begin{equation}
	\int_{\mathbf x \in \mathcal X} D^{(0,1)}f^M(\mathbf x, \bm \theta^*)\delta(\mathbf x) d\mathbf x=\mathbf 0,
	\label{equ:const}
\end{equation}
where $D^{(0,1)}f^M(\mathbf x, \bm \theta_{L_2})$ is the derivative of $f^M(\cdot,\cdot)$ with regard to $\bm \theta$, evaluated at {$ \xbf $ and} $\bm \theta=\bm \theta_{L_2}$. In \cite{plumlee2016bayesian}, it is shown that if the discrepancy is modeled by a zero-mean orthogonal GaSP for any $\mathbf x, \mathbf x' \in \mathcal X$

\[\delta_{O-GaSP}(\cdot) \sim \mbox{GaSP} (0, \sigma^2_{\delta} c^{\delta}_{O-GaSP}(\cdot, \cdot)),\]
where
\begin{equation}
	\label{equ:kernel_trans}
	c^{\delta}_{O-GaSP}(\mathbf x, \mathbf x') =c^{\delta}(\mathbf x, \mathbf x')- g_{O-GaSP}(\mathbf x)^T G_{O-GaSP}^{-1} g_{O-GaSP}(\mathbf x'),
\end{equation}
for any positive definite covariance function $c^{\delta}(\mathbf x, \mathbf x')$ with
\[g_{O-GaSP}(\mathbf x)=\int_{\bm{\xi} \in \mathcal X}D^{(0,1)}f^M(\bm \xi, \bm \theta)c^{\delta}(\mathbf x, \bm \xi) d \bm \xi,\]
and
\[G_{O-GaSP}=\int_{\bm{\xi}' \in \mathcal X}\int_{\bm{\xi} \in \mathcal X} D^{(0,1)}f^M(\bm \xi, \bm \theta)\left[D^{(0,1)}f^M(\bm \xi', \bm \theta)\right]^T c^{\delta}(\bm \xi, \bm \xi') d \bm \xi d \bm \xi', \]
the \cref{equ:const} holds with probability 1. 

The constraint in \cref{equ:const} is essentially the first-order optimality condition. As in the case of all the optimization problems, \cref{equ:const} holds not only for the $L_2$ minimizer $\bm \theta_{L_2}$, but also for many other $\bm \theta$. In particular, any local maximum of the $L_2$ loss also satisfies the constraint in \cref{equ:const}, which makes the likelihood favor the local maximizers as well. {As $f^M(\mathbf x, \bm \theta)$ is typically a nonlinear function of $\bm \theta$, the $L_2$ loss is often multi-dimensional and likely to have multiple extreme values. Hence, in practice, using the O-GaSP also favors the local maximizers of the $L_2$ loss, giving undesired results. We illustrate this problem in Example \ref{eg:nonlinear} below.




\begin{example}
	\label{eg:nonlinear}
	Suppose $y^F(x)=xcos(3x/2)+x+\epsilon$ with $x\in [0,5]$ and $\epsilon \sim N(0,0.2^2)$. The computer model is $f^M(x,\theta)=sin(\theta x)+x$ for $\theta \in [0,3]$.  Fifteen observations, denoted as $y^F(x_i)$ for $i=1,...,15$, are collected with $x_i$ equally spaced in $[0,5]$.
\end{example}


\begin{figure}[t]
	\centering
	\subfloat{\includegraphics[width=0.5\textwidth]{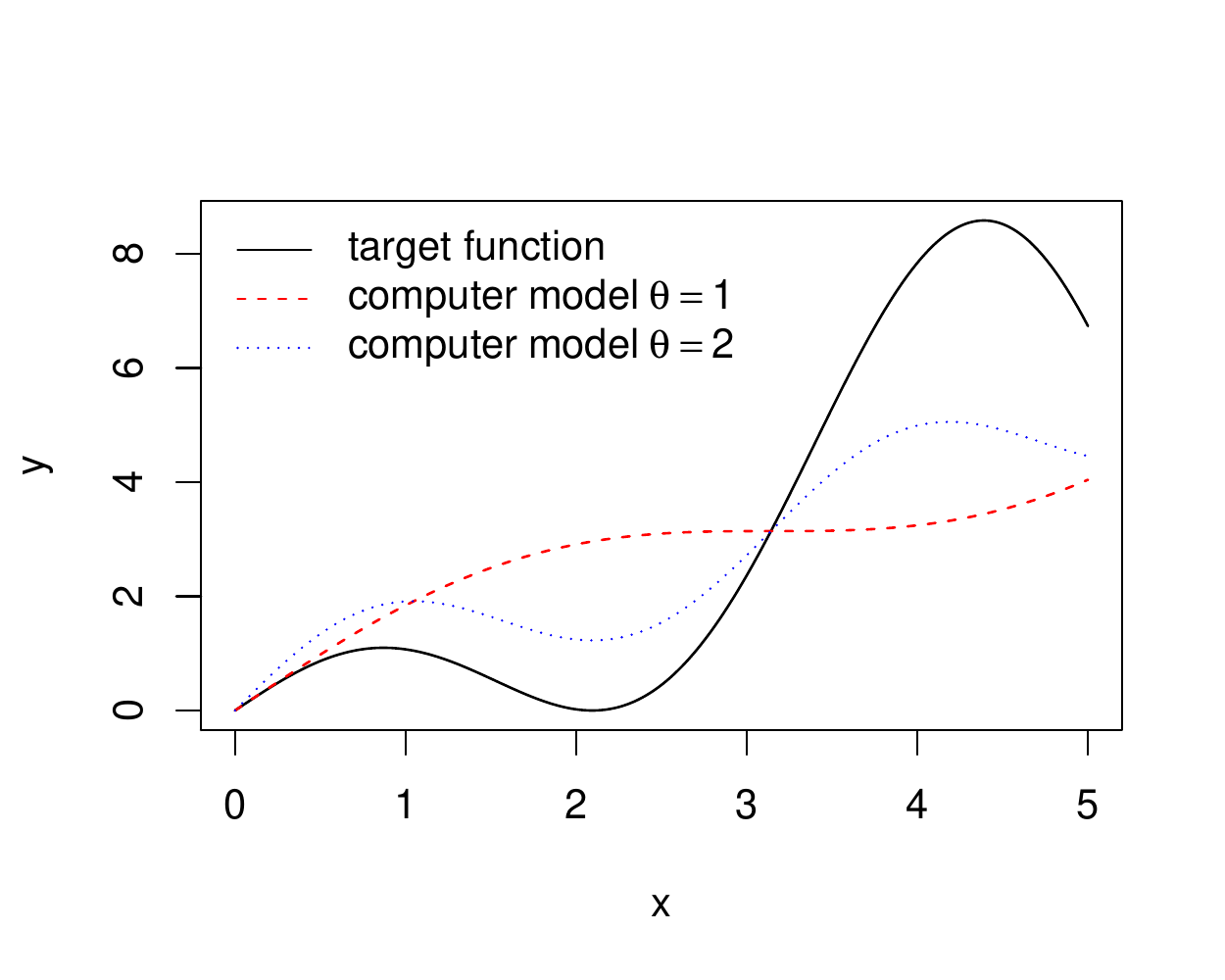}}
	\subfloat{\includegraphics[width=0.5\textwidth]{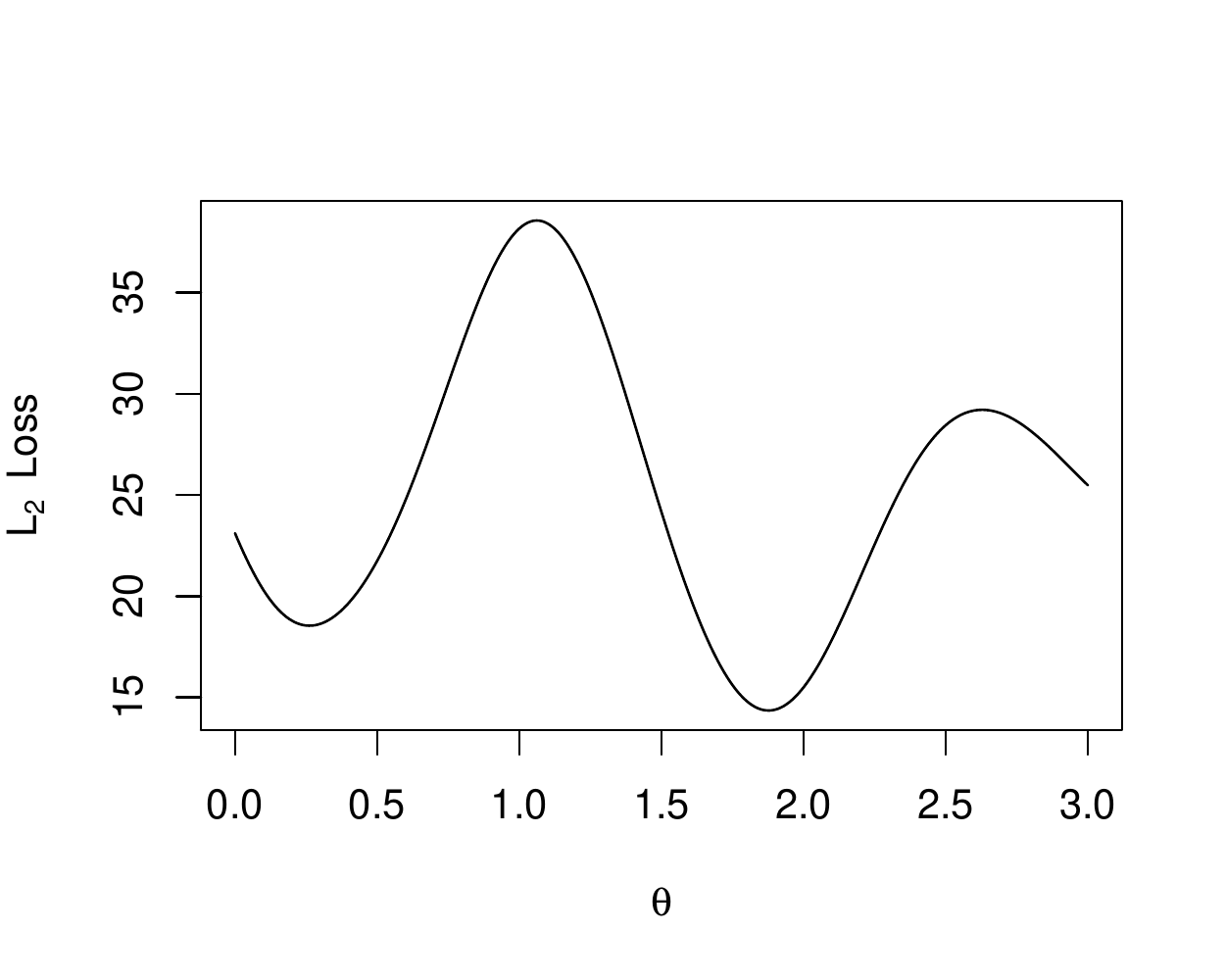}} \vspace{-.4in}\\
	\subfloat{\includegraphics[width=0.33\textwidth]{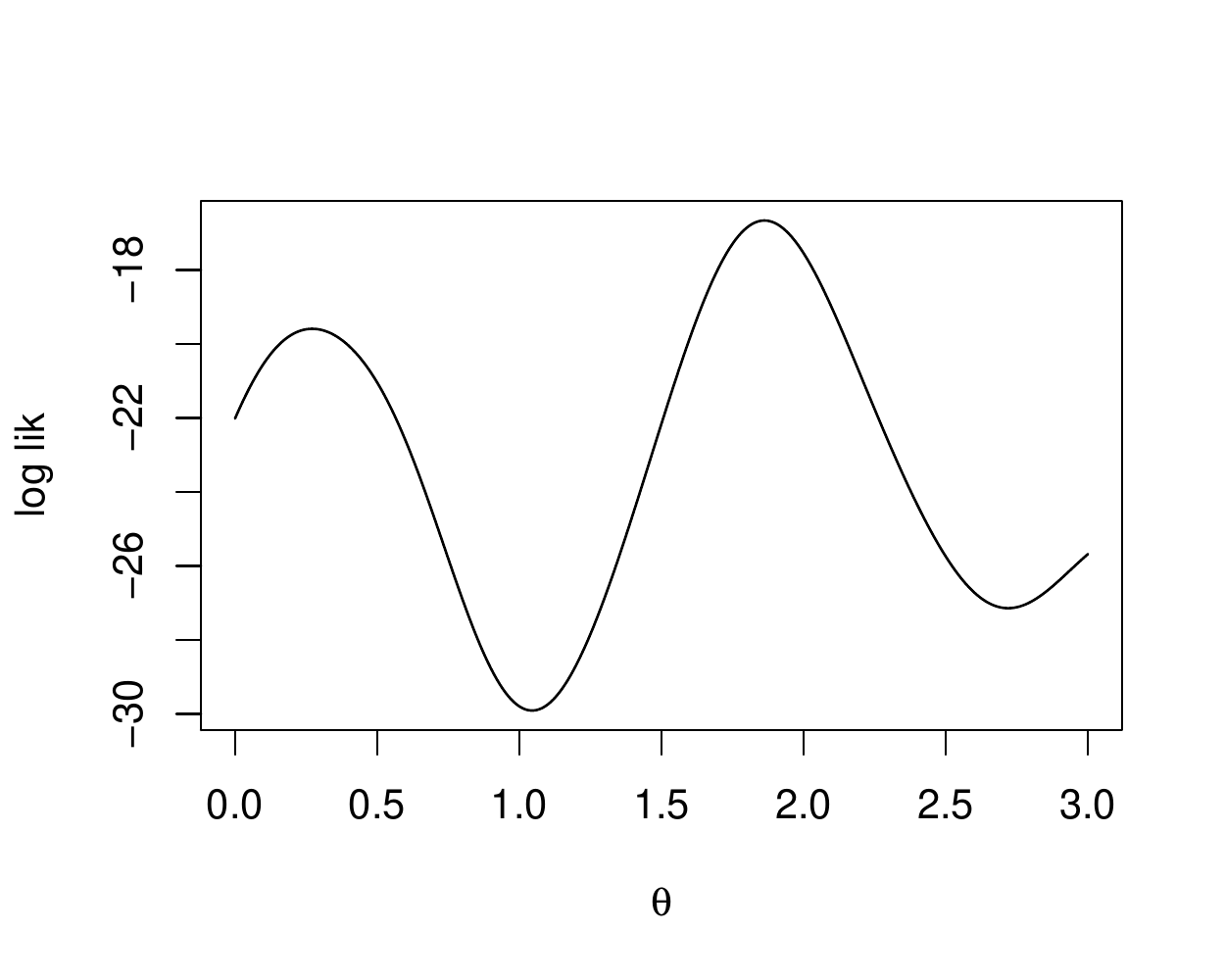}}
	\subfloat{\includegraphics[width=0.33\textwidth]{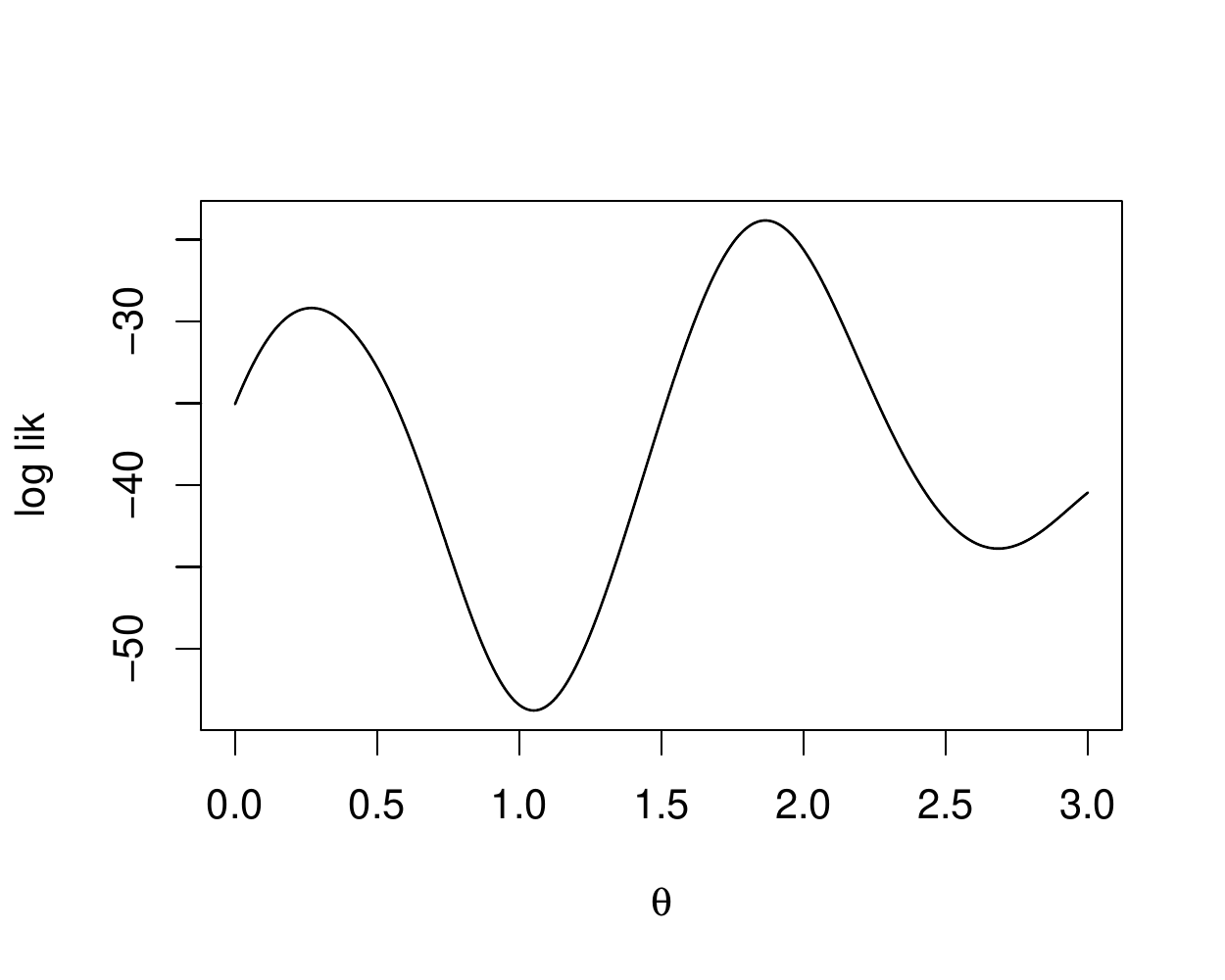}}
	\subfloat{\includegraphics[width=0.33\textwidth]{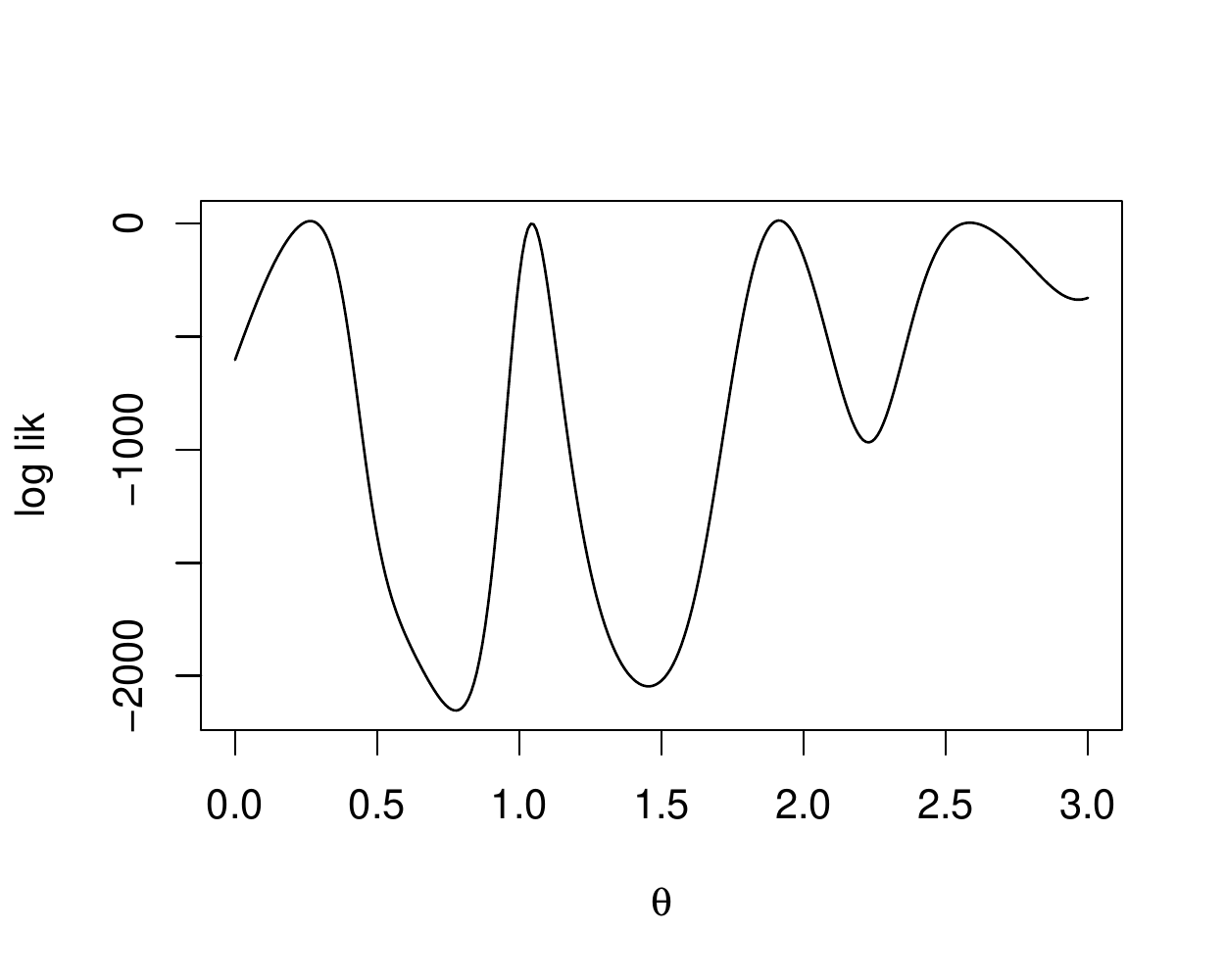}}
	\caption{The target function and computer model outputs at two different $\theta$ are graphed in the upper left panel in the first row. The $L_2$ loss function in \cref{eg:nonlinear} is graphed in the upper right panel. The log-likelihoods of the GaSP, discretized S-GaSP and O-GaSP are graphed in the left, middle, and right panels in the second row, respectively. For all models, the Mat{\'e}rn covariance in \cref{equ:Matern} with $\sigma^2_{\delta}=1$, $\gamma^{\delta}=1/2$ and $\eta=0.01$ is used. In the discretized S-GaSP calibration model, $\mathbf x^C_i=\mathbf x_i$ for $i=1,...,N_C$, $N_C=n$ and $\lambda=n/2$ are assumed. }
	\label{fig:eg1}
\end{figure}


The $L_2$ loss over $\theta \in [0,3] $ for \Cref{eg:nonlinear} is graphed in the upper right panel in \cref{fig:eg1}, which contains the global minimum at $\theta\approx 1.88$, a local minimum at $\theta\approx 0.26$ and two local maxima at $\theta\approx 1.06$ and $\theta\approx 2.62$, respectively. The log-likelihoods of GaSP, S-GaSP and O-GaSP models are plotted in the lower left, middle and right panels in \cref{fig:eg1}, respectively. Not surprisingly, the log-likelihood of the O-GaSP model has modes at all the extreme values in the $L_2$ loss function, including both local minima and maxima. Note that this phenomenon is not caused by the choice of the covariance function, but by the constraint in \cref{equ:const}. Since all {the} extreme values in the $L_2$ loss function satisfy this constraint, the likelihood function of the O-GaSP model is inevitably large at the extreme values as a consequence. Unlike the O-GaSP model, the information in the $L_2$ loss is faithfully expressed in the log likelihoods of the GaSP and S-GaSP. A full numerical comparison between the GaSP and S-GaSP calibrations can be found in the supplementary materials. 

Although \cref{eg:nonlinear} is artificial, this scenario is not unusual in real applications, {where} $\bm \theta$ is often multi-dimensional and the $L_2$ loss function is likely to have multiple extreme values. The likelihood {function} shown in \cref{fig:eg1} by the O-GaSP model is not satisfying, as there is no reason for the likelihood function to favor those local maxima in the $L_2$ loss. To overcome the issue, one could possibly explore {the second-order optimality condition} of the $L_2$ loss function with regard to $\bm \theta$. However, such extension is complicated, as there are $p_x^2$ inequality constraints on the Hessian matrix. It also may not be feasible in practice, as some computer models do not even have the second derivative. 

\section{Calibration and prediction for slow computer models}
\label{sec:calibrate_slow}

This section handles the case where a computer model is slow. When the computer model is computationally expensive to evaluate, the common approach is to use an {emulator}, i.e., a statistical model that {can accurately} approximate the computer model and can be run very fast. Using a GaSP model as the emulator is well-studied in recent literature \cite{bayarri2009using,Gu2016PPGaSP,sacks1989design}. To construct an emulator, we first run the computer model on $D$ design points, denoted as $((\mathbf x^{\mathscr D}_1,\bm \theta^{\mathscr D}_1);...;(\mathbf x^{\mathscr D}_D,\bm \theta^{\mathscr D}_D)) $, usually sampled from the Latin hypercube design \cite{santner2013design}. 
The computer model outputs evaluated at the design points are denoted as $(f^M(\mathbf x^{\mathscr D}_1,\bm \theta^{\mathscr D}_1),...,f^M(\mathbf x^{\mathscr D}_D,\bm \theta^{\mathscr D}_D))^T$. 
We model $f^M(\cdot, \cdot)$ as an unknown function via a GaSP with a mean function $ \mu^M(\cdot,\cdot) $ and a covariance function $ c^M((\cdot,\cdot), (\cdot,\cdot)) $. For any $(\mathbf x, \bm \theta)$, the mean function is still modeled via regression
\[\mu^M(\mathbf x, \bm \theta)= \mathbf h^M(\mathbf x, \bm \theta) \bm \beta^M =\sum^{q_M}_{i=1} h^M_i(\mathbf x, \bm \theta) \beta^M_i, \]
where $\mathbf h^M(\mathbf x, \bm \theta)=(h^M_1(\mathbf x, \bm \theta),..., h^M_{q_M}(\mathbf x, \bm \theta)) $ is the mean vector of basis functions and $ \beta^M_i$ is the $i^{th}$ regression parameter for $h^M_i(\mathbf x, \bm \theta) $, $i=1,...,q_M$. For any two inputs $ (\mathbf x_a, \bm \theta_a)$ and $ (\mathbf x_b, \bm \theta_b)$, the covariance function is again assumed to have a product form 
\begin{equation}
	\sigma^2_M c^{M}( (\mathbf x_a, \bm \theta_a), (\mathbf x_b, \bm \theta_b) )= \sigma^2_M \prod_{i=1}^{p_x }c^{M}_i( x_{ai}, x_{bi})\prod_{j=1}^{p_{\theta}}c^{M}_j( \theta_{aj}, \theta_{bj}),
	\label{equ:product_c_slow}
\end{equation} 
where $c^{M}_i( x_{ai}, x_{bi})$ and $c^{M}_j( \theta_{aj}, \theta_{bj})$ are one-dimensional correlation functions each having an unknown range parameter $\gamma^M_{i,j}$, for $i=1,...,p_x$ and $j=1,...,p_{\theta}$.

The additional parameters introduced by the GaSP emulator are $\sigma^2_M,\, \bm \beta^M$ and $\bm \gamma^M$. To explore the uncertainty of these parameters, a full Bayesian approach can be adopted. However, since the field data is typically much noisier than the computer model outputs, an additional identifiable issue might be caused by combining the emulator and calibration model \cite{bayarri2007framework}. To overcome this issue, a modular approach is often used, which requires the uncertainties of the emulator parameters to be handled only using outputs from the computer model \cite{liu2009modularization}.

Estimating the parameters in an emulator is not trivial, and some routinely used estimators, such as the maximum likelihood estimator, have been widely recognized to be unstable in previous studies \cite{Gu2016thesis,lopes2011development,oakley1999bayesian}. We assume an objective prior for {the} parameters in the GaSP emulator, $\pi(\sigma^2_M, \bm \beta^M, \bm \gamma^M) \propto \pi^R(\bm \gamma^M) / \sigma^2_M$, where $\pi^R(\bm \gamma^M)$ is the reference prior for the range parameters \cite{berger2001objective}. $\sigma^2_M$ and $\bm \beta^M$ can be marginalized out analytically, and $\bm \gamma^M$ is estimated by the marginal posterior mode with the robust parameterization in \cite{Gu2018robustness}. The predictive distribution at any $(\mathbf x^*, \bm \theta^*)$ follows a Student's t-distribution. We omit the details of implementing the emulator due to the limitation of the space. The theoretical justification of the emulator is discussed in {\cite{Gu2018robustness}}, and it is implemented in an R package \cite{gu2018robustgasp_package}. In calibration, we draw from the predictive distribution $p( f^M(\mathbf x^*, \bm \theta^*) \mid f^M(\mathbf x^{\mathscr D}_1,\bm \theta^{\mathscr D}_1),...,f^M(\mathbf x^{\mathscr D}_D,\bm \theta^{\mathscr D}_D)) $ when we need to evaluate a computationally expensive computer {model} at $(\mathbf x^*, \bm \theta^*)$.

\section{Parameter estimation and computation}
\label{sec:computation}

Here we introduce a Bayesian paradigm to assess the parameters uncertainties in both GaSP and S-GaSP models. The shared parameters in both models are $\bm \Theta= [\bm \theta; \bm \beta^{\delta}; \bm \gamma^{\delta}; \sigma^2_{\delta}; \sigma^2_0]$. We first do a transformation {to} define a nugget-variance ratio parameter $\eta=\sigma^2_0/\sigma^2_{\delta}$ and inverse range parameter $\bm \psi^{\delta}_i={1/\bm \gamma^{\delta}_i}$ for $i=1,...,p_x$. The transformed parameters are $\bm {\tilde \Theta}=[\bm \theta; \bm \beta^{\delta}; \bm \psi^{\delta}; \sigma^2_{\delta}; \eta]$. We assume the following prior for $\bm {\tilde \Theta}$ 
\begin{equation}
	\pi(\bm {\tilde \Theta} )\propto \frac{\pi(\bm \theta) \pi(\bm \psi^{\delta},\eta ) }{\sigma^2_{\delta}}, 
	\label{equ:prior_ref}
\end{equation}
where $\pi(\bm \theta)$ and $\pi(\bm \psi^{\delta},\eta )$ are both proper priors.


The prior for the calibration parameters $\bm \theta$ should be chosen based on expert knowledge, as these parameters have real meanings in a computer model. Thus, we do not introduce any specific form herein. The objective priors of the parameters in the covariance matrix of the GaSP model have been studied extensively (see e.g. \cite{berger2001objective,de2007objective,Gu2018robustness,paulo2005default}). However the objective {priors} for the S-GaSP model {have} not been studied. Here we use the jointly robust prior 
\[ \pi(\bm \psi^{\delta},\eta )\propto \parenth{\sum^{p_x}_{i=1} C_i \psi^{\delta}_i+\eta}^a \exp\parenth{-b\parenth{\sum^{p_x}_{i=1} C_i \psi^{\delta}_i+\eta}}, \]
with $a>-p_x-1$, $b>0$ and $C_i>0$ being the prior parameters, for $i=1,...,p_x$. This jointly robust prior is introduced in \cite{gu2018JRprior}. The default choices of the prior parameters for a rectangle $\mathcal X$ are $C_i=|\mathcal X_i| n^{-1/p_x}$, with $|\mathcal X_i|$ being the length of the space of the $i^{th}$ dimension of the variable input, $a=1/2 - p_x$ and $b=1$. The jointly robust prior with the default prior parameters has a moderate penalty on the large correlation in the covariance matrix, which is more helpful for the identifiability problem than the reference prior in the GaSP model \cite{gu2018JRprior}.  	



In the S-GaSP model, the parameter $\lambda$ controls how similar the process to the GaSP model. In all the numerical studies, we let the constraint points be the inputs of the observed data, i.e. $\mathbf x^C_i=\mathbf x_i$, for $i=1,...,n$, and set $\lambda=n/2$ with $n$ being the number of observations. The choice of $\lambda$ is ad-hoc but the S-GaSP model with this choice seems to perform reasonably well in numerical studies. One may develop a prior for the uncertainty in $\lambda$, but we do not pursue this direction in this work.  


With the likelihood in \cref{equ:marginal_lik_scaled} and the prior in \cref{equ:prior_ref}, we use the Metropolis-Hasting algorithm to sample from the posterior distribution. Compare with the GaSP, the additional operations in the S-GaSP are from $\mathbf R_z$ in \cref{equ:R_a_z}, which cost $O(N^3_C)$ and $O(nN^2_C)$ due to the matrix inversion and multiplication, respectively. By choosing $ \mathbf x^C_i= \mathbf x_i $ for $ i = 1, ..., n $, the total number of additional operations becomes $ O(n^3) $. Since the total number of operations in GaSP model is also $ O(n^3) $, the GaSP and S-GaSP models have the same order of computational complexity.

\section{Numerical comparison}
\label{sec:numerical}


In this section, we provide numerical comparisons among several approaches for calibration and prediction. To maintain a fair comparison, the mean function, correlation function, prior distribution, as well as the initial values in the MCMC algorithm are all set to be the same. Specifically, the prior in \cref{equ:prior_ref} with a constant prior for $\bm \theta$, zero mean discrepancy $ \mu^\delta(\xbf) = 0 $, and the product-form correlation function in \cref{equ:product_c} with the Mat{\'e}rn correlation in \cref{equ:Matern} are used. $ S=50,000 $ posterior samples of the parameters $\bm {\tilde \Theta}=(\bm \theta, \bm \beta^{\delta}, \bm \psi^{\delta}, \sigma^2_{\delta}, \eta)$ are generated with the first $ S_0=10,000 $ being the burn-in samples. The code is implemented in the RobustCalibration R Package \cite{Gu2018RobustCalibrationpackage}.



\Cref{subsec:simulation} compares different two-step approaches with the GaSP and S-GaSP calibrations. We denote GaSP+$L_2$ as the two-step $ L_2 $ approach in \cite{tuo2015efficient}, where the reality is first modeled by a GaSP and then $ \thetabm $ is estimated by minimizing the $L_2$ loss. The two-step LS approach in \cite{wong2017frequentist} is denoted as LS+GaSP, where $ \thetabm $ is first estimated by \cref{equ:least_squares} and then the residual is modeled by a GaSP. All optimizations are made based on the low-storage quasi-Newton method with 10 different initializations \cite{nocedal1980updating}. The RobustGaSP package \cite{gu2018robustgasp_package} is used for fitting the GaSP model and another widely-used R package \cite{roustant2012dicekriging} is also included for comparisons in implementing these two-step approaches. A real example of calibrating a geophysical model for the Kilauea Volcano is provided in \Cref{sec:real} to compare the performances between the GaSP and S-GaSP calibrations.


\subsection{Simulated example}
\label{subsec:simulation}
\begin{example}
$y^F( x)=y^R( x) +\epsilon$, where $y^R( x)=sin(10\pi x )+sin(\pi x)$, $f^M( x,\theta)=sin(\theta x )$ and $\epsilon \sim N(0,0.3^2)$. $x_i$ is equally spaced from $[0,1]$ for $i=1,...,n$. 
\label{eg:two_steps_not_working}
\end{example}

	 	 \begin{figure}[t]
		\begin{tabular}{ccc}
			\includegraphics[height=0.3\textwidth,width=.3\textwidth]{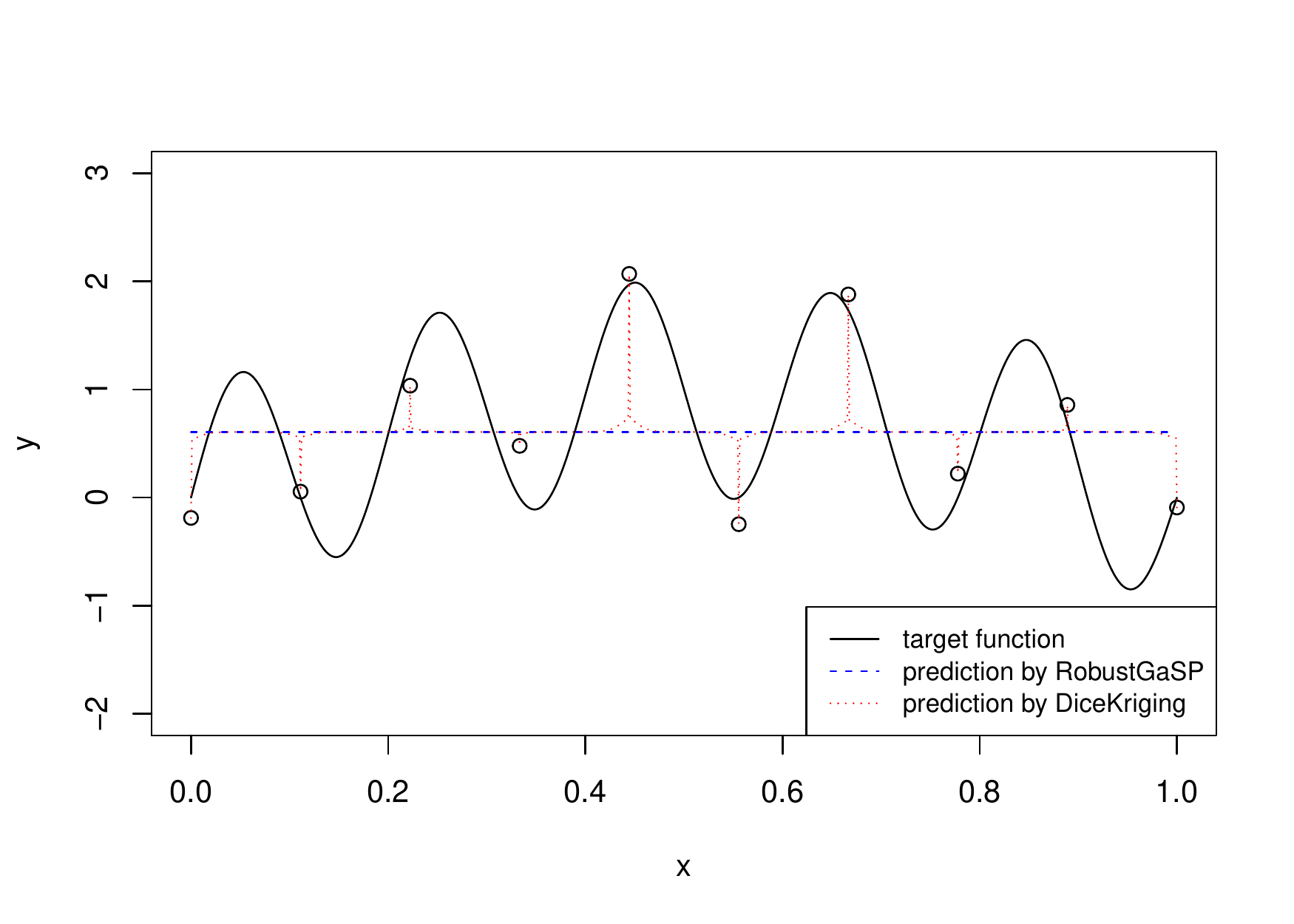}
		    \includegraphics[height=0.3\textwidth,width=.3\textwidth]{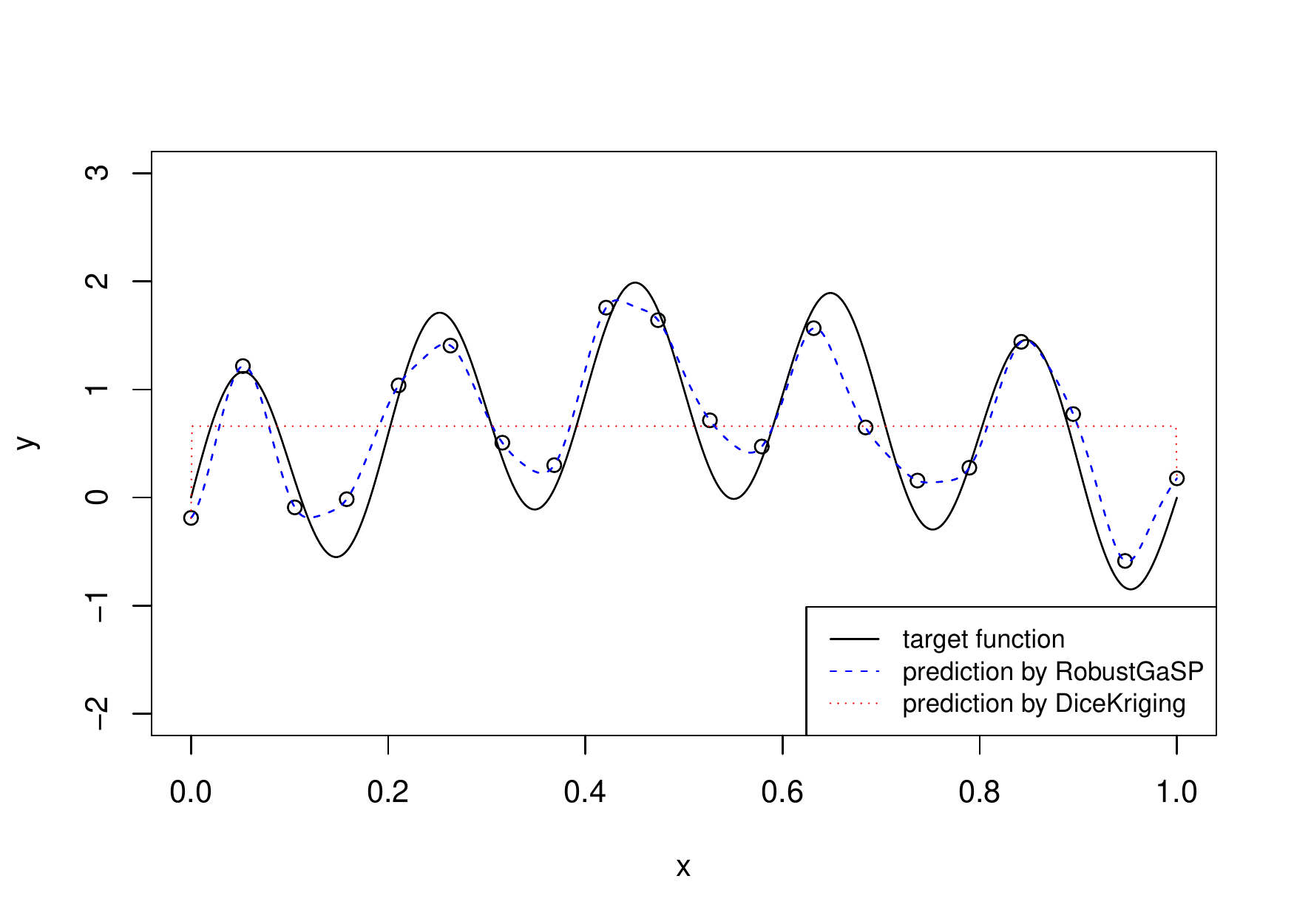}	 
		\includegraphics[height=0.3\textwidth,width=.3\textwidth]{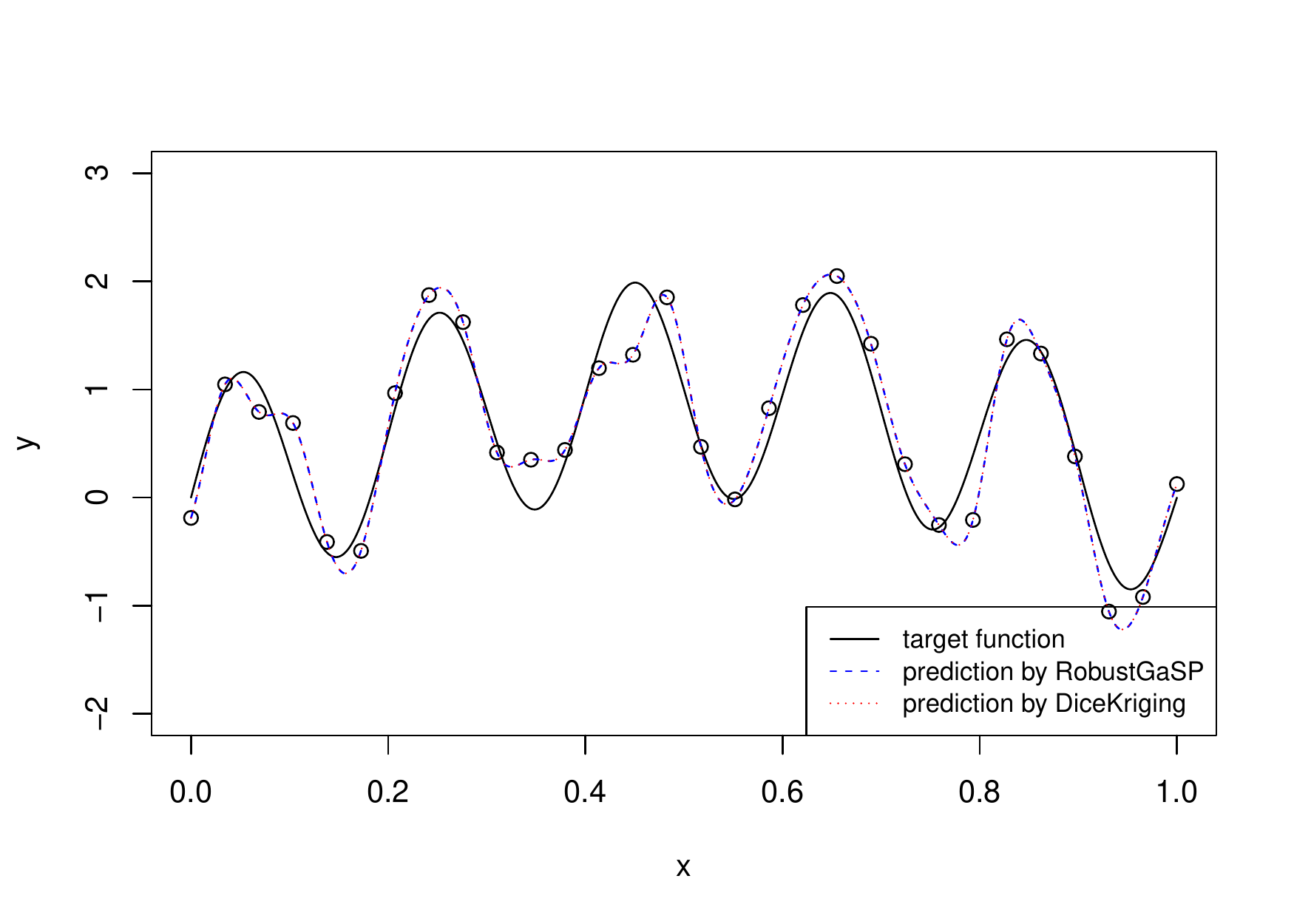} \vspace{-.35in} \\
					\includegraphics[height=0.3\textwidth,width=.3\textwidth]{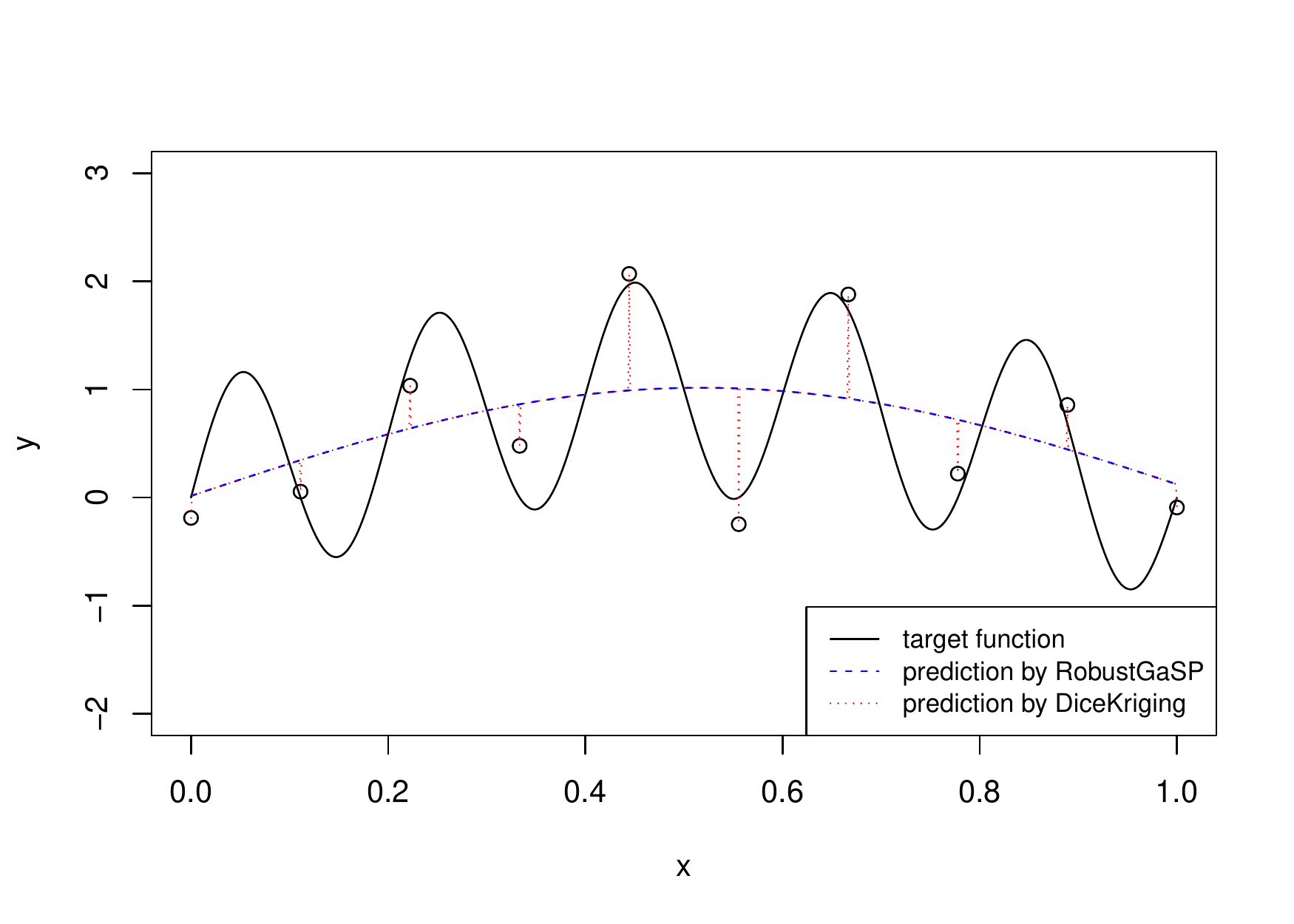}
		    \includegraphics[height=0.3\textwidth,width=.3\textwidth]{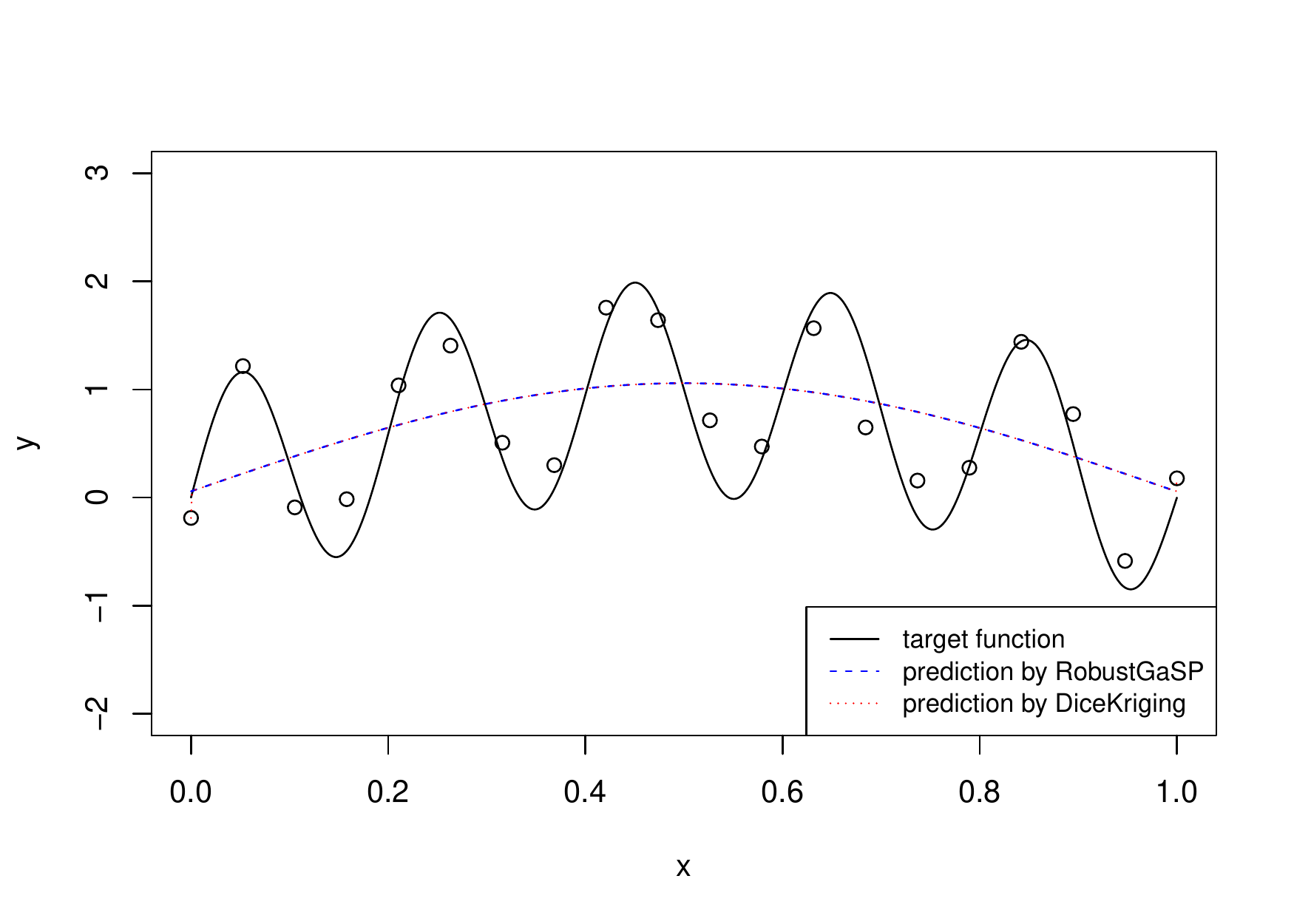}	 
		\includegraphics[height=0.3\textwidth,width=.3\textwidth]{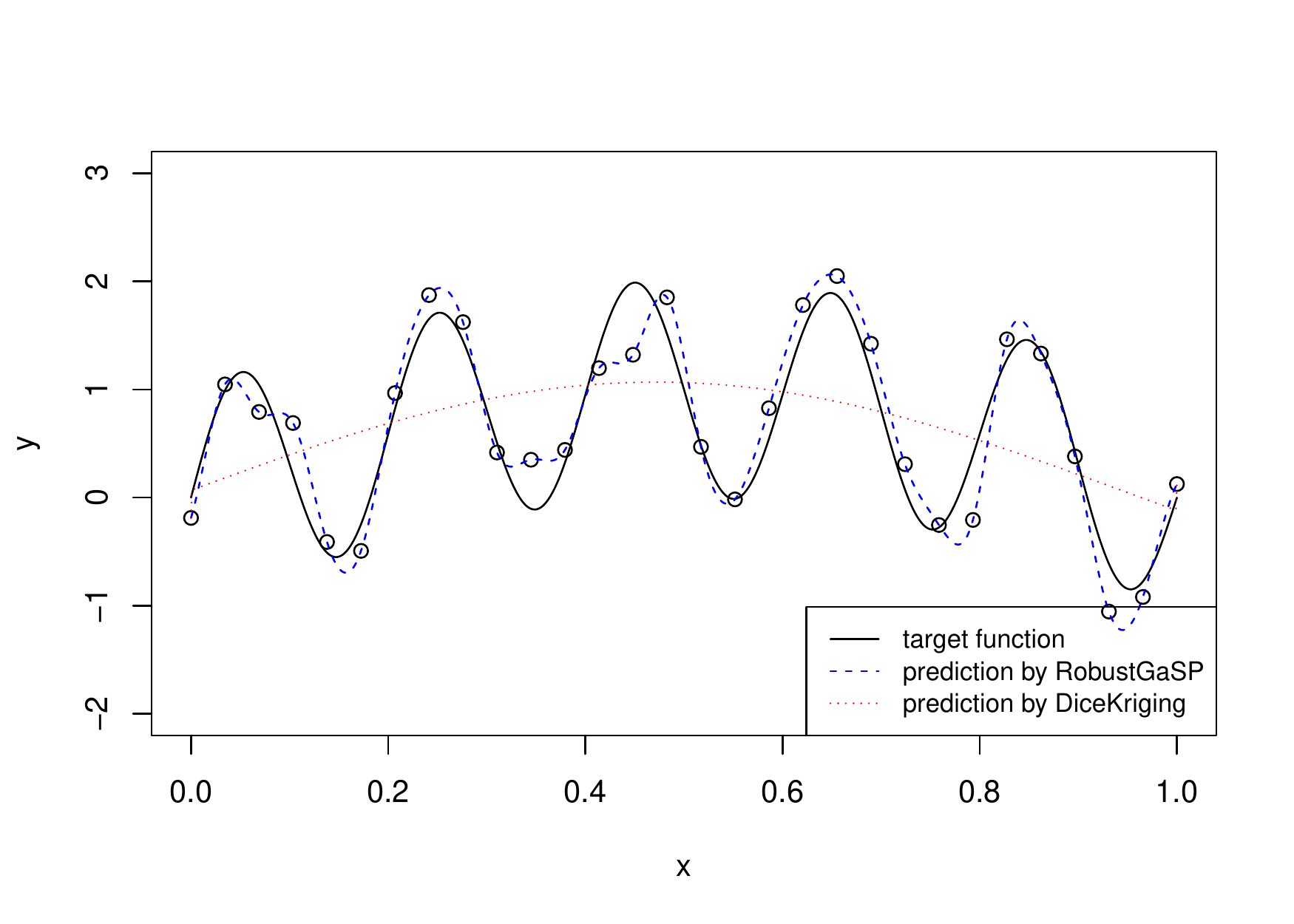} \vspace{-.35in} \\
		\includegraphics[height=0.3\textwidth,width=.3\textwidth]{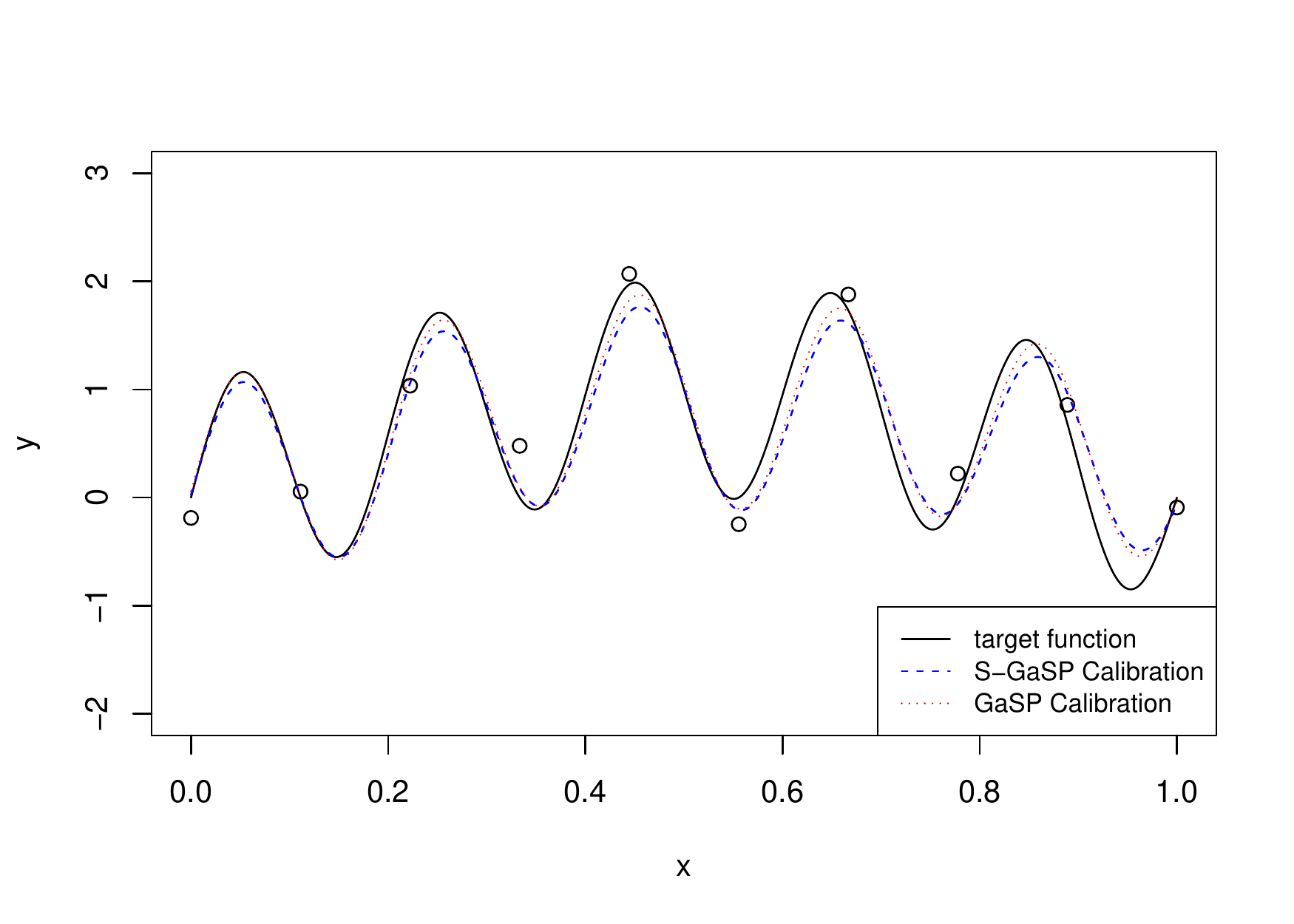}
	\includegraphics[height=0.3\textwidth,width=.3\textwidth]{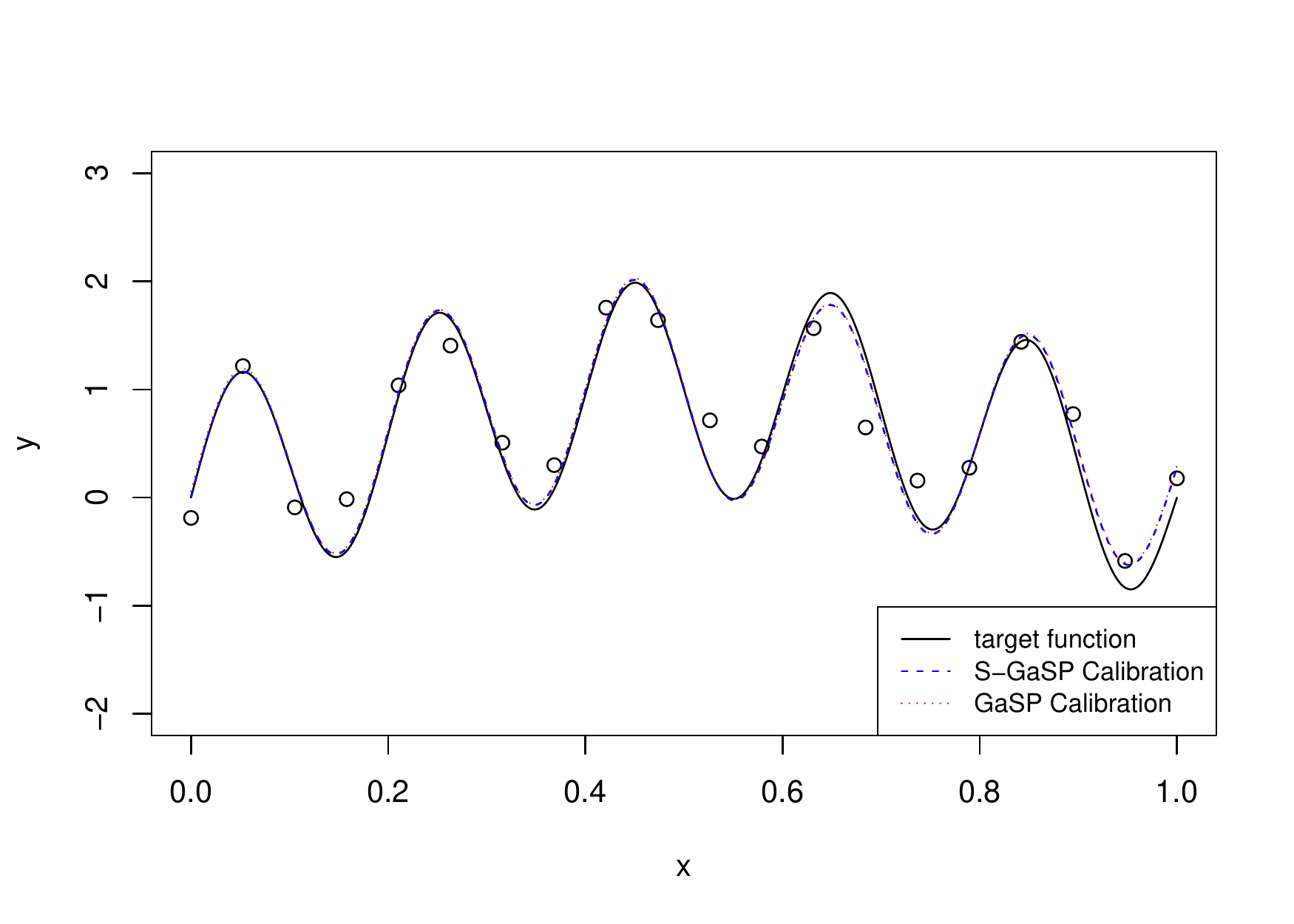}
		\includegraphics[height=0.3\textwidth,width=.3\textwidth]{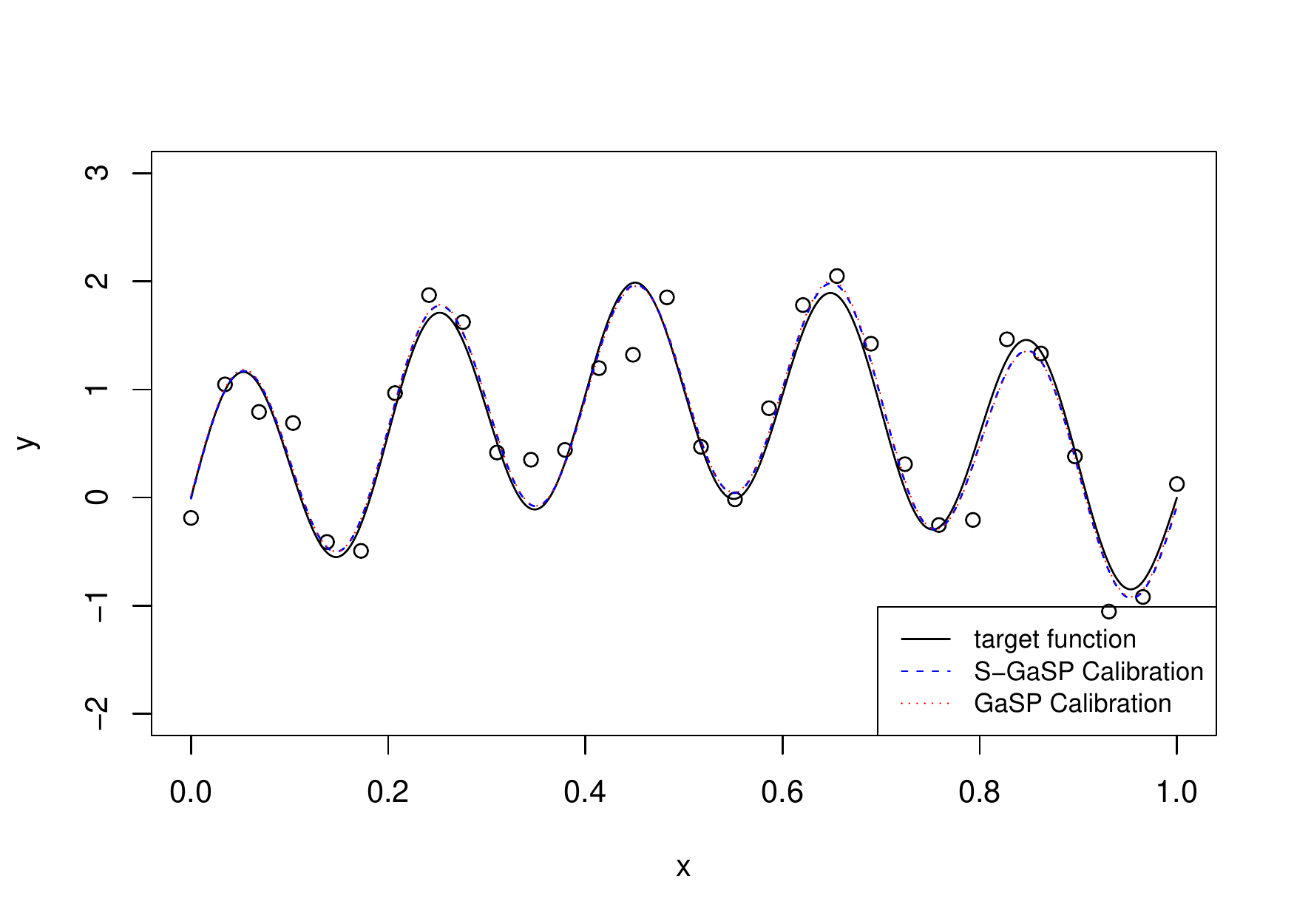}
		\end{tabular}
		\caption{Predictions for Example~\ref{eg:two_steps_not_working}, where the target function is graphed as the black curves. $n=10$, $n=20$ and $n=30$ observations are graphed as the black circles in the left, middle and right panels, respectively. The predictions using the GaSP model without the computer model are graphed in the first row of panels. In the the second row of panels, the calibration parameter is first estimated by the LS estimator and the predictions are from the GaSP model based on the residuals. Two GaSP models are implemented using the RobustGaSP R package \cite{gu2018robustgasp_package} and DiceKriging R package \cite{roustant2012dicekriging}. In the third row of panels,  the predictions combine the mathematical model and discrepancy function modeled as the GaSP and S-GaSP using RobustCalibration R package \cite{Gu2018RobustCalibrationpackage}.}
		\label{fig:eg_two_step_not_working}
	\end{figure}

	\begin{table}[h]
		\caption{Predictive mean squared errors and parameter estimations for Example~\ref{eg:two_steps_not_working}. MSE$_{f^M}$ denotes the mean squared error using only the calibrated computer model and MSE$_{f^M + \delta}$ denotes the mean squared error using both the calibrated computer model and the discrepancy function for prediction.  $\hat \theta$ is the posterior median in the GaSP calibration and S-GaSP calibration. }	
		\label{tab:eg_two_steps_not_working}
		\centering
		\begin{tabular}{lccc}
			\hline
		$n=10$	& MSE$_{f^M}$ & MSE$_{f^M+\delta}$  & $\hat \theta$ \\
			\hline  
	   GaSP $+\mbox{L}_2$ 	& $0.72$  & / & $1.1$  \\
		 $\mbox{LS}+$ GaSP 	& $0.51$  & $0.51$ & $3.0$  \\

		 GaSP calibration	& $0.51$  & $3.4\times 10^{-2}$ & $31$  \\
		 S-GaSP calibration & $0.47$  & $4.9\times 10^{-2}$ & $31$ \\
     \hline
     		$n=20$	& MSE$_{f^M}$ & MSE$_{f^M+\delta}$  & $\hat \theta$ \\
			\hline  
	  GaSP $+\mbox{L}_2$	& $0.50$  & / & $3.2$  \\
		 $\mbox{LS}+$ GaSP 	& $0.50$  & $8.9\times 10^{-2}$ & $3.1$  \\

		 GaSP calibration	& $0.50$  & $8.1\times 10^{-3}$ & $31$  \\
		 S-GaSP calibration & $0.50$  & $7.1\times 10^{-3}$ & $31$ \\
     \hline
     		$n=30$	& MSE$_{f^M}$ & MSE$_{f^M+\delta}$  & $\hat \theta$ \\
			\hline  
	   GaSP $+\mbox{L}_2$	& $0.49$  & / & $3.4$  \\
		 $\mbox{LS}+$ GaSP 	& $0.50$  & $6.8\times 10^{-2}$ & $3.3$  \\

		 GaSP calibration	& $0.50$  & $4.6\times 10^{-3}$ & $31$  \\
		 S-GaSP calibration & $0.50$  & $3.8\times 10^{-3}$ & $31$ \\
		 \hline 
		\end{tabular}
	\end{table}


Similar to \Cref{eg:eg_variance}, the goal in \Cref{eg:two_steps_not_working} is to estimate $\theta$ and predict $y^F(x^*_i)$ at held-out $x^*_i$, uniformly sampled from $[0, 1]$ for $ i = 1, ..., 1000 $. Under various sample sizes, the results of calibration and prediction are presented in \Cref{tab:eg_two_steps_not_working} and \Cref{fig:eg_two_step_not_working}, respectively. When sample size is small, since the truth contains a high frequency term $ \sin(10\pi x) $ that is difficult to be captured by the GaSP without the computer model, the GaSP+$L_2$ approach performs poorly in prediction, shown in the first row of \Cref{fig:eg_two_step_not_working}. Because the loss of predictive accuracy in GaSP+$L_2$, the estimated $ \hat{\theta} $ is around $ \pi $ for $n=20$ and $n=30$ shown in \Cref{tab:eg_two_steps_not_working}. 


The other two-step approach in \cite{wong2017frequentist} estimates the calibration parameter by minimizing \cref{equ:least_squares}. Since this approach does not penalize the model complexity, $\theta$ is estimated close to $\pi$ rather than $10\pi$, shown in Table~\ref{tab:eg_two_steps_not_working}, which makes the GaSP model imprecise for the residuals. As shown in the second row of \Cref{fig:eg_two_step_not_working}, the predictions are not satisfactory. 

In contrast, both the GaSP and S-GaSP estimate $ \theta $ close to $ 10\pi $ as shown in \Cref{tab:eg_two_steps_not_working}, allowing the high frequency term to be explained by the computer model. Consequently, the predictions of GaSP and S-GaSP are accurate even when the sample size is small, shown in the last row of \Cref{fig:eg_two_step_not_working}. In practice, the computer model is developed for reproducing the reality, so combining it with the discrepancy function typically improves the predictive accuracy \cite{kennedy2001bayesian}.

\subsection{Real example: calibration of the geophysical model for the Kilauea Volcano}
\label{sec:real}
\begin{figure}[tbhp]
	\centering
	\includegraphics[height=.3\textwidth,width=\textwidth]{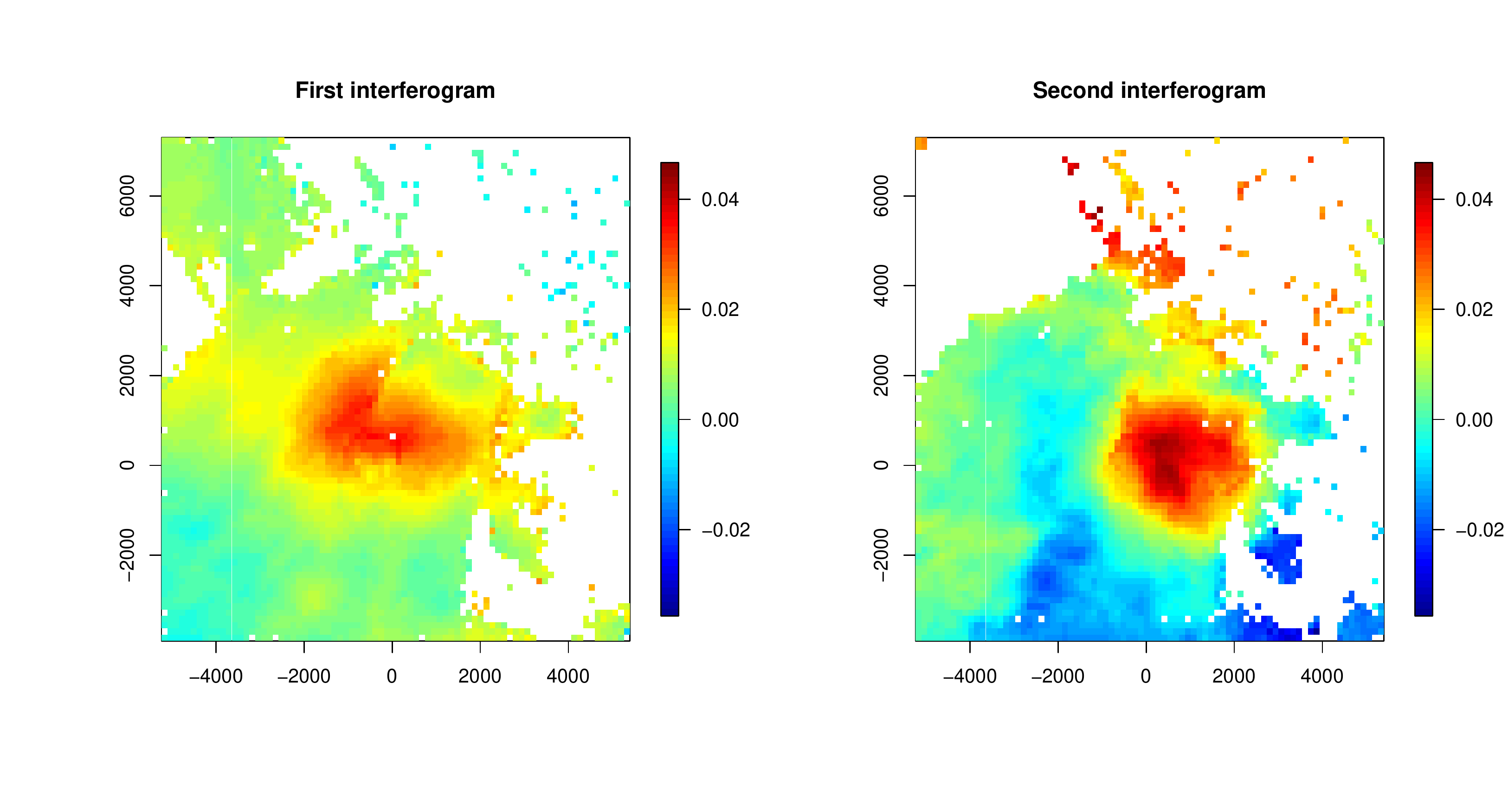} \vspace{-.4in} \\
		\includegraphics[height=.3\textwidth,width=\textwidth]{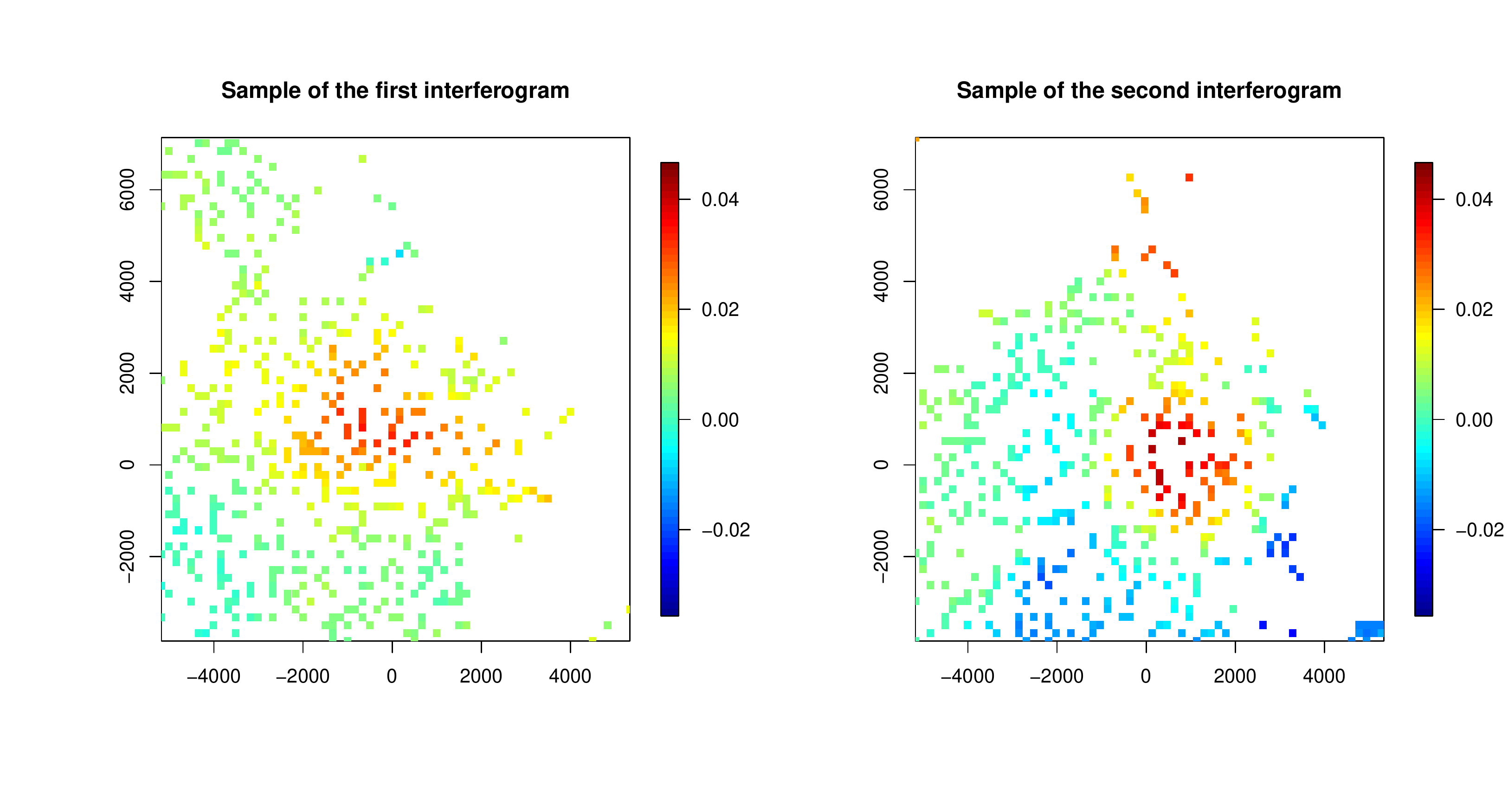}
			\vspace{-.2in}
	\caption{Two interferograms of Kilauea Volcano are graphed in the first row and 500 uniform samples for each interferogram are graphed in the second row.}
	\label{fig:Kilauea_t3_data}
\end{figure}

We consider the geophysical model of Kilauea Volcano introduced in \cite{anderson2016bayesian, anderson2017abundant} for the study of the magma supply rate for Kilauea and the carbon concentration in Earth's mantle. Several different types of data were used in this study, including the $\mbox{SO}_2$ and $\mbox{CO}_2$ emission data, and Interferometric synthetic aperture radar (InSAR) data, a radar technique to estimate the ground deformation in centimeters \cite{massonnet1998radar}. In \cite{anderson2016bayesian}, the parameters in the geophysical model are estimated using a Bayesian method without assuming a discrepancy function. 



For the demonstration purpose, we limit ourselves to calibrate one part of the geophysical model in \cite{anderson2016bayesian} -- the displacement of the ground's surface caused by addition of magma to a spherical reservoir. We use one ascending-mode and one descending-mode COSMO-SkyMed interferogram shown in \cite{anderson2016bayesian}, spanning the period of time from October 21, 2011 to May 16, 2012 and October 20, 2011 to May 15, 2012, respectively. These interferograms are graphed in the first row in \cref{fig:Kilauea_t3_data}, showing that the ground deformation caused by the volcano is between 0 to 4 centimeters during this period of time. As the number of pixels in the InSAR data is large, a Quadtree algorithm is used for downsampling before calibration \cite{anderson2016bayesian}. The Quadtree algorithm converts each interferogram to an image with only several hundred boxes, where the size of the box is small if the values within the box change rapidly corresponding to where the deformation gradients are high. The Quadtree algorithm is designed for visualization purposes, however, as it favors those area with high ground deformations, it may cause potential bias in the calibration. Here we uniformly sample 500 points from each interferogram and use them as our training data for the calibration and prediction. Two variable inputs and five calibration parameters of this geophysical model are provided in \cref{tab:description}.
\begin{table}[tbhp]
	\caption{Input variables and calibration parameters of the geophysical model for Kilauea Volcano in 2011 to 2012.}
	\label{tab:description}
	\centering
	\begin{tabular}{lll} \hline
		Variable Input ($\mathbf{x}$) & Name & Description \\ \hline
		${x}_1$ & Latitude & Spatial coordinate\\
		${x}_2$ & Longitude & Spatial coordinate \\ \hline
		Parameter ($\bm{\theta}$) & Name & Description \\ \hline
		${\theta}_1 \in [-2000,3000]$ & Chamber east ($m$) & Spatial coordinate for the chamber\\
		${\theta}_2 \in [-2000,5000]$ & Chamber north ($m$) & Spatial coordinate for the chamber\\
		${\theta}_3 \in [500,6000]$ & Chamber depth ($m$) & Depth of the chamber\\
		${\theta}_4 \in [0,0.15]$ & Res. vol. change rate ($m^3/s$) & Volume change rate of the reservoir\\
		${\theta}_5 \in [0.25,0.33]$ & Poisson's ratio & Host rock property \\ \hline
	\end{tabular}
\end{table}

\begin{figure}[tbhp]
	\centering
	\includegraphics[height=.5\textwidth,width=\textwidth]{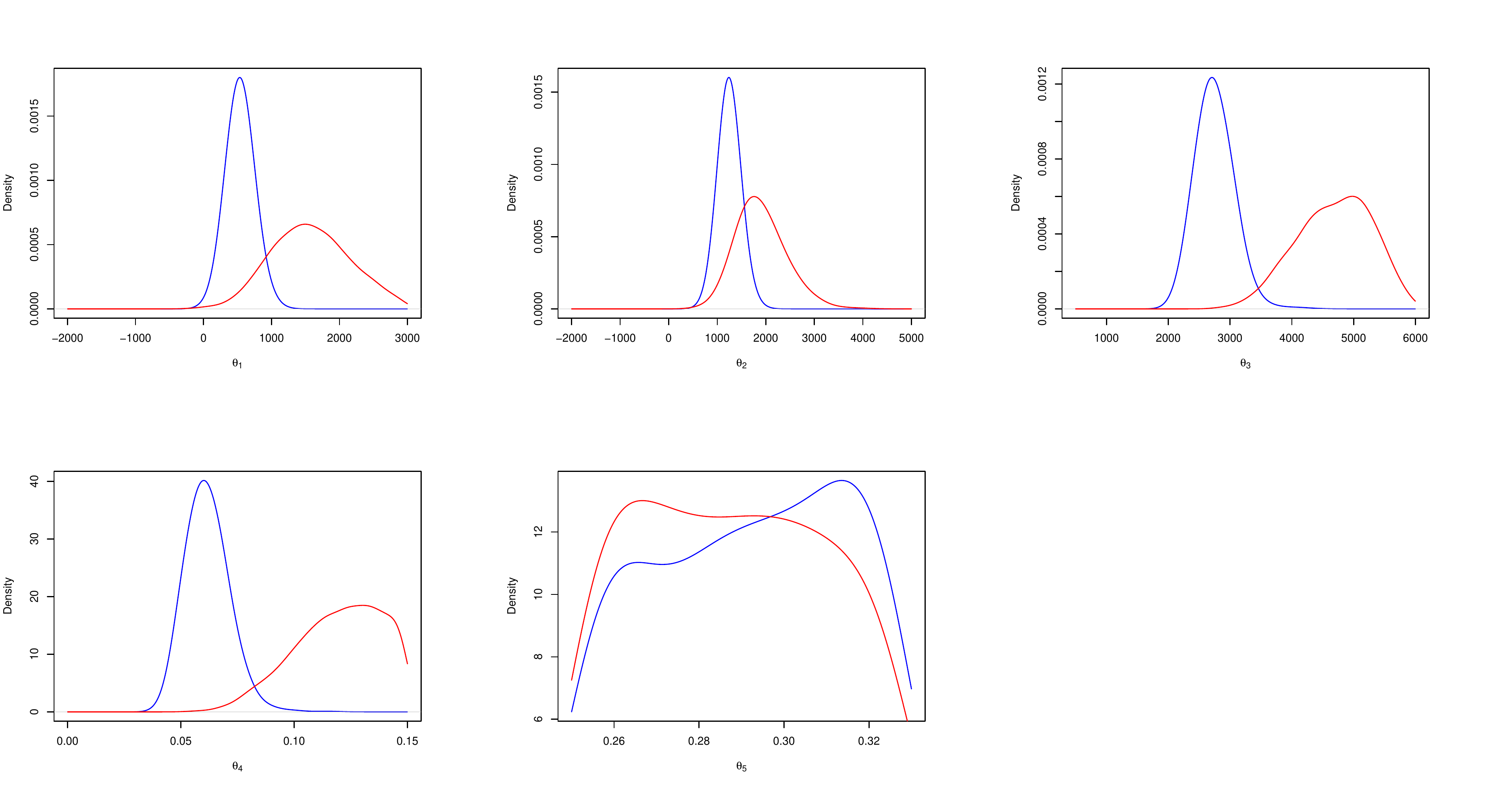}
	\caption{Marginal posterior densities of $\bm \theta$ from the GaSP calibration (red curves) and S-GaSP calibration (blue curve). }
	\label{fig:marginal_post}
\end{figure}

The marginal posterior densities of $\bm \theta$ by the GaSP and S-GaSP calibration are graphed as the red and blue curves in \cref{fig:marginal_post}, respectively. The posterior mass of the GaSP spreads widely throughout its domain, and a geophysical model with a deep chamber and high volume change rate of the reservoir is preferred. The uncertainties of location of the chamber from the GaSP calibration also seem quite large. In comparison, the S-GaSP suggests a geophysical model with a much smaller chamber depth and a low reservoir volume change rate. In particular, the posterior medians of the depth of the chamber and reservoir volume change rate are around 2700 meters and 0.06 $m^3/s$, respectively. Both values are close to the results in \cite{anderson2016bayesian}, which reports around 2300 meters for the depth of the chamber and 0.05 $m^3/s$ for the reservoir volume change rate.



\begin{figure}[tbhp]
	\centering
	\includegraphics[height=.3\textwidth,width=1\textwidth]{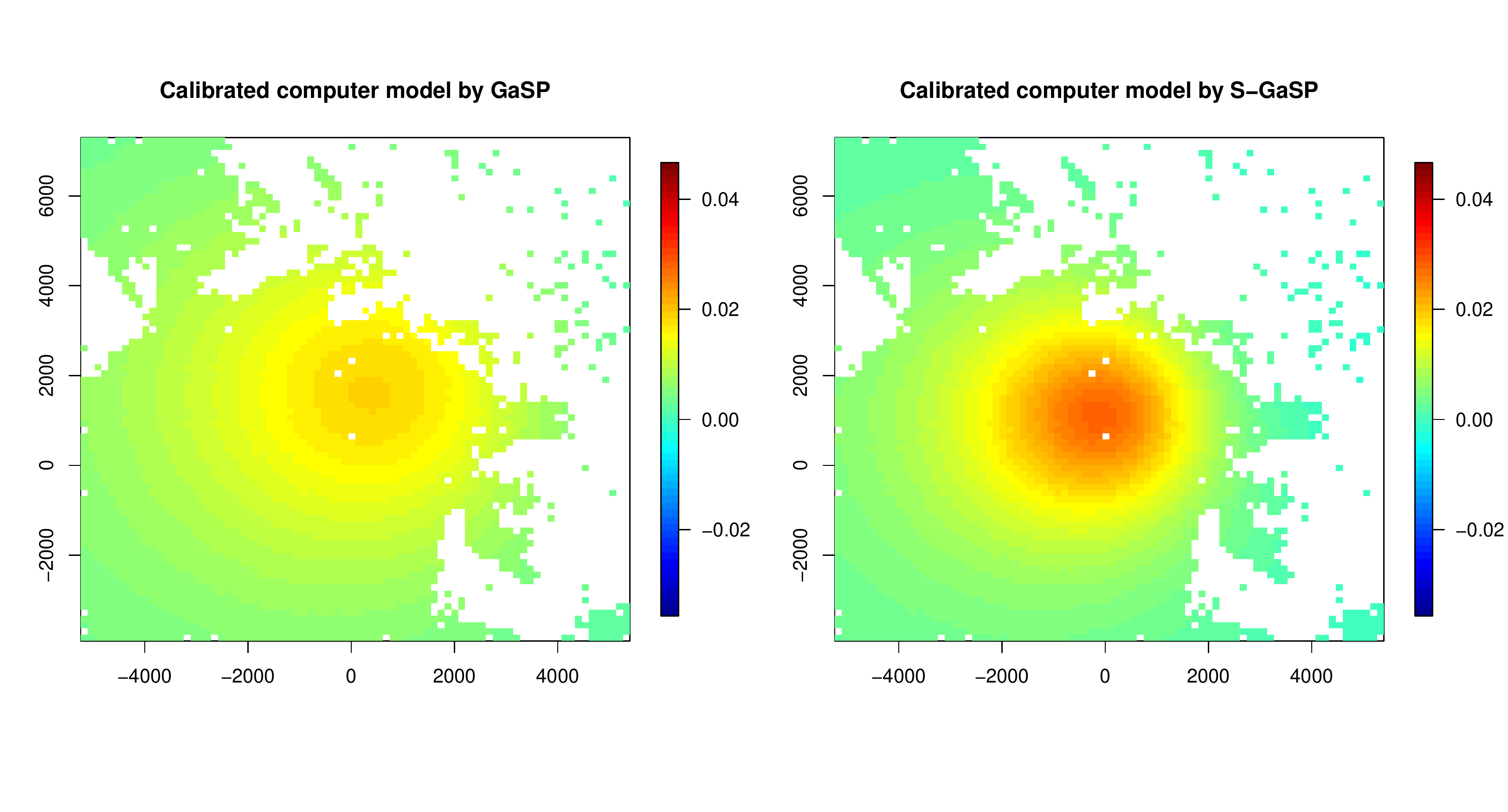}	\vspace{-.4in} \\
		\includegraphics[height=.3\textwidth,width=1\textwidth]{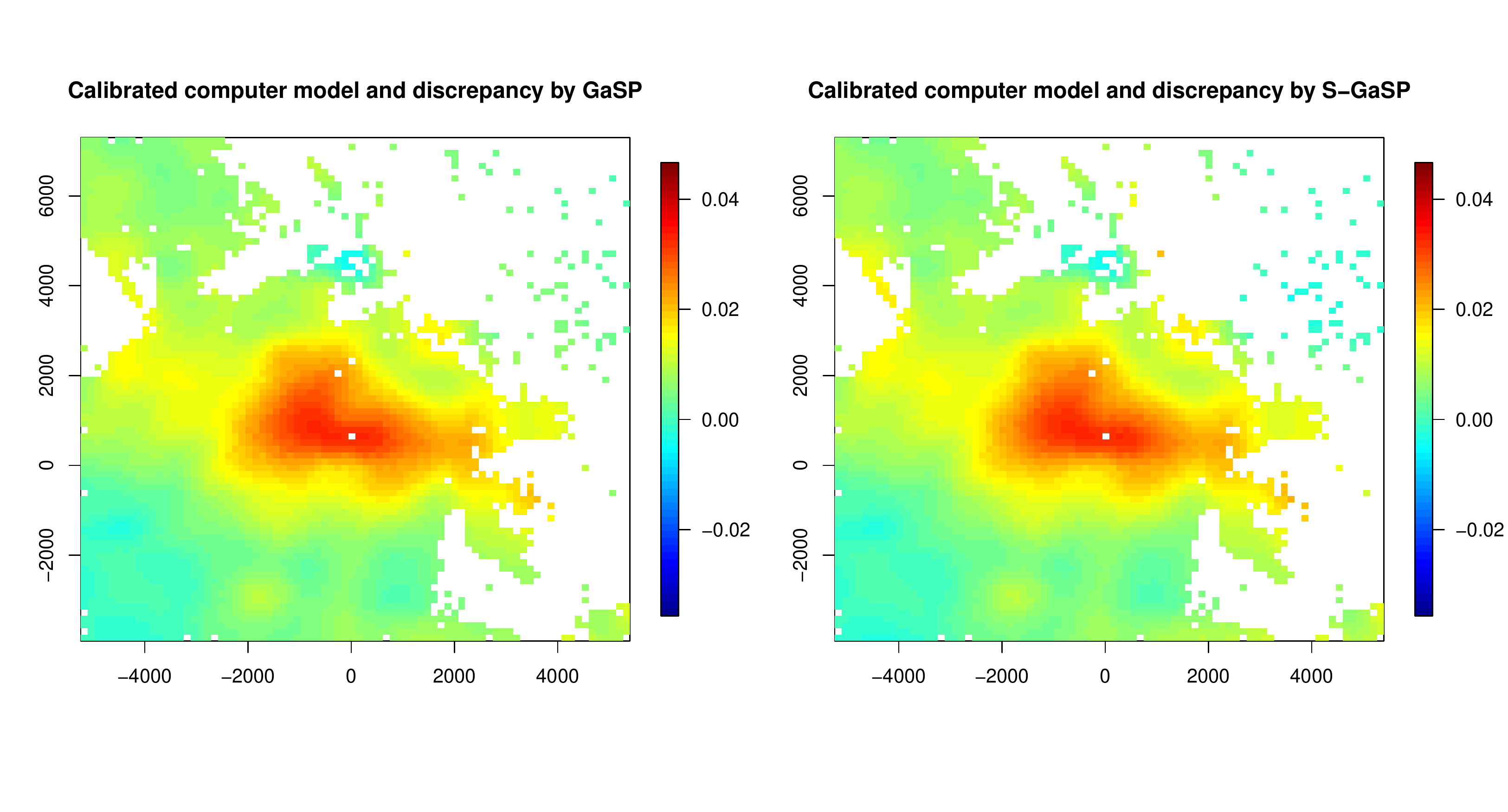}
			\vspace{-.2in}	
	\caption{Predictions by the GaSP and S-GaSP models for the first interferogram. The predictions of the calibrated geophysical model by the GaSP and S-GaSP calibration are graphed in the upper left and upper right panel, respectively. The predictions of the calibrated geophysical model and discrepancy function are graphed in the lower panels. }
	\label{fig:pred_first_image}
	
\end{figure}

\begin{figure}[tbhp]
	\centering
	\includegraphics[height=.3\textwidth,width=1\textwidth]{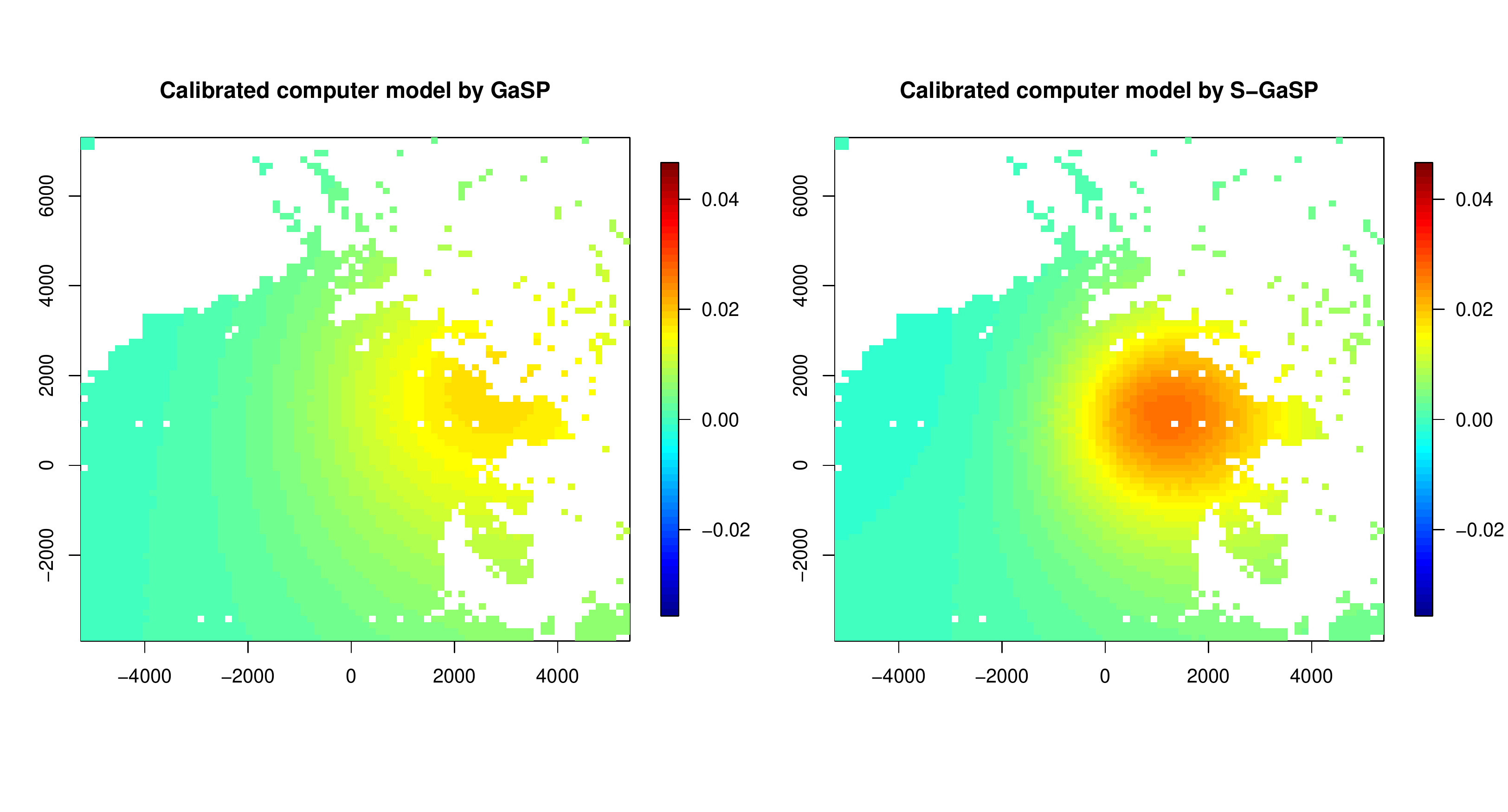} 	\vspace{-.4in} \\
				\includegraphics[height=.3\textwidth,width=1\textwidth]{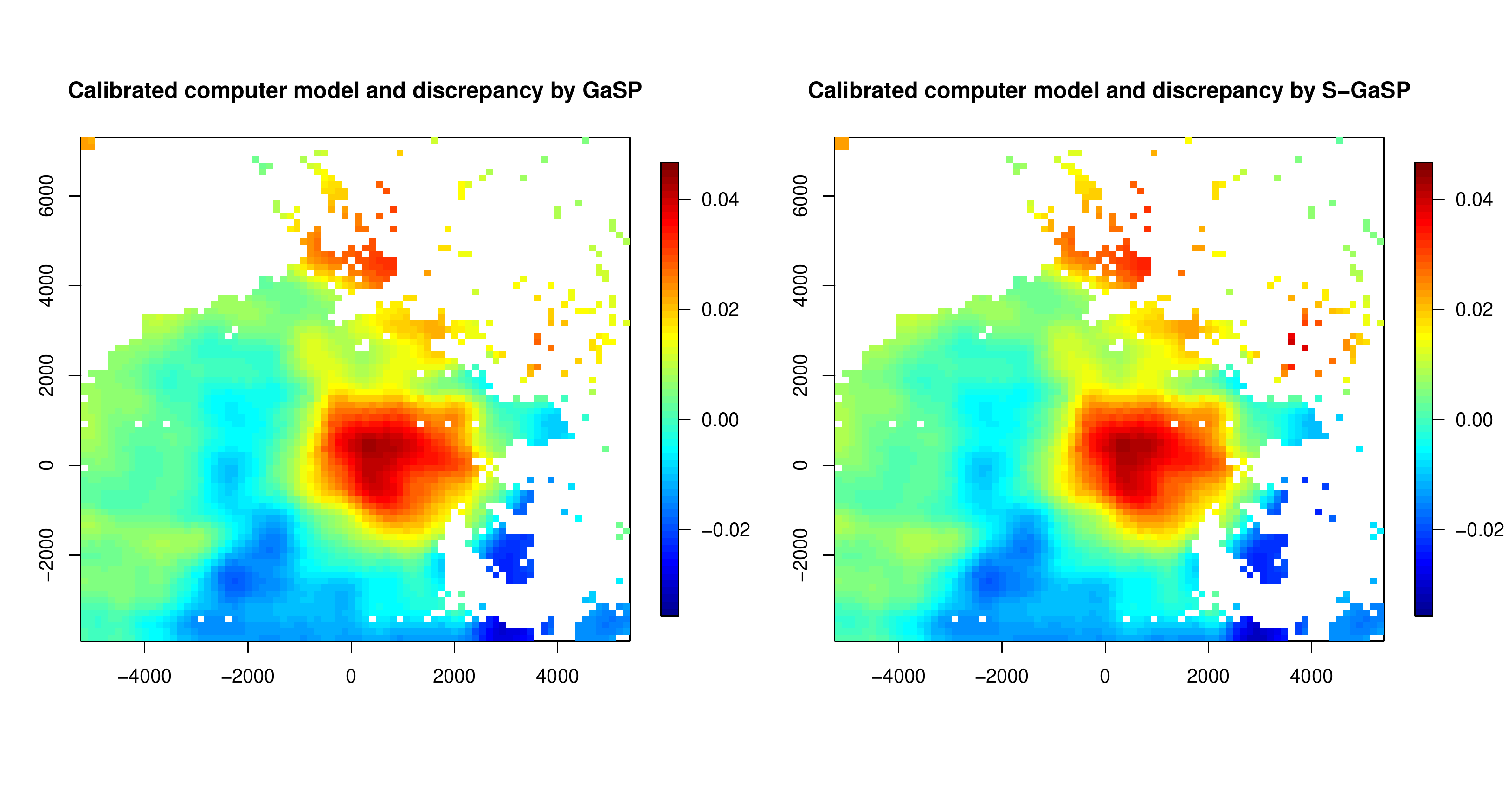}	
					\vspace{-.2in}
	\caption{Predictions by the GaSP and S-GaSP models for the second interferogram. The predictions of the calibrated geophysical model by the GaSP and S-GaSP calibration are graphed in the upper left and upper right panel, respectively. The predictions of the calibrated geophysical model and discrepancy function are graphed in the lower panels.}
	\label{fig:pred_second_image}
\end{figure}

The differences of the marginal posterior distributions between the GaSP and S-GaSP calibration result in large differences in predictions by the calibrated geophysical model, shown in the upper panels in \cref{fig:pred_first_image} and \cref{fig:pred_second_image} for the first and second interferograms, respectively. As the calibrated geophysical model from the GaSP calibration has a deep chamber of the volcano, the ground deformation caused by this geophysical model is very small, with the maximum change being less than 2 centimeters. The predictive mean squared errors by the GaSP calibration are $2.8\times 10^{-5}$ and $1.5\times 10^{-4}$ for the first and second interferograms, respectively.
Alternatively, the maximum ground deformation reported in the S-GaSP calibration is around 3 centimeters, which is larger than the one by the GaSP calibration, since a comparatively shallow chamber of the reservoir can cause a larger ground displacement in a smaller area. The predictive mean squared errors by the S-GaSP calibration are $2.4\times 10^{-5}$ and $1.2\times 10^{-4}$, which are both smaller than the ones by the GaSP model. 

 
Note the maximum ground deformations by the S-GaSP calibration are still a little smaller than maximum values in the held-out images, which might be caused by the flaws in the interferograms. For example, the ground deformation in the northern part of the second interferogram is very large, which can be caused by the factors other than the volcanic activities, such as the air and ground conditions. This phenomenon is common in the InSAR data, and manually deleting these flawed pixels may be possible but costly. The S-GaSP calibration is more robust than the GaSP calibration using the imperfect data.



%


As shown in the lower panels in \cref{fig:pred_first_image} and \cref{fig:pred_second_image}, for both the GaSP and S-GaSP, the predictions by the calibrated geophysical model and discrepancy are better than using the calibrated computer model alone. The predictive mean squared errors by the GaSP calibration are $3.4 \times 10^{-6}$ and $7.4\times 10^{-6}$, while the predictive mean squared errors by the S-GaSP calibration are $3.5 \times 10^{-6}$ and $7.4\times 10^{-6}$ for the first and second interferograms, respectively. The predictive errors by both models are very small, compared to the errors by the calibrated geophysical model alone. 

\section{Concluding remarks}
\label{sec:conclusion}
 %

We have introduced a new approach, called the scaled Gaussian stochastic process (S-GaSP), for modeling the discrepancy function in calibrating imperfect computer models. The new approach bridges the gap between the $L_2$ calibration and GaSP calibration. Unlike the GaSP model, the calibrated computer model can predict the reality reasonably well, even when the field data are strongly correlated. We also show numerically that the S-GaSP model is as good as the GaSP model in prediction by combining the calibrated computer model and discrepancy function. Under the Bayesian framework, a computationally feasible approach is proposed for the S-GaSP calibration and its computational complexity is shown to be the same as the GaSP calibration. Both simulated and real examples demonstrate the benefits of using S-GaSP in calibration and prediction.

\appendix
\section{Proofs}
\label{sec:appendix}
\begin{proof}[proof of \cref{lemma:chi-square}]
	First by Karhunen-Lo\`{e}ve expansion, one has 
	\begin{equation}
		\frac{\delta(\mathbf x)- \mu(\mathbf x) }{\sigma}= \sum_{i=1}^{\infty} \sqrt{\lambda_i} Z_i \phi_i(\mathbf x), 
		\label{equ:KL}
	\end{equation}
	where $Z_i$ is a mean-zero unit-variance normal random variable, $\lambda_i$ and $\phi_i$ are the $i^{th}$ eigenvalue and {normalized} eigenfunction of the kernel $ c(\cdot,\cdot)$ with regard to Lebesgue measure, respectively. $\{\phi_i\}^{\infty}_{i=1}$ is the orthonormal basis functions of the space of $L_2$ integrable functions defined on $\mathcal X$. As $||\mu(\cdot) ||_{L_2(\mathcal X) }<\infty$, one has 
	\begin{equation}
		\mu(\mathbf x)=\sum^{\infty}_{i=1} \mu_i \phi_i(\mathbf x), 
		\label{equ:mu}
	\end{equation}
	for some coefficients $\mu_i$, since
	\begin{align}
		\langle \mu(\cdot), \phi_i(\cdot) \rangle_{L_2(\mathcal X)}
		&=\left\langle \sum^{\infty}_{i=1} \mu_i \phi_i(\cdot), \phi_i(\cdot) \right\rangle_{L_2(\mathcal X)} \nonumber \\
		&=\mu_i \langle \phi_i(\cdot), \phi_i(\cdot) \rangle_{L_2(\mathcal X)} \nonumber\\
		&=\mu_i,
		\label{equ:mu_i_1}
	\end{align}
	where $ \langle f(\cdot), g(\cdot) \rangle_{L_2(\mathcal X)}= \int_{\mathcal X} f(x)g(x)dx $ for any $L_2$ integrable function $f(\cdot)$ and $g(\cdot)$ on $\mathcal X$. 
	{The last two equalities are due to the linearity of $L_2$-inner product and orthonormality of $\{\phi_i\}^{\infty}_{i=1} $, respectively.}
	
	Plugging \cref{equ:mu} into \cref{equ:KL}, one has
	\begin{equation}
		\delta(\mathbf x)
		={\sigma\sum_{i=1}^{\infty} \parenth{\sqrt{\lambda_i} Z_i \phi_i(\mathbf x) + \frac{\mu_i\phi_i(\xbf)}{\sigma}}
			=} \sigma \sum_{i=1}^{\infty} \sqrt{\lambda_i} \phi_i(\mathbf x)\left(Z_i+ \frac{\mu_i}{\sqrt{\lambda_i} \sigma}\right) , 
		\label{equ:result_chi_square}
	\end{equation}
	where $\mu_i$ is given in \cref{equ:mu_i_1}. The result thus follows.
\end{proof}

\begin{proof}[proof of \cref{lemma:connection}]   
	Assume $ f_Z(z) = f_Z(z|\Thetabm) = C > 0 $, then \cref{equ:p_z} becomes
	\begin{equation}
		p_{\delta_z}(z\mid\Thetabm)
		= \frac{C\cdot p_\delta\parenth{Z=z \mid\Thetabm} }{\int_0^\infty C\cdot p_\delta\parenth{Z=t \mid\Thetabm}dt} = p_\delta\parenth{\smallintdelta\mid\Thetabm}.
		\label{equ:p_z_c}
	\end{equation}
	Plugging \cref{equ:p_z_c} into \cref{equ:integrate_z}, one has
	\begin{align}
		p_{\delta_z}(\deltabmz\mid\Thetabm)
		&= p_\delta(\deltabmz\mid\Thetabm) \int_0^\infty \frac{p_\delta\parenth{Z=z\mid\deltabmz,\Thetabm}}{p_\delta\parenth{Z=z\mid\Thetabm}} p_\delta\parenth{\smallintdelta\mid\Thetabm} dz \nonumber\\
		&= p_\delta(\deltabmz\mid\Thetabm) \int_0^\infty p_\delta\parenth{Z=z\mid\deltabmz,\Thetabm} dz \nonumber\\
		&= p_\delta(\deltabmz\mid\Thetabm),
	\end{align}
	from which the results follow.
\end{proof}

\begin{proof}[proof of \cref{lemma:density_delta_z}]   
	Since $ \delta(\cdot)\mid\deltabmz,\Thetabm\sim\GaSP({\mu}^{*\delta}(\cdot), \sigma^2_{\delta} {c}^{*\delta}(\cdot, \cdot)) $, from \cref{lemma:chi-square} one has
	\begin{equation}
		\int_{\xbf\in\Xcal} \delta(\xbf)^2d\xbf \mid\deltabmz, \Thetabm \sim \sigma_\delta^2\sum_{i=1}^\infty\lambda_i^*\chi^2_{a_i^*}(1),
	\end{equation}
	where $ a_i^* = (\mu_i^*)^2/(\lambda_i^*\sigma_\delta^2) $ with $ \mu_i^* = \int_{\xbf\in\Xcal} \mu^{*\delta}(\xbf)\phi_i^*(\xbf)d\xbf $. Denote $ M_X(t) $ as the moment generating function for $ X $, i.e., $ M_X(t) = \E\sqbracket{e^{tX}} $. Assume $ X_i $ follows a non-central chi-squared distribution with 1 degree of freedom and the non-central parameter $ a_i^* $, then $ M_{X_i}(t) = (1-2t)^{-1/2}e^{a_i^*t/(1-2t)} $ for $ t < 1/2 $. Moreover, let $ S = \sigma_\delta^2 \sum_{i=1}^\infty \lambda_i^* X_i $, then
	\begin{equation}
		M_S(t) = \E\sqbracket{e^{\sigma_\delta^2t \sum_{i=1}^\infty \lambda_i^*X_i}} = \prod_{i=1}^\infty \E\sqbracket{e^{\sigma_\delta^2t\lambda_i^*X_i}} = \prod_{i=1}^\infty M_{X_i}(\sigma_\delta^2t\lambda_i^*).
		\label{equ:S_mgf}
	\end{equation}
	{Let $\tilde \lambda = \lambda/(2\sigma^2_{\delta} \Vol(\mathcal X))$. Using \cref{equ:S_mgf}, $ \bone $ can be computed as follows 
		\begin{align}
			\bone
			&= \int_0^\infty p_\delta\parenth{Z=z\ mid\deltabmz,\Thetabm} \tilde \lambda e^{-\tilde \lambda z}dz \nonumber\\
			&= \tilde \lambda M_{S}(-\tilde \lambda) \nonumber\\
			&= \tilde \lambda \prod_{i=1}^\infty M_{X_i}(-\sigma_\delta^2\tilde \lambda\lambda_i^*) \nonumber\\
			&= \tilde \lambda \sqbracket{\prod_{i=1}^\infty (1+2\sigma_\delta^2\tilde \lambda\lambda_i^*)^{-1/2}} e^{-\sum_{i=1}^\infty\frac{a_i^*\sigma_\delta^2 \tilde \lambda\lambda_i^*}{1+2\sigma_\delta^2\tilde \lambda\lambda_i^*}}.
			\label{equ:bone_mgf}
	\end{align}}
	
	For the following term in \cref{equ:bone_mgf}, one has 
	\begin{align}
		\label{equ:a_sigma_lambda_i}
		a_i^*\sigma_\delta^2\lambda_i^*
		&= \sqbracket{\int_{\xbf\in\Xcal} \mu^{*\delta}(\xbf)\phi_i^*(\xbf)d\xbf}^T \sqbracket{\int_{\xbf\in\Xcal} \mu^{*\delta}(\xbf)\phi_i^*(\xbf)d\xbf} \nonumber\\
		&= \sqbracket{\int_{\xbf\in\Xcal} \rbf^{\delta}(\xbf)^T \left(\Rbf^\delta\right)^{-1} \deltabmz\phi_i^*(\xbf)d\xbf}^T \sqbracket{\int_{\xbf\in\Xcal} \rbf^{\delta}(\xbf)^T \left(\Rbf^\delta\right)^{-1} \deltabmz\phi_i^*(\xbf)d\xbf} \nonumber\\
		&= \deltabmz^T \left(\Rbf^\delta\right)^{-1} \sqbracket{\int_{\xbf\in\Xcal} \rbf^{\delta}(\xbf)\phi_i^*(\xbf)d\xbf} \sqbracket{\int_{\xbf\in\Xcal} \rbf^{\delta}(\xbf) \phi_i^*(\xbf)d\xbf}^T \left(\Rbf^\delta\right)^{-1} \deltabmz,
	\end{align}
	where the first equality is by $a_i^* = (\mu_i^*)^2/(\lambda_i^*\sigma_\delta^2) $ and $ \mu_i^* = \int_{\xbf\in\Xcal} \mu^{*\delta}(\xbf)\phi_i^*(\xbf)d\xbf $; the second equality follows from \cref{equ:pred_mean_gasp}.

	From \cref{equ:integrate_z_assume_f}, \cref{equ:bone_mgf} and \cref{equ:a_sigma_lambda_i}, one has
	{\begin{align}
			p_{\delta_z}(\deltabmz\mid\Thetabm)
			&\propto \exp\left\{-\frac{1}{2}\deltabmz^T \sqbracket{\sigma_\delta^{-2} \Bbf + \left(\sigma_\delta^2\Rbf^\delta\right)^{-1}} \deltabmz\right\} = \exp\left\{-\frac{1}{2\sigma_\delta^2}\deltabmz^T \sqbracket{\Bbf + \left(\Rbf^\delta\right)^{-1}} \deltabmz\right\},
	\end{align}}
	where
	\begin{equation}
		{\mathbf B= (\mathbf R^{\delta})^{-1} \left\{\sum^{\infty}_{i=1} \frac{\lambda}{\Vol(\mathcal X)+\lambda^*_i \lambda } \left( \int_{{\xbf\in\Xcal}} \mathbf r^{\delta}(\mathbf x) \phi^*_i(\mathbf x) d\mathbf x \right)\left( \int_{{\xbf\in\Xcal}} \mathbf r^{\delta}(\mathbf x) \phi^*_i(\mathbf x) d\mathbf x\right)^T \right\} (\mathbf R^{\delta})^{-1},}
	\end{equation}
	Hence, $ \deltabmz $ follows a multivariate normal distribution
	\begin{equation}
		\deltabmz\mid\Thetabm\sim MN\parenth{\zerobm, \sigma_\delta^2\Rbf_z},
	\end{equation}
	with {$ \Rbf_z = \sqbracket{\Bbf + \left(\Rbf^\delta\right)^{-1}}^{-1} $}.
\end{proof}

\begin{proof}[proof of \cref{lemma:approx_delta_z}]
	For \cref{equ:scaled_GP_approx}, one has
	\begin{equation}
		p_{\delta^a_z}(\bm \delta^a_z \mid\Thetabm) = \frac{\bonea}{\bzeroa} p_\delta(\bm \delta^a_z \mid\Thetabm),
		\label{equ:delta_z_a}
	\end{equation}
	where
	\begin{align}
		\bzeroa&= \int_0^\infty p_\delta\parenth{\sumdeltat\mid\Thetabm} f_Z( t\mid\Thetabm) dt,\\
		\bonea&= \int_0^\infty p_\delta\parenth{\sumdelta\mid\deltabmz,\Thetabm} f_Z(z\mid\bm \Theta) dz.
	\end{align}
	Applying equations \cref{equ:pred_mean_gasp} and \cref{equ:c_star_gasp} on $ \xbf_1^C, \dots, \xbf_{N_C}^C $, we have
	\begin{equation}
		\parenth{\delta(\xbf_1^C), \dots, \delta(\xbf_{N_C}^C)}^T \mid \deltabmz,\Thetabm \sim MN(\mubm^{*C}, \sigma_\delta^2\Rbf^{*C}),
	\end{equation}
	where
	\begin{equation}
		\mubm^{*C} = \rbf^C \parenth{\Rbf^\delta}^{-1} \deltabmz,
		\label{equ:mu_star_C}
	\end{equation}
	and
	\begin{equation}
		\Rbf^{*C} = \Rbf^C - \rbf^C \parenth{\Rbf^\delta}^{-1} \trans{\rbf^C}.
	\end{equation}
	Denote the eigen-decomposition of $ \Rbf^{*C} $ as $ \Rbf^{*C} = \Ubf^C\Lambdabm^C\parenth{\Ubf^C}^T $, where the $i$-th diagonal element of $ \Lambdabm^C $ is $ \lambda_i^C $. From the proof of \cref{lemma:chi-square}, one has
	\begin{equation}
		\sum_{i=1}^{N_C}\delta\parenth{\xbf_i^C}^2\mid\deltabmz,\Thetabm \sim \sigma^2_\delta\sum_{i=1}^{N_C} \lambda_i^C\chi^2_{a_i^C}(1),
	\end{equation}
	where $ a_i^C = \sqbracket{\parenth{\Ubf_i^C}^T\mubm^{*C}}^2 / \parenth{\lambda_i^C\sigma_\delta^2} $ and $ \Ubf_i^C $ is the $i$-th column of $ \Ubf^C $. Similar to the proof in \cref{lemma:density_delta_z}, defining $ \tilde{\lambda}^C = \Delta x \tilde{\lambda} = \lambda / (2\sigma_\delta^2 N_C) $, we can write $ \bonea $ as
	\begin{equation}
			\bonea = \tilde{\lambda}^C M_{S^C}(-\tilde{\lambda}^C) =\frac{\lambda}{2\sigma_\delta^2N_C} \sqbracket{\prod_{i=1}^{{N_C}} \left(1+ { \frac{\lambda}{N_C}} \lambda_i^C\right)^{-1/2}} e^{-\sum_{i=1}^{{N_C}} \frac{a_i^C{\lambda}\lambda_i^C}{2( N_C+\lambda\lambda_i^C)}},
			\label{equ:bonea_mfg}
	\end{equation}
	where $ S^C = \sigma_\delta^2 \sum_{i=1}^{N_C} \lambda_i^C X_i^C $ with $X_i^C $ following a non-central chi-squared distribution with 1 degree of freedom and the non-central parameter $ a_i^C $.
	
	Define a diagonal matrix $ \Lambdabm^{*C} $ with elements
	{\begin{equation}
			\sqbracket{\Lambdabm^{*C}}_{ii} = \frac{\lambda}{N_C+\lambda\lambda_i^C},
	\end{equation}}
	then the following term in \cref{equ:bonea_mfg} becomes
	{\begin{equation}
			-\sum_{i=1}^{{N_C}} \frac{a_i^C{\lambda}\lambda_i^C}{2( N_C+\lambda\lambda_i^C)} = -\frac{1}{2\sigma_\delta^2} \parenth{\mubm^{*C}}^T \Ubf^C\Lambdabm^{*C}\trans{\Ubf^C} \mubm^{*C}.
			\label{equ:idk}
	\end{equation}}
	Combining \cref{equ:delta_z_a}, \cref{equ:mu_star_C}, \cref{equ:bonea_mfg} and \cref{equ:idk}, it is easy to see that
	\begin{equation}
		p_{\delta_z}(\deltabmz^a\mid\Thetabm) \propto \exp\left\{-\frac{1}{2\sigma_\delta^2}\deltabmz^T\sqbracket{\inv{\Rbf^\delta} \trans{\rbf^C} \Ubf^C\Lambdabm^{*C}\trans{\Ubf^C} \rbf^C \inv{\Rbf^\delta}+\inv{\Rbf^\delta}} \deltabmz\right\}.
	\end{equation}
	Applying Woodbury matrix identity gives
	\begin{align}
		&\sqbracket{\inv{\Rbf^\delta} \trans{\rbf^C} \Ubf^C\Lambdabm^{*C}\trans{\Ubf^C} \rbf^C \inv{\Rbf^\delta}+\inv{\Rbf^\delta}}^{-1} \nonumber\\
		&=\Rbf^\delta - \trans{\rbf^C}\Ubf^C \sqbracket{\inv{\Lambdabm^{*C}} + \trans{\Ubf^C}{\rbf^C}\inv{\Rbf^\delta}\trans{\rbf^C}\Ubf^C}^{-1} \trans{\Ubf^C} {\rbf^C}.
		\label{equ:delta_z_a_var_woodbury}
	\end{align}
	Note that {$ \inv{\Lambdabm^{*C}} = N_C \Ibf_{N_C} / \lambda + \Lambdabm^C $}. Hence, \cref{equ:delta_z_a_var_woodbury} becomes
	{\begin{align}
			&\Rbf^\delta - \trans{\rbf^C}\Ubf^C \sqbracket{\frac{N_C}{\lambda}\Ibf_{N_C} + \Lambdabm^C + \trans{\Ubf^C}{\rbf^C}\inv{\Rbf^\delta}\trans{\rbf^C}\Ubf^C}^{-1} \trans{\Ubf^C} {\rbf^C} \nonumber\\
			&=\Rbf^\delta - \trans{\rbf^C}\Ubf^C \sqbracket{\frac{N_C}{\lambda}\Ibf_{N_C} + \trans{\Ubf^C}\parenth{\Rbf^{*C}+{\rbf^C}\inv{\Rbf^\delta}\trans{\rbf^C}}\Ubf^C}^{-1}\trans{\Ubf^C} {\rbf^C} \nonumber\\
			&=\Rbf^\delta - \trans{\rbf^C}\Ubf^C \sqbracket{\frac{N_C}{\lambda}\Ibf_{N_C} + \trans{\Ubf^C}\Rbf^C\Ubf^C}^{-1}\trans{\Ubf^C} {\rbf^C} \nonumber\\
			&=\Rbf^\delta - \trans{\rbf^C} \sqbracket{\Ubf^C\parenth{\frac{N_C}{\lambda}\Ibf_{N_C} + \trans{\Ubf^C}\Rbf^C\Ubf^C}\trans{\Ubf^C}}^{-1} {\rbf^C} \nonumber\\
			&=\Rbf^\delta - \trans{\rbf^C}\parenth{\frac{N_C}{\lambda}\Ibf_{N_C} + \Rbf^C}^{-1} {\rbf^C} \nonumber\\
			&= \Rbf_z^a
	\end{align}}
	Therefore, we conclude that
	\begin{equation}
		\deltabmz^a \mid \Thetabm \sim MN(\zerobm, \sigma_\delta^2\Rbf_z^a).
	\end{equation}
\end{proof}

\section*{{Acknowledgment}} The authors would like to thank Kyle Anderson for providing the real example in the numerical study. The authors thank the editor, the associate editor and two referees for their comments that substantially improve the article.

\bibliographystyle{siamplain}
\bibliography{references}
\end{document}